\documentclass[sigconf,final,screen,nonacm]{acmart}

\AtBeginDocument{%
  }

\setcopyright{none} 

\usepackage{ifdraft}
\usepackage{amsmath,amsfonts}

\usepackage{algorithmic}
\usepackage{graphicx}
\graphicspath{{./images/}}
\usepackage[abbreviations]{foreign}

\usepackage{textcomp}
\usepackage{color}
\usepackage{xcolor,colortbl}
\usepackage{enumitem}
\usepackage[htt]{hyphenat}
\usepackage{xspace}
\usepackage{floatflt}
\usepackage{placeins}
\newcommand{\hdr}[1]{\textbf{#1.}\xspace}
\usepackage{svg} 

\usepackage[scaled=0.95]{newtxtt}
\usepackage[T1]{fontenc}
\usepackage{fontawesome}
\usepackage{pifont}

\usepackage[font=bf]{caption}
\usepackage{subcaption}
\usepackage{graphicx}
\graphicspath{{./images/}}
\usepackage{rotating}

\usepackage{tabularx}
\usepackage{threeparttable}
\usepackage{makecell}

\usepackage{booktabs,nicematrix} 
\usepackage{multirow}
\usepackage{wrapfig}
\usepackage[noend,ruled,vlined]{algorithm2e}

\definecolor{cites}{RGB}{150,0,0}
\definecolor{links}{RGB}{0,150,0}

\hypersetup{
    colorlinks=true,
    linkcolor={links},
    citecolor={cites},
  }

\usepackage{cleveref}
\crefformat{section}{#2§#1#3}
\crefname{section}{§}{§§}
\Crefname{section}{§}{§§}

\usepackage{glossaries-extra}
\glsdisablehyper
\setabbreviationstyle[acronym]{long-short}

\usepackage{soul}
\usepackage{outlines}
\usepackage{circledsteps} 

\usepackage[commandnameprefix=ifneeded,todonotes={textsize=tiny}]{changes}
\newcommand{\hlblue}[1]{{\tt\color{white}\sethlcolor{ssblue}\hl{#1}}}
\newcommand{\hlgray}[1]{{\tt\sethlcolor{lgray}\hl{#1}}}

\definechangesauthor[color=blue]{kha}
\definechangesauthor[color=teal]{noh}
\definechangesauthor[color=ssblue]{hj}
\definechangesauthor[color=purple]{kw}
\definechangesauthor[color=ssblue]{keypoint}

\definechangesauthor[color=orange]{wip}
\definechangesauthor[color=red]{issue}
\definechangesauthor[color=blue]{commit}

\newcommand{\keypoint}[1]{\added[id=keypoint]{#1}}

\newcommand{\issue}[1]{\chcomment[id=issue]{#1}}

\newcommand{\kha}[1]{\chcomment[id=kha]{#1}}

\newcolumntype{Y}{>{\centering\arraybackslash}X}
\newcolumntype{Z}{>{\raggedleft\arraybackslash}X}

\usepackage{balance}
\usepackage{xspace}

\usepackage{listings}

\definecolor{Ambient}{RGB}{240,240,240}
\definecolor{Private}{RGB}{42,100,168}

\usepackage{pifont}
\usepackage{multirow}
\usepackage{diagbox} 
\usepackage{fontawesome} 
\usepackage[htt]{hyphenat} 
\usepackage{makecell} 

\usepackage{multirow}

\usepackage{glossaries}
\usepackage{glossaries-extra}
\glsdisablehyper
\setabbreviationstyle[acronym]{long-short}

\usepackage{outlines}

\newcommand\circledalph[2][]{\ifmmode
\Circled[fill color=black,inner color=white,#1]{\mathsf{#2}}
\else
\Circled[fill color=black,inner color=white,#1]{\sffamily#2}
\fi
}

\newcommand\circlednum[2][]{\ifmmode
\Circled[fill color=black,inner color=white,#1]{\mathsf{#2}}
\else
\Circled[fill color=black,inner color=white,#1]{\sffamily#2}
\fi
}

\usepackage{listings}
\definecolor{clr-background}{RGB}{255,255,255}
\definecolor{clr-text}{RGB}{0,0,0}
\definecolor{clr-string}{RGB}{163,21,21}
\definecolor{clr-namespace}{RGB}{0,0,0}
\definecolor{clr-preprocessor}{RGB}{128,128,128}
\definecolor{clr-keyword}{RGB}{0,0,255}
\definecolor{clr-type}{RGB}{43,145,175}
\definecolor{clr-variable}{RGB}{0,0,0}
\definecolor{clr-constant}{RGB}{171,0,118} 
\definecolor{clr-comment}{RGB}{0,128,0}

\lstdefinestyle{sslab-wasm}{
language=WebAssembly,
  sensitive=true,
    morestring=[d]{'}, 
    morestring=*[d]{"},
    morestring=[s][]{\#\{}{\}},
  stringstyle=\color{string},
  morekeywords=[1]{i32,f32,i64,f64},
  keywordstyle={[1]\color{magenta}},
  morekeywords=[5]{add, load, get, const, module, func, param, result, global, get_global, mut, set_global, export, import, memory, data, get_local, set_local, elem, table, call,call_indirect, type},
  keywordstyle={[5]\color{keyword}},
  morecomment=**[l][\color{clr-comment}]{;;},
 }

\lstdefinestyle{sslab-c}{
language=C,
 sensitive=true,
  basicstyle=\linespread{0.8}\small\ttfamily,
  rulesepcolor=\color{gray},
  numberstyle=\scriptsize\color{gray},
  keywordstyle=\color{clr-keyword},
  morekeywords={capac_fd_t,capac_open,capac_exit, capac_enter,capac_alloc, DOM_PRIV_STACK, DOM_PRIV_PTR,DOM_PRIV, DOM_ENCRYPT,cap_fd_t,cap_ptr_t,cap_path_t,domain_signature_t}, stringstyle=\color{clr-constant},
  identifierstyle=\color{clr-variable},
  commentstyle=\color{clr-comment},
  emphstyle=\color{clr-type},
  keywordstyle=[3]{\color{magenta}},
  morekeywords=[3]{net_device_ops},
 }

 \lstdefinestyle{sslab-mini-c}{
  language=C,
  sensitive=true,
  basicstyle=\linespread{0.8}\footnotesize\ttfamily,
  rulesepcolor=\color{gray},
  numberstyle=\scriptsize\color{gray},
  keywordstyle=\color{clr-keyword},
  morekeywords={capac_fd_t,capac_open,capac_exit, capac_enter,capac_alloc,  DOM_PRIV_STACK, DOM_PRIV_PTR, CAPAC_DOM,DOM_ENCRYPT, domain_signature_t},
  stringstyle=\color{clr-constant},
  identifierstyle=\color{clr-variable},
  commentstyle=\color{clr-comment},
  emphstyle=\color{clr-type},
  escapeinside={/*!}{!*/},
}

\lstdefinestyle{sslab-capacity-c}{
  language=C,
  sensitive=true,
  basicstyle=\linespread{0.8}\footnotesize\ttfamily,
  rulesepcolor=\color{gray},
  numberstyle=\scriptsize\color{gray},
  keywordstyle=\bfseries,
  morekeywords=[9]{capac_fd_t,capac_open,capac_malloc, capac_free, capac_delegate_ptr, capac_modifier, capac_delegate_fd, capac_limit_fd, capac_init, capac_exit, capac_enter,capac_alloc,DOM_PRIV,  DOM_PRIV_STACK, DOM_PRIV_PTR, CAPAC_DOM,DOM_ENCRYPT, domain_signature_t},
  keywordstyle=[9]\color{blue},
  stringstyle=\color{clr-constant},
  identifierstyle=\color{clr-variable},
  commentstyle=\color{clr-comment},
  emphstyle=\color{clr-type},
  escapeinside={/*!}{!*/},
  }

\lstdefinelanguage[ARM]{Assembler}%
   {morekeywords={b.ne,ret,lsl,adc,adcal,adcals,adccc,adcccs,adccs,adccss,adceq,adceqs,    %
      adcge,adcges,adcgt,adcgts,adchi,adchis,adchs,adchss,adcle,adcles,       %
      adclo,adclos,adcls,adclss,adclt,adclts,adcmi,adcmis,adcne,adcnes,       %
      adcpl,adcpls,adcs,adcvc,adcvcs,adcvs,adcvss,add,addal,addals,addcc,     %
      addccs,addcs,addcss,addeq,addeqs,addge,addges,addgt,addgts,addhi,       %
      addhis,addhs,addhss,addle,addles,addlo,addlos,addls,addlss,addlt,       %
      addlts,addmi,addmis,addne,addnes,addpl,addpls,adds,addvc,addvcs,addvs,  %
      addvss,and,andal,andals,andcc,andccs,andcs,andcss,andeq,andeqs,andge,   %
      andges,andgt,andgts,andhi,andhis,andhs,andhss,andle,andles,andlo,       %
      andlos,andls,andlss,andlt,andlts,andmi,andmis,andne,andnes,andpl,       %
      andpls,ands,andvc,andvcs,andvs,andvss,b,bal,bcc,bcs,beq,bge,bgt,bhi,    %
      bhs,bic,bical,bicals,biccc,bicccs,biccs,biccss,biceq,biceqs,bicge,      %
      bicges,bicgt,bicgts,bichi,bichis,bichs,bichss,bicle,bicles,biclo,       %
      biclos,bicls,biclss,biclt,biclts,bicmi,bicmis,bicne,bicnes,bicpl,       %
      bicpls,bics,bicvc,bicvcs,bicvs,bicvss,bkpt,bl,blal,blcc,blcs,ble,bleq,  %
      blge,blgt,blhi,blhs,blle,bllo,blls,bllt,blmi,blne,blo,blpl,bls,blt,     %
      blvc,blvs,blx,blxal,blxcc,blxcs,blxeq,blxge,blxgt,blxhi,blxhs,blxle,    %
      blxlo,blxls,blxlt,blxmi,blxne,blxpl,blxvc,blxvs,bmi,bne,bpl,bvc,bvs,    %
      bx,bxal,bxcc,bxcs,bxeq,bxge,bxgt,bxhi,bxhs,bxj,bxjal,bxjcc,bxjcs,       %
      bxjeq,bxjge,bxjgt,bxjhi,bxjhs,bxjle,bxjlo,bxjls,bxjlt,bxjmi,bxjne,      %
      bxjpl,bxjvc,bxjvs,bxle,bxlo,bxls,bxlt,bxmi,bxne,bxpl,bxvc,bxvs,cdp,     %
      cdp2,cdpal,cdpcc,cdpcs,cdpeq,cdpge,cdpgt,cdphi,cdphs,cdple,cdplo,       %
      cdpls,cdplt,cdpmi,cdpne,cdppl,cdpvc,cdpvs,clz,clzal,clzcc,clzcs,clzeq,  %
      clzge,clzgt,clzhi,clzhs,clzle,clzlo,clzls,clzlt,clzmi,clzne,clzpl,      %
      clzvc,clzvs,cmn,cmnal,cmncc,cmncs,cmneq,cmnge,cmngt,cmnhi,cmnhs,cmnle,  %
      cmnlo,cmnls,cmnlt,cmnmi,cmnne,cmnpl,cmnvc,cmnvs,cmp,cmpal,cmpcc,cmpcs,  %
      cmpeq,cmpge,cmpgt,cmphi,cmphs,cmple,cmplo,cmpls,cmplt,cmpmi,cmpne,      %
      cmppl,cmpvc,cmpvs,cps,cpsid,cpsie,cpy,cpyal,cpycc,cpycs,cpyeq,cpyge,    %
      cpygt,cpyhi,cpyhs,cpyle,cpylo,cpyls,cpylt,cpymi,cpyne,cpypl,cpyvc,      %
      cpyvs,eor,eoral,eorals,eorcc,eorccs,eorcs,eorcss,eoreq,eoreqs,eorge,    %
      eorges,eorgt,eorgts,eorhi,eorhis,eorhs,eorhss,eorle,eorles,eorlo,       %
      eorlos,eorls,eorlss,eorlt,eorlts,eormi,eormis,eorne,eornes,eorpl,       %
      eorpls,eors,eorvc,eorvcs,eorvs,eorvss,ldc,ldc2,ldcal,ldccc,ldccs,       %
      ldceq,ldcge,ldcgt,ldchi,ldchs,ldcle,ldclo,ldcls,ldclt,ldcmi,ldcne,      %
      ldcpl,ldcvc,ldcvs,ldmalda,ldmaldb,ldmalea,ldmaled,ldmalfa,ldmalfd,      %
      ldmalia,ldmalib,ldmccda,ldmccdb,ldmccea,ldmcced,ldmccfa,ldmccfd,        %
      ldmccia,ldmccib,ldmcsda,ldmcsdb,ldmcsea,ldmcsed,ldmcsfa,ldmcsfd,        %
      ldmcsia,ldmcsib,ldmda,ldmdb,ldmea,ldmed,ldmeqda,ldmeqdb,ldmeqea,        %
      ldmeqed,ldmeqfa,ldmeqfd,ldmeqia,ldmeqib,ldmfa,ldmfd,ldmgeda,ldmgedb,    %
      ldmgeea,ldmgeed,ldmgefa,ldmgefd,ldmgeia,ldmgeib,ldmgtda,ldmgtdb,        %
      ldmgtea,ldmgted,ldmgtfa,ldmgtfd,ldmgtia,ldmgtib,ldmhida,ldmhidb,        %
      ldmhiea,ldmhied,ldmhifa,ldmhifd,ldmhiia,ldmhiib,ldmhsda,ldmhsdb,        %
      ldmhsea,ldmhsed,ldmhsfa,ldmhsfd,ldmhsia,ldmhsib,ldmia,ldmib,ldmleda,    %
      ldmledb,ldmleea,ldmleed,ldmlefa,ldmlefd,ldmleia,ldmleib,ldmloda,        %
      ldmlodb,ldmloea,ldmloed,ldmlofa,ldmlofd,ldmloia,ldmloib,ldmlsda,        %
      ldmlsdb,ldmlsea,ldmlsed,ldmlsfa,ldmlsfd,ldmlsia,ldmlsib,ldmltda,        %
      ldmltdb,ldmltea,ldmlted,ldmltfa,ldmltfd,ldmltia,ldmltib,ldmmida,        %
      ldmmidb,ldmmiea,ldmmied,ldmmifa,ldmmifd,ldmmiia,ldmmiib,ldmneda,        %
      ldmnedb,ldmneea,ldmneed,ldmnefa,ldmnefd,ldmneia,ldmneib,ldmplda,        %
      ldmpldb,ldmplea,ldmpled,ldmplfa,ldmplfd,ldmplia,ldmplib,ldmvcda,        %
      ldmvcdb,ldmvcea,ldmvced,ldmvcfa,ldmvcfd,ldmvcia,ldmvcib,ldmvsda,        %
      ldmvsdb,ldmvsea,ldmvsed,ldmvsfa,ldmvsfd,ldmvsia,ldmvsib,ldr,ldral,      %
      ldralb,ldralbt,ldrald,ldralh,ldralsb,ldralsh,ldralt,ldrb,ldrbt,ldrcc,   %
      ldrccb,ldrccbt,ldrccd,ldrcch,ldrccsb,ldrccsh,ldrcct,ldrcs,ldrcsb,       %
      ldrcsbt,ldrcsd,ldrcsh,ldrcssb,ldrcssh,ldrcst,ldrd,ldreq,ldreqb,         %
      ldreqbt,ldreqd,ldreqh,ldreqsb,ldreqsh,ldreqt,ldrex,ldrexal,ldrexcc,     %
      ldrexcs,ldrexeq,ldrexge,ldrexgt,ldrexhi,ldrexhs,ldrexle,ldrexlo,        %
      ldrexls,ldrexlt,ldrexmi,ldrexne,ldrexpl,ldrexvc,ldrexvs,ldrge,ldrgeb,   %
      ldrgebt,ldrged,ldrgeh,ldrgesb,ldrgesh,ldrget,ldrgt,ldrgtb,ldrgtbt,      %
      ldrgtd,ldrgth,ldrgtsb,ldrgtsh,ldrgtt,ldrh,ldrhi,ldrhib,ldrhibt,ldrhid,  %
      ldrhih,ldrhisb,ldrhish,ldrhit,ldrhs,ldrhsb,ldrhsbt,ldrhsd,ldrhsh,       %
      ldrhssb,ldrhssh,ldrhst,ldrle,ldrleb,ldrlebt,ldrled,ldrleh,ldrlesb,      %
      ldrlesh,ldrlet,ldrlo,ldrlob,ldrlobt,ldrlod,ldrloh,ldrlosb,ldrlosh,      %
      ldrlot,ldrls,ldrlsb,ldrlsbt,ldrlsd,ldrlsh,ldrlssb,ldrlssh,ldrlst,       %
      ldrlt,ldrltb,ldrltbt,ldrltd,ldrlth,ldrltsb,ldrltsh,ldrltt,ldrmi,        %
      ldrmib,ldrmibt,ldrmid,ldrmih,ldrmisb,ldrmish,ldrmit,ldrne,ldrneb,       %
      ldrnebt,ldrned,ldrneh,ldrnesb,ldrnesh,ldrnet,ldrpl,ldrplb,ldrplbt,      %
      ldrpld,ldrplh,ldrplsb,ldrplsh,ldrplt,ldrsb,ldrsh,ldrt,ldrvc,ldrvcb,     %
      ldrvcbt,ldrvcd,ldrvch,ldrvcsb,ldrvcsh,ldrvct,ldrvs,ldrvsb,ldrvsbt,      %
      ldrvsd,ldrvsh,ldrvssb,ldrvssh,ldrvst,mar,maral,marcc,marcs,mareq,       %
      marge,margt,marhi,marhs,marle,marlo,marls,marlt,marmi,marne,marpl,      %
      marvc,marvs,mcr,mcr2,mcral,mcrcc,mcrcs,mcreq,mcrge,mcrgt,mcrhi,mcrhs,   %
      mcrle,mcrlo,mcrls,mcrlt,mcrmi,mcrne,mcrpl,mcrr,mcrr2,mcrral,mcrrcc,     %
      mcrrcs,mcrreq,mcrrge,mcrrgt,mcrrhi,mcrrhs,mcrrle,mcrrlo,mcrrls,mcrrlt,  %
      mcrrmi,mcrrne,mcrrpl,mcrrvc,mcrrvs,mcrvc,mcrvs,mia,miaal,miacc,miacs,   %
      miaeq,miage,miagt,miahi,miahs,miale,mialo,mials,mialt,miami,miane,      %
      miaph,miaphal,miaphcc,miaphcs,miapheq,miaphge,miaphgt,miaphhi,miaphhs,  %
      miaphle,miaphlo,miaphls,miaphlt,miaphmi,miaphne,miaphpl,miaphvc,        %
      miaphvs,miapl,miavc,miavs,miaxy,miaxyal,miaxycc,miaxycs,miaxyeq,        %
      miaxyge,miaxygt,miaxyhi,miaxyhs,miaxyle,miaxylo,miaxyls,miaxylt,        %
      miaxymi,miaxyne,miaxypl,miaxyvc,miaxyvs,mla,mlaal,mlaals,mlacc,mlaccs,  %
      mlacs,mlacss,mlaeq,mlaeqs,mlage,mlages,mlagt,mlagts,mlahi,mlahis,       %
      mlahs,mlahss,mlale,mlales,mlalo,mlalos,mlals,mlalss,mlalt,mlalts,       %
      mlami,mlamis,mlane,mlanes,mlapl,mlapls,mlas,mlavc,mlavcs,mlavs,mlavss,  %
      mov,moval,movals,movcc,movccs,movcs,movcss,moveq,moveqs,movge,movges,   %
      movgt,movgts,movhi,movhis,movhs,movhss,movle,movles,movlo,movlos,       %
      movls,movlss,movlt,movlts,movmi,movmis,movne,movnes,movpl,movpls,movs,  %
      movvc,movvcs,movvs,movvss,mra,mraal,mracc,mracs,mraeq,mrage,mragt,      %
      mrahi,mrahs,mrale,mralo,mrals,mralt,mrami,mrane,mrapl,mravc,mravs,mrc,  %
      mrc2,mrcal,mrccc,mrccs,mrceq,mrcge,mrcgt,mrchi,mrchs,mrcle,mrclo,       %
      mrcls,mrclt,mrcmi,mrcne,mrcpl,mrcvc,mrcvs,mrrc,mrrc2,mrrcal,mrrccc,     %
      mrrccs,mrrceq,mrrcge,mrrcgt,mrrchi,mrrchs,mrrcle,mrrclo,mrrcls,mrrclt,  %
      mrrcmi,mrrcne,mrrcpl,mrrcvc,mrrcvs,mrs,mrsal,mrscc,mrscs,mrseq,mrsge,   %
      mrsgt,mrshi,mrshs,mrsle,mrslo,mrsls,mrslt,mrsmi,mrsne,mrspl,mrsvc,      %
      mrsvs,msr,msral,msrcc,msrcs,msreq,msrge,msrgt,msrhi,msrhs,msrle,msrlo,  %
      msrls,msrlt,msrmi,msrne,msrpl,msrvc,msrvs,mul,mulal,mulals,mulcc,       %
      mulccs,mulcs,mulcss,muleq,muleqs,mulge,mulges,mulgt,mulgts,mulhi,       %
      mulhis,mulhs,mulhss,mulle,mulles,mullo,mullos,mulls,mullss,mullt,       %
      mullts,mulmi,mulmis,mulne,mulnes,mulpl,mulpls,muls,mulvc,mulvcs,mulvs,  %
      mulvss,mvn,mvnal,mvnals,mvncc,mvnccs,mvncs,mvncss,mvneq,mvneqs,mvnge,   %
      mvnges,mvngt,mvngts,mvnhi,mvnhis,mvnhs,mvnhss,mvnle,mvnles,mvnlo,       %
      mvnlos,mvnls,mvnlss,mvnlt,mvnlts,mvnmi,mvnmis,mvnne,mvnnes,mvnpl,       %
      mvnpls,mvns,mvnvc,mvnvcs,mvnvs,mvnvss,nop,orral,orrals,orrcc,       %
      orrccs,orrcs,orrcss,orreq,orreqs,orrge,orrges,orrgt,orrgts,orrhi,       %
      orrhis,orrhs,orrhss,orrle,orrles,orrlo,orrlos,orrls,orrlss,orrlt,       %
      orrlts,orrmi,orrmis,orrne,orrnes,orrpl,orrpls,orrs,orrvc,orrvcs,orrvs,  %
      orrvss,pkhbt,pkhbtal,pkhbtcc,pkhbtcs,pkhbteq,pkhbtge,pkhbtgt,pkhbthi,   %
      pkhbths,pkhbtle,pkhbtlo,pkhbtls,pkhbtlt,pkhbtmi,pkhbtne,pkhbtpl,        %
      pkhbtvc,pkhbtvs,pkhtb,pkhtbal,pkhtbcc,pkhtbcs,pkhtbeq,pkhtbge,pkhtbgt,  %
      pkhtbhi,pkhtbhs,pkhtble,pkhtblo,pkhtbls,pkhtblt,pkhtbmi,pkhtbne,        %
      pkhtbpl,pkhtbvc,pkhtbvs,pld,pop,popal,popcc,popcs,popeq,popge,popgt,    %
      pophi,pophs,pople,poplo,popls,poplt,popmi,popne,poppl,popvc,popvs,      %
      push,pushal,pushcc,pushcs,pusheq,pushge,pushgt,pushhi,pushhs,pushle,    %
      pushlo,pushls,pushlt,pushmi,pushne,pushpl,pushvc,pushvs,qadd,qadd16,    %
      qadd16al,qadd16cc,qadd16cs,qadd16eq,qadd16ge,qadd16gt,qadd16hi,         %
      qadd16hs,qadd16le,qadd16lo,qadd16ls,qadd16lt,qadd16mi,qadd16ne,         %
      qadd16pl,qadd16vc,qadd16vs,qadd8,qadd8al,qadd8cc,qadd8cs,qadd8eq,       %
      qadd8ge,qadd8gt,qadd8hi,qadd8hs,qadd8le,qadd8lo,qadd8ls,qadd8lt,        %
      qadd8mi,qadd8ne,qadd8pl,qadd8vc,qadd8vs,qaddal,qaddcc,qaddcs,qaddeq,    %
      qaddge,qaddgt,qaddhi,qaddhs,qaddle,qaddlo,qaddls,qaddlt,qaddmi,qaddne,  %
      qaddpl,qaddsubx,qaddsubxal,qaddsubxcc,qaddsubxcs,qaddsubxeq,            %
      qaddsubxge,qaddsubxgt,qaddsubxhi,qaddsubxhs,qaddsubxle,qaddsubxlo,      %
      qaddsubxls,qaddsubxlt,qaddsubxmi,qaddsubxne,qaddsubxpl,qaddsubxvc,      %
      qaddsubxvs,qaddvc,qaddvs,qdadd,qdaddal,qdaddcc,qdaddcs,qdaddeq,         %
      qdaddge,qdaddgt,qdaddhi,qdaddhs,qdaddle,qdaddlo,qdaddls,qdaddlt,        %
      qdaddmi,qdaddne,qdaddpl,qdaddvc,qdaddvs,qdsub,qdsubal,qdsubcc,qdsubcs,  %
      qdsubeq,qdsubge,qdsubgt,qdsubhi,qdsubhs,qdsuble,qdsublo,qdsubls,        %
      qdsublt,qdsubmi,qdsubne,qdsubpl,qdsubvc,qdsubvs,qsub,qsub16,qsub16al,   %
      qsub16cc,qsub16cs,qsub16eq,qsub16ge,qsub16gt,qsub16hi,qsub16hs,         %
      qsub16le,qsub16lo,qsub16ls,qsub16lt,qsub16mi,qsub16ne,qsub16pl,         %
      qsub16vc,qsub16vs,qsub8,qsub8al,qsub8cc,qsub8cs,qsub8eq,qsub8ge,        %
      qsub8gt,qsub8hi,qsub8hs,qsub8le,qsub8lo,qsub8ls,qsub8lt,qsub8mi,        %
      qsub8ne,qsub8pl,qsub8vc,qsub8vs,qsubaddx,qsubaddxal,qsubaddxcc,         %
      qsubaddxcs,qsubaddxeq,qsubaddxge,qsubaddxgt,qsubaddxhi,qsubaddxhs,      %
      qsubaddxle,qsubaddxlo,qsubaddxls,qsubaddxlt,qsubaddxmi,qsubaddxne,      %
      qsubaddxpl,qsubaddxvc,qsubaddxvs,qsubal,qsubcc,qsubcs,qsubeq,qsubge,    %
      qsubgt,qsubhi,qsubhs,qsuble,qsublo,qsubls,qsublt,qsubmi,qsubne,qsubpl,  %
      qsubvc,qsubvs,rev,rev16,rev16al,rev16cc,rev16cs,rev16eq,rev16ge,        %
      rev16gt,rev16hi,rev16hs,rev16le,rev16lo,rev16ls,rev16lt,rev16mi,        %
      rev16ne,rev16pl,rev16vc,rev16vs,reval,revcc,revcs,reveq,revge,revgt,    %
      revhi,revhs,revle,revlo,revls,revlt,revmi,revne,revpl,revsh,revshal,    %
      revshcc,revshcs,revsheq,revshge,revshgt,revshhi,revshhs,revshle,        %
      revshlo,revshls,revshlt,revshmi,revshne,revshpl,revshvc,revshvs,revvc,  %
      revvs,rfeda,rfedb,rfeea,rfeed,rfefa,rfefd,rfeia,rfeib,rsb,rsbal,        %
      rsbals,rsbcc,rsbccs,rsbcs,rsbcss,rsbeq,rsbeqs,rsbge,rsbges,rsbgt,       %
      rsbgts,rsbhi,rsbhis,rsbhs,rsbhss,rsble,rsbles,rsblo,rsblos,rsbls,       %
      rsblss,rsblt,rsblts,rsbmi,rsbmis,rsbne,rsbnes,rsbpl,rsbpls,rsbs,rsbvc,  %
      rsbvcs,rsbvs,rsbvss,rsc,rscal,rscals,rsccc,rscccs,rsccs,rsccss,rsceq,   %
      rsceqs,rscge,rscges,rscgt,rscgts,rschi,rschis,rschs,rschss,rscle,       %
      rscles,rsclo,rsclos,rscls,rsclss,rsclt,rsclts,rscmi,rscmis,rscne,       %
      rscnes,rscpl,rscpls,rscs,rscvc,rscvcs,rscvs,rscvss,sadd16,sadd16al,     %
      sadd16cc,sadd16cs,sadd16eq,sadd16ge,sadd16gt,sadd16hi,sadd16hs,         %
      sadd16le,sadd16lo,sadd16ls,sadd16lt,sadd16mi,sadd16ne,sadd16pl,         %
      sadd16vc,sadd16vs,sadd8,sadd8al,sadd8cc,sadd8cs,sadd8eq,sadd8ge,        %
      sadd8gt,sadd8hi,sadd8hs,sadd8le,sadd8lo,sadd8ls,sadd8lt,sadd8mi,        %
      sadd8ne,sadd8pl,sadd8vc,sadd8vs,saddsubx,saddsubxal,saddsubxcc,         %
      saddsubxcs,saddsubxeq,saddsubxge,saddsubxgt,saddsubxhi,saddsubxhs,      %
      saddsubxle,saddsubxlo,saddsubxls,saddsubxlt,saddsubxmi,saddsubxne,      %
      saddsubxpl,saddsubxvc,saddsubxvs,sbc,sbcal,sbcals,sbccc,sbcccs,sbccs,   %
      sbccss,sbceq,sbceqs,sbcge,sbcges,sbcgt,sbcgts,sbchi,sbchis,sbchs,       %
      sbchss,sbcle,sbcles,sbclo,sbclos,sbcls,sbclss,sbclt,sbclts,sbcmi,       %
      sbcmis,sbcne,sbcnes,sbcpl,sbcpls,sbcs,sbcvc,sbcvcs,sbcvs,sbcvss,sel,    %
      selal,selcc,selcs,seleq,selge,selgt,selhi,selhs,selle,sello,sells,      %
      sellt,selmi,selne,selpl,selvc,selvs,setend,shadd16,shadd16al,           %
      shadd16cc,shadd16cs,shadd16eq,shadd16ge,shadd16gt,shadd16hi,shadd16hs,  %
      shadd16le,shadd16lo,shadd16ls,shadd16lt,shadd16mi,shadd16ne,shadd16pl,  %
      shadd16vc,shadd16vs,shadd8,shadd8al,shadd8cc,shadd8cs,shadd8eq,         %
      shadd8ge,shadd8gt,shadd8hi,shadd8hs,shadd8le,shadd8lo,shadd8ls,         %
      shadd8lt,shadd8mi,shadd8ne,shadd8pl,shadd8vc,shadd8vs,shaddsubx,        %
      shaddsubxal,shaddsubxcc,shaddsubxcs,shaddsubxeq,shaddsubxge,            %
      shaddsubxgt,shaddsubxhi,shaddsubxhs,shaddsubxle,shaddsubxlo,            %
      shaddsubxls,shaddsubxlt,shaddsubxmi,shaddsubxne,shaddsubxpl,            %
      shaddsubxvc,shaddsubxvs,shsub16,shsub16al,shsub16cc,shsub16cs,          %
      shsub16eq,shsub16ge,shsub16gt,shsub16hi,shsub16hs,shsub16le,shsub16lo,  %
      shsub16ls,shsub16lt,shsub16mi,shsub16ne,shsub16pl,shsub16vc,shsub16vs,  %
      shsub8,shsub8al,shsub8cc,shsub8cs,shsub8eq,shsub8ge,shsub8gt,shsub8hi,  %
      shsub8hs,shsub8le,shsub8lo,shsub8ls,shsub8lt,shsub8mi,shsub8ne,         %
      shsub8pl,shsub8vc,shsub8vs,shsubaddx,shsubaddxal,shsubaddxcc,           %
      shsubaddxcs,shsubaddxeq,shsubaddxge,shsubaddxgt,shsubaddxhi,            %
      shsubaddxhs,shsubaddxle,shsubaddxlo,shsubaddxls,shsubaddxlt,            %
      shsubaddxmi,shsubaddxne,shsubaddxpl,shsubaddxvc,shsubaddxvs,smlad,      %
      smladal,smladcc,smladcs,smladeq,smladge,smladgt,smladhi,smladhs,        %
      smladle,smladlo,smladls,smladlt,smladmi,smladne,smladpl,smladvc,        %
      smladvs,smladx,smladxal,smladxcc,smladxcs,smladxeq,smladxge,smladxgt,   %
      smladxhi,smladxhs,smladxle,smladxlo,smladxls,smladxlt,smladxmi,         %
      smladxne,smladxpl,smladxvc,smladxvs,smlal,smlalal,smlalals,smlalcc,     %
      smlalccs,smlalcs,smlalcss,smlald,smlaldal,smlaldcc,smlaldcs,smlaldeq,   %
      smlaldge,smlaldgt,smlaldhi,smlaldhs,smlaldle,smlaldlo,smlaldls,         %
      smlaldlt,smlaldmi,smlaldne,smlaldpl,smlaldvc,smlaldvs,smlaldx,          %
      smlaldxal,smlaldxcc,smlaldxcs,smlaldxeq,smlaldxge,smlaldxgt,smlaldxhi,  %
      smlaldxhs,smlaldxle,smlaldxlo,smlaldxls,smlaldxlt,smlaldxmi,smlaldxne,  %
      smlaldxpl,smlaldxvc,smlaldxvs,smlaleq,smlaleqs,smlalge,smlalges,        %
      smlalgt,smlalgts,smlalhi,smlalhis,smlalhs,smlalhss,smlalle,smlalles,    %
      smlallo,smlallos,smlalls,smlallss,smlallt,smlallts,smlalmi,smlalmis,    %
      smlalne,smlalnes,smlalpl,smlalpls,smlals,smlalvc,smlalvcs,smlalvs,      %
      smlalvss,smlalxy,smlalxyal,smlalxycc,smlalxycs,smlalxyeq,smlalxyge,     %
      smlalxygt,smlalxyhi,smlalxyhs,smlalxyle,smlalxylo,smlalxyls,smlalxylt,  %
      smlalxymi,smlalxyne,smlalxypl,smlalxyvc,smlalxyvs,smlawy,smlawyal,      %
      smlawycc,smlawycs,smlawyeq,smlawyge,smlawygt,smlawyhi,smlawyhs,         %
      smlawyle,smlawylo,smlawyls,smlawylt,smlawymi,smlawyne,smlawypl,         %
      smlawyvc,smlawyvs,smlaxy,smlaxyal,smlaxycc,smlaxycs,smlaxyeq,smlaxyge,  %
      smlaxygt,smlaxyhi,smlaxyhs,smlaxyle,smlaxylo,smlaxyls,smlaxylt,         %
      smlaxymi,smlaxyne,smlaxypl,smlaxyvc,smlaxyvs,smlsd,smlsdal,smlsdcc,     %
      smlsdcs,smlsdeq,smlsdge,smlsdgt,smlsdhi,smlsdhs,smlsdle,smlsdlo,        %
      smlsdls,smlsdlt,smlsdmi,smlsdne,smlsdpl,smlsdvc,smlsdvs,smlsdx,         %
      smlsdxal,smlsdxcc,smlsdxcs,smlsdxeq,smlsdxge,smlsdxgt,smlsdxhi,         %
      smlsdxhs,smlsdxle,smlsdxlo,smlsdxls,smlsdxlt,smlsdxmi,smlsdxne,         %
      smlsdxpl,smlsdxvc,smlsdxvs,smlsld,smlsldal,smlsldcc,smlsldcs,smlsldeq,  %
      smlsldge,smlsldgt,smlsldhi,smlsldhs,smlsldle,smlsldlo,smlsldls,         %
      smlsldlt,smlsldmi,smlsldne,smlsldpl,smlsldvc,smlsldvs,smlsldx,          %
      smlsldxal,smlsldxcc,smlsldxcs,smlsldxeq,smlsldxge,smlsldxgt,smlsldxhi,  %
      smlsldxhs,smlsldxle,smlsldxlo,smlsldxls,smlsldxlt,smlsldxmi,smlsldxne,  %
      smlsldxpl,smlsldxvc,smlsldxvs,smmla,smmlaal,smmlacc,smmlacs,smmlaeq,    %
      smmlage,smmlagt,smmlahi,smmlahs,smmlale,smmlalo,smmlals,smmlalt,        %
      smmlami,smmlane,smmlapl,smmlar,smmlaral,smmlarcc,smmlarcs,smmlareq,     %
      smmlarge,smmlargt,smmlarhi,smmlarhs,smmlarle,smmlarlo,smmlarls,         %
      smmlarlt,smmlarmi,smmlarne,smmlarpl,smmlarvc,smmlarvs,smmlavc,smmlavs,  %
      smmls,smmlsal,smmlscc,smmlscs,smmlseq,smmlsge,smmlsgt,smmlshi,smmlshs,  %
      smmlsle,smmlslo,smmlsls,smmlslt,smmlsmi,smmlsne,smmlspl,smmlsr,         %
      smmlsral,smmlsrcc,smmlsrcs,smmlsreq,smmlsrge,smmlsrgt,smmlsrhi,         %
      smmlsrhs,smmlsrle,smmlsrlo,smmlsrls,smmlsrlt,smmlsrmi,smmlsrne,         %
      smmlsrpl,smmlsrvc,smmlsrvs,smmlsvc,smmlsvs,smmul,smmulal,smmulcc,       %
      smmulcs,smmuleq,smmulge,smmulgt,smmulhi,smmulhs,smmulle,smmullo,        %
      smmulls,smmullt,smmulmi,smmulne,smmulpl,smmulr,smmulral,smmulrcc,       %
      smmulrcs,smmulreq,smmulrge,smmulrgt,smmulrhi,smmulrhs,smmulrle,         %
      smmulrlo,smmulrls,smmulrlt,smmulrmi,smmulrne,smmulrpl,smmulrvc,         %
      smmulrvs,smmulvc,smmulvs,smuad,smuadal,smuadcc,smuadcs,smuadeq,         %
      smuadge,smuadgt,smuadhi,smuadhs,smuadle,smuadlo,smuadls,smuadlt,        %
      smuadmi,smuadne,smuadpl,smuadvc,smuadvs,smuadx,smuadxal,smuadxcc,       %
      smuadxcs,smuadxeq,smuadxge,smuadxgt,smuadxhi,smuadxhs,smuadxle,         %
      smuadxlo,smuadxls,smuadxlt,smuadxmi,smuadxne,smuadxpl,smuadxvc,         %
      smuadxvs,smull,smullal,smullals,smullcc,smullccs,smullcs,smullcss,      %
      smulleq,smulleqs,smullge,smullges,smullgt,smullgts,smullhi,smullhis,    %
      smullhs,smullhss,smullle,smullles,smulllo,smulllos,smullls,smulllss,    %
      smulllt,smulllts,smullmi,smullmis,smullne,smullnes,smullpl,smullpls,    %
      smulls,smullvc,smullvcs,smullvs,smullvss,smulwy,smulwyal,smulwycc,      %
      smulwycs,smulwyeq,smulwyge,smulwygt,smulwyhi,smulwyhs,smulwyle,         %
      smulwylo,smulwyls,smulwylt,smulwymi,smulwyne,smulwypl,smulwyvc,         %
      smulwyvs,smulxy,smulxyal,smulxycc,smulxycs,smulxyeq,smulxyge,smulxygt,  %
      smulxyhi,smulxyhs,smulxyle,smulxylo,smulxyls,smulxylt,smulxymi,         %
      smulxyne,smulxypl,smulxyvc,smulxyvs,smusd,smusdal,smusdcc,smusdcs,      %
      smusdeq,smusdge,smusdgt,smusdhi,smusdhs,smusdle,smusdlo,smusdls,        %
      smusdlt,smusdmi,smusdne,smusdpl,smusdvc,smusdvs,smusdx,smusdxal,        %
      smusdxcc,smusdxcs,smusdxeq,smusdxge,smusdxgt,smusdxhi,smusdxhs,         %
      smusdxle,smusdxlo,smusdxls,smusdxlt,smusdxmi,smusdxne,smusdxpl,         %
      smusdxvc,smusdxvs,srsda,srsdb,srsea,srsed,srsfa,srsfd,srsia,srsib,      %
      ssat,ssat16,ssat16al,ssat16cc,ssat16cs,ssat16eq,ssat16ge,ssat16gt,      %
      ssat16hi,ssat16hs,ssat16le,ssat16lo,ssat16ls,ssat16lt,ssat16mi,         %
      ssat16ne,ssat16pl,ssat16vc,ssat16vs,ssatal,ssatcc,ssatcs,ssateq,        %
      ssatge,ssatgt,ssathi,ssaths,ssatle,ssatlo,ssatls,ssatlt,ssatmi,ssatne,  %
      ssatpl,ssatvc,ssatvs,ssub16,ssub16al,ssub16cc,ssub16cs,ssub16eq,        %
      ssub16ge,ssub16gt,ssub16hi,ssub16hs,ssub16le,ssub16lo,ssub16ls,         %
      ssub16lt,ssub16mi,ssub16ne,ssub16pl,ssub16vc,ssub16vs,ssub8,ssub8al,    %
      ssub8cc,ssub8cs,ssub8eq,ssub8ge,ssub8gt,ssub8hi,ssub8hs,ssub8le,        %
      ssub8lo,ssub8ls,ssub8lt,ssub8mi,ssub8ne,ssub8pl,ssub8vc,ssub8vs,        %
      ssubaddx,ssubaddxal,ssubaddxcc,ssubaddxcs,ssubaddxeq,ssubaddxge,        %
      ssubaddxgt,ssubaddxhi,ssubaddxhs,ssubaddxle,ssubaddxlo,ssubaddxls,      %
      ssubaddxlt,ssubaddxmi,ssubaddxne,ssubaddxpl,ssubaddxvc,ssubaddxvs,stc,  %
      stc2,stcal,stccc,stccs,stceq,stcge,stcgt,stchi,stchs,stcle,stclo,       %
      stcls,stclt,stcmi,stcne,stcpl,stcvc,stcvs,stmalda,stmaldb,stmalea,      %
      stmaled,stmalfa,stmalfd,stmalia,stmalib,stmccda,stmccdb,stmccea,        %
      stmcced,stmccfa,stmccfd,stmccia,stmccib,stmcsda,stmcsdb,stmcsea,        %
      stmcsed,stmcsfa,stmcsfd,stmcsia,stmcsib,stmda,stmdb,stmea,stmed,        %
      stmeqda,stmeqdb,stmeqea,stmeqed,stmeqfa,stmeqfd,stmeqia,stmeqib,stmfa,  %
      stmfd,stmgeda,stmgedb,stmgeea,stmgeed,stmgefa,stmgefd,stmgeia,stmgeib,  %
      stmgtda,stmgtdb,stmgtea,stmgted,stmgtfa,stmgtfd,stmgtia,stmgtib,        %
      stmhida,stmhidb,stmhiea,stmhied,stmhifa,stmhifd,stmhiia,stmhiib,        %
      stmhsda,stmhsdb,stmhsea,stmhsed,stmhsfa,stmhsfd,stmhsia,stmhsib,stmia,  %
      stmib,stmleda,stmledb,stmleea,stmleed,stmlefa,stmlefd,stmleia,stmleib,  %
      stmloda,stmlodb,stmloea,stmloed,stmlofa,stmlofd,stmloia,stmloib,        %
      stmlsda,stmlsdb,stmlsea,stmlsed,stmlsfa,stmlsfd,stmlsia,stmlsib,        %
      stmltda,stmltdb,stmltea,stmlted,stmltfa,stmltfd,stmltia,stmltib,        %
      stmmida,stmmidb,stmmiea,stmmied,stmmifa,stmmifd,stmmiia,stmmiib,        %
      stmneda,stmnedb,stmneea,stmneed,stmnefa,stmnefd,stmneia,stmneib,        %
      stmplda,stmpldb,stmplea,stmpled,stmplfa,stmplfd,stmplia,stmplib,        %
      stmvcda,stmvcdb,stmvcea,stmvced,stmvcfa,stmvcfd,stmvcia,stmvcib,        %
      stmvsda,stmvsdb,stmvsea,stmvsed,stmvsfa,stmvsfd,stmvsia,stmvsib,str,    %
      stral,stralb,stralbt,strald,stralh,stralt,strb,strbt,strcc,strccb,      %
      strccbt,strccd,strcch,strcct,strcs,strcsb,strcsbt,strcsd,strcsh,        %
      strcst,strd,streq,streqb,streqbt,streqd,streqh,streqt,strex,strexal,    %
      strexcc,strexcs,strexeq,strexge,strexgt,strexhi,strexhs,strexle,        %
      strexlo,strexls,strexlt,strexmi,strexne,strexpl,strexvc,strexvs,strge,  %
      strgeb,strgebt,strged,strgeh,strget,strgt,strgtb,strgtbt,strgtd,        %
      strgth,strgtt,strh,strhi,strhib,strhibt,strhid,strhih,strhit,strhs,     %
      strhsb,strhsbt,strhsd,strhsh,strhst,strle,strleb,strlebt,strled,        %
      strleh,strlet,strlo,strlob,strlobt,strlod,strloh,strlot,strls,strlsb,   %
      strlsbt,strlsd,strlsh,strlst,strlt,strltb,strltbt,strltd,strlth,        %
      strltt,strmi,strmib,strmibt,strmid,strmih,strmit,strne,strneb,strnebt,  %
      strned,strneh,strnet,strpl,strplb,strplbt,strpld,strplh,strplt,strt,    %
      strvc,strvcb,strvcbt,strvcd,strvch,strvct,strvs,strvsb,strvsbt,strvsd,  %
      strvsh,strvst,sub,subal,subals,subcc,subccs,subcs,subcss,subeq,subeqs,  %
      subge,subges,subgt,subgts,subhi,subhis,subhs,subhss,suble,subles,       %
      sublo,sublos,subls,sublss,sublt,sublts,submi,submis,subne,subnes,       %
      subpl,subpls,subs,subvc,subvcs,subvs,subvss,swi,swial,swicc,swics,      %
      swieq,swige,swigt,swihi,swihs,swile,swilo,swils,swilt,swimi,swine,      %
      swipl,swivc,swivs,swp,swpal,swpalb,swpb,swpcc,swpccb,swpcs,swpcsb,      %
      swpeq,swpeqb,swpge,swpgeb,swpgt,swpgtb,swphi,swphib,swphs,swphsb,       %
      swple,swpleb,swplo,swplob,swpls,swplsb,swplt,swpltb,swpmi,swpmib,       %
      swpne,swpneb,swppl,swpplb,swpvc,swpvcb,swpvs,swpvsb,sxtab,sxtab16,      %
      sxtab16al,sxtab16cc,sxtab16cs,sxtab16eq,sxtab16ge,sxtab16gt,sxtab16hi,  %
      sxtab16hs,sxtab16le,sxtab16lo,sxtab16ls,sxtab16lt,sxtab16mi,sxtab16ne,  %
      sxtab16pl,sxtab16vc,sxtab16vs,sxtabal,sxtabcc,sxtabcs,sxtabeq,sxtabge,  %
      sxtabgt,sxtabhi,sxtabhs,sxtable,sxtablo,sxtabls,sxtablt,sxtabmi,        %
      sxtabne,sxtabpl,sxtabvc,sxtabvs,sxtah,sxtahal,sxtahcc,sxtahcs,sxtaheq,  %
      sxtahge,sxtahgt,sxtahhi,sxtahhs,sxtahle,sxtahlo,sxtahls,sxtahlt,        %
      sxtahmi,sxtahne,sxtahpl,sxtahvc,sxtahvs,sxtb,sxtb16,sxtb16al,sxtb16cc,  %
      sxtb16cs,sxtb16eq,sxtb16ge,sxtb16gt,sxtb16hi,sxtb16hs,sxtb16le,         %
      sxtb16lo,sxtb16ls,sxtb16lt,sxtb16mi,sxtb16ne,sxtb16pl,sxtb16vc,         %
      sxtb16vs,sxtbal,sxtbcc,sxtbcs,sxtbeq,sxtbge,sxtbgt,sxtbhi,sxtbhs,       %
      sxtble,sxtblo,sxtbls,sxtblt,sxtbmi,sxtbne,sxtbpl,sxtbvc,sxtbvs,sxth,    %
      sxthal,sxthcc,sxthcs,sxtheq,sxthge,sxthgt,sxthhi,sxthhs,sxthle,sxthlo,  %
      sxthls,sxthlt,sxthmi,sxthne,sxthpl,sxthvc,sxthvs,teq,teqal,teqcc,       %
      teqcs,teqeq,teqge,teqgt,teqhi,teqhs,teqle,teqlo,teqls,teqlt,teqmi,      %
      teqne,teqpl,teqvc,teqvs,tst,tstal,tstcc,tstcs,tsteq,tstge,tstgt,tsthi,  %
      tsths,tstle,tstlo,tstls,tstlt,tstmi,tstne,tstpl,tstvc,tstvs,uadd16,     %
      uadd16al,uadd16cc,uadd16cs,uadd16eq,uadd16ge,uadd16gt,uadd16hi,         %
      uadd16hs,uadd16le,uadd16lo,uadd16ls,uadd16lt,uadd16mi,uadd16ne,         %
      uadd16pl,uadd16vc,uadd16vs,uadd8,uadd8al,uadd8cc,uadd8cs,uadd8eq,       %
      uadd8ge,uadd8gt,uadd8hi,uadd8hs,uadd8le,uadd8lo,uadd8ls,uadd8lt,        %
      uadd8mi,uadd8ne,uadd8pl,uadd8vc,uadd8vs,uaddsubx,uaddsubxal,            %
      uaddsubxcc,uaddsubxcs,uaddsubxeq,uaddsubxge,uaddsubxgt,uaddsubxhi,      %
      uaddsubxhs,uaddsubxle,uaddsubxlo,uaddsubxls,uaddsubxlt,uaddsubxmi,      %
      uaddsubxne,uaddsubxpl,uaddsubxvc,uaddsubxvs,uhadd16,uhadd16al,          %
      uhadd16cc,uhadd16cs,uhadd16eq,uhadd16ge,uhadd16gt,uhadd16hi,uhadd16hs,  %
      uhadd16le,uhadd16lo,uhadd16ls,uhadd16lt,uhadd16mi,uhadd16ne,uhadd16pl,  %
      uhadd16vc,uhadd16vs,uhadd8,uhadd8al,uhadd8cc,uhadd8cs,uhadd8eq,         %
      uhadd8ge,uhadd8gt,uhadd8hi,uhadd8hs,uhadd8le,uhadd8lo,uhadd8ls,         %
      uhadd8lt,uhadd8mi,uhadd8ne,uhadd8pl,uhadd8vc,uhadd8vs,uhaddsubx,        %
      uhaddsubxal,uhaddsubxcc,uhaddsubxcs,uhaddsubxeq,uhaddsubxge,            %
      uhaddsubxgt,uhaddsubxhi,uhaddsubxhs,uhaddsubxle,uhaddsubxlo,            %
      uhaddsubxls,uhaddsubxlt,uhaddsubxmi,uhaddsubxne,uhaddsubxpl,            %
      uhaddsubxvc,uhaddsubxvs,uhsub16,uhsub16al,uhsub16cc,uhsub16cs,          %
      uhsub16eq,uhsub16ge,uhsub16gt,uhsub16hi,uhsub16hs,uhsub16le,uhsub16lo,  %
      uhsub16ls,uhsub16lt,uhsub16mi,uhsub16ne,uhsub16pl,uhsub16vc,uhsub16vs,  %
      uhsub8,uhsub8al,uhsub8cc,uhsub8cs,uhsub8eq,uhsub8ge,uhsub8gt,uhsub8hi,  %
      uhsub8hs,uhsub8le,uhsub8lo,uhsub8ls,uhsub8lt,uhsub8mi,uhsub8ne,         %
      uhsub8pl,uhsub8vc,uhsub8vs,uhsubaddx,uhsubaddxal,uhsubaddxcc,           %
      uhsubaddxcs,uhsubaddxeq,uhsubaddxge,uhsubaddxgt,uhsubaddxhi,            %
      uhsubaddxhs,uhsubaddxle,uhsubaddxlo,uhsubaddxls,uhsubaddxlt,            %
      uhsubaddxmi,uhsubaddxne,uhsubaddxpl,uhsubaddxvc,uhsubaddxvs,umaal,      %
      umaalal,umaalcc,umaalcs,umaaleq,umaalge,umaalgt,umaalhi,umaalhs,        %
      umaalle,umaallo,umaalls,umaallt,umaalmi,umaalne,umaalpl,umaalvc,        %
      umaalvs,umlal,umlalal,umlalals,umlalcc,umlalccs,umlalcs,umlalcss,       %
      umlaleq,umlaleqs,umlalge,umlalges,umlalgt,umlalgts,umlalhi,umlalhis,    %
      umlalhs,umlalhss,umlalle,umlalles,umlallo,umlallos,umlalls,umlallss,    %
      umlallt,umlallts,umlalmi,umlalmis,umlalne,umlalnes,umlalpl,umlalpls,    %
      umlals,umlalvc,umlalvcs,umlalvs,umlalvss,umull,umullal,umullals,        %
      umullcc,umullccs,umullcs,umullcss,umulleq,umulleqs,umullge,umullges,    %
      umullgt,umullgts,umullhi,umullhis,umullhs,umullhss,umullle,umullles,    %
      umulllo,umulllos,umullls,umulllss,umulllt,umulllts,umullmi,umullmis,    %
      umullne,umullnes,umullpl,umullpls,umulls,umullvc,umullvcs,umullvs,      %
      umullvss,uqadd16,uqadd16al,uqadd16cc,uqadd16cs,uqadd16eq,uqadd16ge,     %
      uqadd16gt,uqadd16hi,uqadd16hs,uqadd16le,uqadd16lo,uqadd16ls,uqadd16lt,  %
      uqadd16mi,uqadd16ne,uqadd16pl,uqadd16vc,uqadd16vs,uqadd8,uqadd8al,      %
      uqadd8cc,uqadd8cs,uqadd8eq,uqadd8ge,uqadd8gt,uqadd8hi,uqadd8hs,         %
      uqadd8le,uqadd8lo,uqadd8ls,uqadd8lt,uqadd8mi,uqadd8ne,uqadd8pl,         %
      uqadd8vc,uqadd8vs,uqaddsubx,uqaddsubxal,uqaddsubxcc,uqaddsubxcs,        %
      uqaddsubxeq,uqaddsubxge,uqaddsubxgt,uqaddsubxhi,uqaddsubxhs,            %
      uqaddsubxle,uqaddsubxlo,uqaddsubxls,uqaddsubxlt,uqaddsubxmi,            %
      uqaddsubxne,uqaddsubxpl,uqaddsubxvc,uqaddsubxvs,uqsub16,uqsub16al,      %
      uqsub16cc,uqsub16cs,uqsub16eq,uqsub16ge,uqsub16gt,uqsub16hi,uqsub16hs,  %
      uqsub16le,uqsub16lo,uqsub16ls,uqsub16lt,uqsub16mi,uqsub16ne,uqsub16pl,  %
      uqsub16vc,uqsub16vs,uqsub8,uqsub8al,uqsub8cc,uqsub8cs,uqsub8eq,         %
      uqsub8ge,uqsub8gt,uqsub8hi,uqsub8hs,uqsub8le,uqsub8lo,uqsub8ls,         %
      uqsub8lt,uqsub8mi,uqsub8ne,uqsub8pl,uqsub8vc,uqsub8vs,uqsubaddx,        %
      uqsubaddxal,uqsubaddxcc,uqsubaddxcs,uqsubaddxeq,uqsubaddxge,            %
      uqsubaddxgt,uqsubaddxhi,uqsubaddxhs,uqsubaddxle,uqsubaddxlo,            %
      uqsubaddxls,uqsubaddxlt,uqsubaddxmi,uqsubaddxne,uqsubaddxpl,            %
      uqsubaddxvc,uqsubaddxvs,usad8,usad8al,usad8cc,usad8cs,usad8eq,usad8ge,  %
      usad8gt,usad8hi,usad8hs,usad8le,usad8lo,usad8ls,usad8lt,usad8mi,        %
      usad8ne,usad8pl,usad8vc,usad8vs,usada8,usada8al,usada8cc,usada8cs,      %
      usada8eq,usada8ge,usada8gt,usada8hi,usada8hs,usada8le,usada8lo,         %
      usada8ls,usada8lt,usada8mi,usada8ne,usada8pl,usada8vc,usada8vs,usat,    %
      usat16,usat16al,usat16cc,usat16cs,usat16eq,usat16ge,usat16gt,usat16hi,  %
      usat16hs,usat16le,usat16lo,usat16ls,usat16lt,usat16mi,usat16ne,         %
      usat16pl,usat16vc,usat16vs,usatal,usatcc,usatcs,usateq,usatge,usatgt,   %
      usathi,usaths,usatle,usatlo,usatls,usatlt,usatmi,usatne,usatpl,usatvc,  %
      usatvs,usub16,usub16al,usub16cc,usub16cs,usub16eq,usub16ge,usub16gt,    %
      usub16hi,usub16hs,usub16le,usub16lo,usub16ls,usub16lt,usub16mi,         %
      usub16ne,usub16pl,usub16vc,usub16vs,usub8,usub8al,usub8cc,usub8cs,      %
      usub8eq,usub8ge,usub8gt,usub8hi,usub8hs,usub8le,usub8lo,usub8ls,        %
      usub8lt,usub8mi,usub8ne,usub8pl,usub8vc,usub8vs,usubaddx,usubaddxal,    %
      usubaddxcc,usubaddxcs,usubaddxeq,usubaddxge,usubaddxgt,usubaddxhi,      %
      usubaddxhs,usubaddxle,usubaddxlo,usubaddxls,usubaddxlt,usubaddxmi,      %
      usubaddxne,usubaddxpl,usubaddxvc,usubaddxvs,uxtab,uxtab16,uxtab16al,    %
      uxtab16cc,uxtab16cs,uxtab16eq,uxtab16ge,uxtab16gt,uxtab16hi,uxtab16hs,  %
      uxtab16le,uxtab16lo,uxtab16ls,uxtab16lt,uxtab16mi,uxtab16ne,uxtab16pl,  %
      uxtab16vc,uxtab16vs,uxtabal,uxtabcc,uxtabcs,uxtabeq,uxtabge,uxtabgt,    %
      uxtabhi,uxtabhs,uxtable,uxtablo,uxtabls,uxtablt,uxtabmi,uxtabne,        %
      uxtabpl,uxtabvc,uxtabvs,uxtah,uxtahal,uxtahcc,uxtahcs,uxtaheq,uxtahge,  %
      uxtahgt,uxtahhi,uxtahhs,uxtahle,uxtahlo,uxtahls,uxtahlt,uxtahmi,        %
      uxtahne,uxtahpl,uxtahvc,uxtahvs,uxtb,uxtb16,uxtb16al,uxtb16cc,          %
      uxtb16cs,uxtb16eq,uxtb16ge,uxtb16gt,uxtb16hi,uxtb16hs,uxtb16le,         %
      uxtb16lo,uxtb16ls,uxtb16lt,uxtb16mi,uxtb16ne,uxtb16pl,uxtb16vc,         %
      uxtb16vs,uxtbal,uxtbcc,uxtbcs,uxtbeq,uxtbge,uxtbgt,uxtbhi,uxtbhs,       %
      uxtble,uxtblo,uxtbls,uxtblt,uxtbmi,uxtbne,uxtbpl,uxtbvc,uxtbvs,uxth,    %
      uxthal,uxthcc,uxthcs,uxtheq,uxthge,uxthgt,uxthhi,uxthhs,uxthle,uxthlo,  %
      uxthls,uxthlt,uxthmi,uxthne,uxthpl,uxthvc,uxthvs,orr},%
    morekeywords=[3]{CAPAC_VAR,CAPAC_STACK,read,encrypt,.prologue,.epilogue,open,capac\_malloc,.encrypt_loop,.auth_failed,.function_start},%
    morekeywords=[2]{pacdb,pacdzb,autdzb,pacdza,autdza,autdb,pacda,pacib,autda,tg,stzg,stg,retab,curr_dom,DST},
    alsoletter={.,0,1,2,3,4,5,6,7,8,9},%
    alsodigit={?},%
    sensitive=false,%
    morestring=[b]",%
    morecomment=[s]{/*}{*/},%
    morecomment=[l]@,%
    morecomment=[l]//,%
    keywordstyle={[3]\color{clr-keyword}},%
    keywordstyle={[4]\color{magenta}},%
    keywordstyle={[2]\bfseries},%
    keywordstyle={\color{black}}%
   }[keywords,comments,strings]

\usepackage{tikz}
\usetikzlibrary{arrows.meta}

\newcounter{tmlistings}

\newcommand\makenode[2]{%
  \tikz[baseline=0pt, remember picture] { \node[fill=gray!50,thick,rounded corners,anchor=base,#1/.try] (listings-\the\value{tmlistings}) {#2}; }%
  \stepcounter{tmlistings}%
}

\tikzset{
  keyword/.style={
    fill=gray!75,
    draw=black
  }
}

\usepackage[most]{tcolorbox}
\tcbuselibrary{listings}

\usetikzlibrary{
  arrows.meta,
  tikzmark,
  calc,
  positioning
}
\usetikzmarklibrary{listings}

\usepackage{arydshln}
\usepackage{tabu}

\newcommand*{\thename}[0]{\textsc{Capacity}\xspace}
\newcommand*{\thenameplain}[0]{\textsc{Capacity}\xspace}
\newcommand*{\libcapac}[0]{\texttt{libcapacity}\xspace}

\newacronym{capability}{capability-based access control}{capability}
\newacronym{syscall}{syscall}{system call}
\newacronym{pa}{PA}{Pointer Authentication}
\newacronym{pac}{PAC}{Pointer Authentication Code}
\newacronym{tbi}{TBI}{Top-Byte Ignore}
\newacronym{ipi}{\textcolor{red}{?????IPI?????}}{Intra-Process Isolation}
\newacronym{pku}{PKU}{Protection Keys for Userspace}
\newacronym{cfi}{CFI}{Control-Flow Integrity}
\newacronym{lsm}{LSM}{Linux Security Modules}
\newacronym{aslr}{ASLR}{Address Space Layout Randomization}
\newacronym{hmac}{HMAC}{Hash-based Message Authentication Code}
\newacronym{mte}{MTE}{Memory Tagging Extension}
\newacronym{acl}{ACL}{Access Control List}
\newacronym{dst}{DST}{Domain Signature Table}
\newacronym{uuid}{UUID}{Uniquely Identifiable IDentification}
\newacronym{sfi}{SFI}{Software Fault Isolation}
\newacronym{ipc}{IPC}{Inter-Process Communication}
\newacronym{pdg}{PDG}{Program Dependence Graph}
\newacronym{iot}{IoT}{Internet of Things}
\newacronym{dpi}{DPI}{Data Pointer Integrity}
\newacronym{mmu}{MMU}{Memory Management Unit}
\newacronym{mpu}{MPU}{Memory Protection Unit}
\newacronym{ac}{AC}{Authentication Code}
\newacronym{ir}{IR}{Intermediate Representation}
\newacronym{mir}{MIR}{Machine IR}

\newcommand*{\pac}[3]{\sloppy\texttt{PAC}({${\textbf{K}_{#1}}$}, {#2}, {#3})}

\newcommand*{\packey}[1]{\sloppy\texttt{PAC}$_{\textbf{K}_{#1}}$}

\newcommand*{\autkey}[1]{\sloppy\texttt{AUT}$_{\textbf{K}_{#1}}$}

\newcommand*{\aut}[3]{\sloppy\texttt{AUT}({${\textbf{K}_{#1}}$}, {#2}, {#3})}

\newcommand*{\tagnum}[1]{\textbf{T}$_{#1}$}

\newcommand*{\tagz}[2]{\texttt{TAG}(#2, $\textbf{T}_{#1}$, \texttt{size})}

\newcommand*{\key}[1]{\textbf{K}$_{#1}$\xspace}

\newcommand*{\fd}[0]{\texttt{FD}\xspace}
\newcommand*{\ptr}[0]{\texttt{PTR}\xspace}
\newcommand*{\fds}[0]{\texttt{FD}s\xspace}

\newcommand*{\cpath}[0]{\texttt{PATH}\xspace}

\newcommand\action[2][]{
\Circled[fill color=ssblue,inner color=white,#1]{\scriptsize\sffamily #2}
}

\newcommand\actionambient[2][]{
\Circled[fill color=lgray!90,inner color=black,#1]{\scriptsize\sffamily #2}
}

\newcommand{\domconn}{\textsc{Connection}\xspace}
\newcommand{\domhs}{\textsc{Handshake}\xspace}
\newcommand{\domses}{\textsc{Session}\xspace}
\newcommand{\domamb}{\textsc{Ambient}\xspace}

\newcommand\secreq[2][]{\ifmmode
\Circled[fill color=ssblue,inner color=white,#1]{\scriptsize\mathsf{#2}}
\else
\Circled[fill color=white,inner color=black,#1]{\scriptsize\bfseries\sffamily#2}
\fi
}

\newcommand{\srimper}[0]{\secreq{R1}}
\newcommand{\srimpers}[0]{\secreq{R1}}
\newcommand{\srcompmed}[0]{\secreq{R2}}
\newcommand{\srreuse}[0]{\secreq{R3}}
\newcommand{\srforge}[0]{\secreq{R4}}

\definecolor{allowed}{RGB}{0,150,0}
\definecolor{kindof}{RGB}{200,200,0}
\definecolor{lgray}{gray}{0.8}
\definecolor{ssblue}{RGB}{42,100,168}

\newcommand*{\pamte}[0]{\gls{pa}+\gls{mte}\xspace}

\begin{document}
\definecolor{cites}{RGB}{150,0,0}
\definecolor{links}{RGB}{0,180,0}

\hypersetup{
    colorlinks=true,
    linkcolor={links},
    citecolor={cites},
    urlcolor={black}
  }


\title{\thenameplain: Cryptographically-Enforced In-Process Capabilities \\ for Modern ARM Architectures \\ (Extended Version)}

\author{Kha Dinh Duy}
\email{khadinh@skku.edu}
\affiliation{
  \institution{Sungkyunkwan University} 
  \country{}
}

\author{Kyuwon Cho}
\email{kyuwon.cho@skku.edu}
\affiliation{
  \institution{Sungkyunkwan University} 
  \country{}
}

\author{Taehyun Noh}
\email{dove0255@skku.edu}
\affiliation{
  \institution{Sungkyunkwan University} 
  \country{}
}

\author{Hojoon Lee}\authornote{Corresponding author}
\email{hojoon.lee@skku.edu}
\affiliation{
  \institution{Sungkyunkwan University} 
  \country{}
}


\begin{abstract}
    In-process compartmentalization and access control have been actively explored to provide in-place and efficient isolation of in-process security domains. Many works have proposed compartmentalization schemes that leverage hardware features.
    Newer ARM architectures introduce Pointer Authentication (PA) and Memory Tagging Extension (MTE), adapting the reference validation model for memory safety and runtime exploit mitigation. 
    Despite their potential, these features are underexplored in the context of userspace program compartmentalization.

   	This paper presents \thename, a novel hardware-assisted intra-process access control design that embraces capability-based security principles.
	\thename coherently incorporates the new hardware security features on ARM, based on the insight that the features already exhibit inherent capability characteristics.
	It supports the life-cycle protection of the domain's sensitive objects -- starting from their import from the file system to their place in memory.
	With intra-process domains authenticated with unique PA keys, \thename transforms file descriptors and memory pointers into cryptographically-authenticated references and completely mediates reference usage with its program instrumentation framework and an efficient system call monitor.
	We evaluate our \thename-enabled NGINX web server prototype and other common applications in which sensitive resources are isolated into different domains. Our evaluation shows that \thename incurs a low-performance overhead of approximately $17\%$ for the single-threaded and $13.54\%$ for the multi-threaded webserver.
\end{abstract}

\maketitle

\glsresetall

\section{Introduction}
\label{sec:intro}
Modern software is often large and complex.
As a result, it suffers from bugs, some of which are security vulnerabilities that concede program control to adversaries or leak sensitive program resources (e.g., cryptographic keys).
Researchers and industry have therefore sought to isolate the monolithic program into multiple process-level compartments~\cite{aces,privman,privtrans,wedge,hakc,muscope}, each ideally performing a specific task (\emph{separation of privilege}) and is given only the essential privileges (\emph{least privilege}).
The process-level compartments of the program must now communicate via \gls{ipc} which accompanies inherent performance overhead.

Many works have proposed in-process and in-place compartmentalization methodologies~\cite{lotrx86,shreds,seimi,erim,hodor,jenny,cerberus,donky} that either re-purposes existing hardware features~\cite{lotrx86,shreds,seimi} or adapts of newly introduced hardware features, most notably using Intel's \gls{pku}~\cite{intel-manual}.
Recent \gls{pku}-based proposals~\cite{pkrusafe,hodor,erim,jenny,cerberus} have demonstrated in-place isolation with limited performance overhead.
Hardware-assisted isolation on ARM was previously explored \cite{shreds} using \emph{domains} memory protection feature that had an uncanny resemblance to \gls{pku}.
However, the domains feature has now been deprecated in AArch64.

We argue that the in-process compartmentalization designs on the modern ARM architectures are currently underexplored.
The ARM processor architecture's recent iterations introduced new hardware-assisted software security features.
\Gls{pa}~\cite{arm-arch,armv8-m}, is a hardware feature in ARMv8 that employs cryptographic \gls{ac} to protect pointers from corruption.
\gls{mte} is another hardware feature included in the ARMv8.5-A architecture that implements a key-and-lock mechanism
to enable tagging of pointers and the 16-byte \emph{pointee} memory blocks.
However, the principles and mechanisms of the ARM's direction in hardware-assisted security are vastly different from those of x86, i.e., \gls{pku}, and open a new design space for program compartmentalization.

In this paper, we propose \thename, a novel OS access control model that revolves around capability security principles~\cite{obj-cap,eros}. 
A plethora of existing research has discussed capability as the ideal scheme for achieving the \emph{least privilege} compartments and eliminating \emph{ambient authority} in OSes~\cite{cheri,capsicum,eros,cheri,cheri-risc,cheriabi}.
We observe that the new ARM hardware extensions, \gls{pa} and \gls{mte}, carry inherent characteristics of \emph{capabilities}.
\thename's design choices fully leverage these features. 
It creates in-process \emph{domains}, subprogram components that are given exclusive access to private resources.
Each domain is identified and authenticated by their \emph{domain authentication keys} (or domain keys), which is a \gls{pa} cryptographic key that \thename reserved for authenticating resources.
Our work retrofits the OS-based access control with its consistent capability scheme that provides \emph{life-cycle} protection of sensitive objects; it transforms not only memory references (\ie, pointers) but also file object references into non-forgeable tokens.

\hdr{In-process capabilities for object life-cycles}
\thename brings a coherent in-process capability for compartments within the process (subjects) and abstract process resources (objects) whose state may alternate between a file or memory content throughout its life cycle.
Capability-based \emph{memory} access control has been explored by many previous works~\cite{cheri,cheri-risc,cheriabi,lowfat-pointer,new-lowfat-pointer}.
Notably, the CHERI architecture~\cite{cheri,cheri-risc,cheriabi} applies memory capabilities to pointers, although the requirement of customized processor architecture limits its applicability.
Process-level capabilities have also been studied from OS design perspectives~\cite{obj-cap,eros,freebsd-capsicum,cap-addressing,capsicum}. 
These works focus on access control of OS resources bestowed on the process often represented in the form of \emph{files}.
However, securing today's large monolithic user programs calls for finer-granular access control on files, not to mention the necessity of consolidating a memory access control for in-process compartments.

\hdr{Simple and efficient reference monitoring} \thename's coherent capability scheme also seamlessly incorporates a simple and efficient reference monitor design for in-process access control.
The necessity and design space of efficient \gls{syscall} reference monitors for in-process domains have been discussed in many previous works~\cite{pku-pitfall,jenny,cerberus,monguard}.
Since the OS kernel is unaware of the in-process compartments, a reference monitor must mediate accesses to process resources among the potentially mutually distrusting compartments in addition to \glspl{syscall} filtering that may undermine the security of the compartments.
We take a different route from previous works to achieve both goals coherently.
\thename's capability-engraved file descriptors carry information for authentication and authorization.
This eliminates the need for separate data structures, i.e., an \gls{acl}, to keep track of each compartment domain's ownership and access rights on file objects.
Also, the validation process of the capability is fast, as it is essentially achieved through a single \gls{pa} instruction.

\issue{Incomplete}

\hdr{Novel domain and \gls{pa} context binding scheme} 
Our way of creating \gls{pa} contexts for in-process compartments is unique and specifically devised for \thename's life-cycle capabilities. 
Previous work has presented a framework for using \gls{pa} and \gls{mte} that establishes kernel compartments with policy-defining \gls{pa} \emph{modifiers}~\cite{hakc}.
However, \thename chooses to assign a unique \emph{\gls{pa} key} for each domain and perform \emph{key switches} during domain transitions and employs the \gls{pa} modifier to isolate references between domain \emph{instances}.
\thename then uses the per-domain key and instance modifier to compute the cryptographic \gls{ac} for the resources and embeds them into the resource handles themselves.
This seemingly small difference is a key design component for \thename.
With a single key switch, \thename can efficiently switch the authentication context for both system resource handles and signed userspace pointers.
This allows \thename to establish arbitrary compartmentalization boundaries within programs and enables a unified and consistent \gls{pa}-based authentication throughout the life cycles of program resources without a complex modifier management scheme.

\hdr{Challenges} Design and implementation of \thename must satisfy security requirements and be mature enough to be adopted to real-world applications.
\thename rigorously assesses and addresses the security requirements for the capability references, namely \emph{non-forgeability} and \emph{non-reusability}.
This effort has been made for every sensitive operation carried out by \thename, from domain transitions to signing and authentication of the capability tokens.
We detail our security considerations as we elaborate on the design and provide a dedicated security analysis.

In addition, \thename implements a robust instrumentation framework for complex user programs from scratch.
\thename's \pamte-accelerated pointer capability requires \emph{complete mediation} of all pointer uses.
Applying such a scheme  to complex programs poses a daunting challenge since a single incorrect instrumentation would inadvertently crash the program. 
Similar \gls{pa}-based complete mediation of pointer uses have been developed by previous works; however, none with the level of maturity of \thename.
For instance, PARTS~\cite{pac-it-up} was only evaluated with a benchmark suite.
Also, \thename incorporates \gls{mte} to inter-domain memory isolation.
HAKC~\cite{hakc} presented \pamte instrumentation for kernel module compartmentalization that only authenticates the cross-domain pointers only once before their first use, while \thename must provide a domain-aware and completely mediated pointer load and store instrumentation.
As we will show through our evaluation, \thename's instrumentation framework is mature enough to compile programs such as NGINX-LibreSSL and OpenSSH SSH client.
In summary, our contributions are as follows:
\vspace{0.3em}
\issue{needs re-work according to new intro}
\begin{itemize}[]
	\setlength\itemsep{0.4em}
	\item We introduce a novel in-process compartmentalization design that adapts hardware-accelerated capabilities for the life-cycle protection of domain resources.
    \item We design a simple and efficient capability-based reference monitor to isolate in-process system resources.
	\item We establish \gls{pa} key-identified domains to efficiently isolate kernel and userspace resource references.
	\item We assess and address the unique security challenges of capability-based isolation that requires complete mediation on reference uses, prevention of impersonation, and protection against forging and reusing of references.
	\item We develop an instrumentation framework that is robust enough to completely mediate complex user programs by addressing compatibility issues.
	\item We evaluate our implementation\footnote{Available at: \url{https://github.com/sslab-skku/capacity}} on real-world applications and report low overheads on the approach.
\end{itemize}

\begin{figure*}[ht!]
	\centering
	\includegraphics[width=0.85\textwidth,viewport=0 0 1398 393]{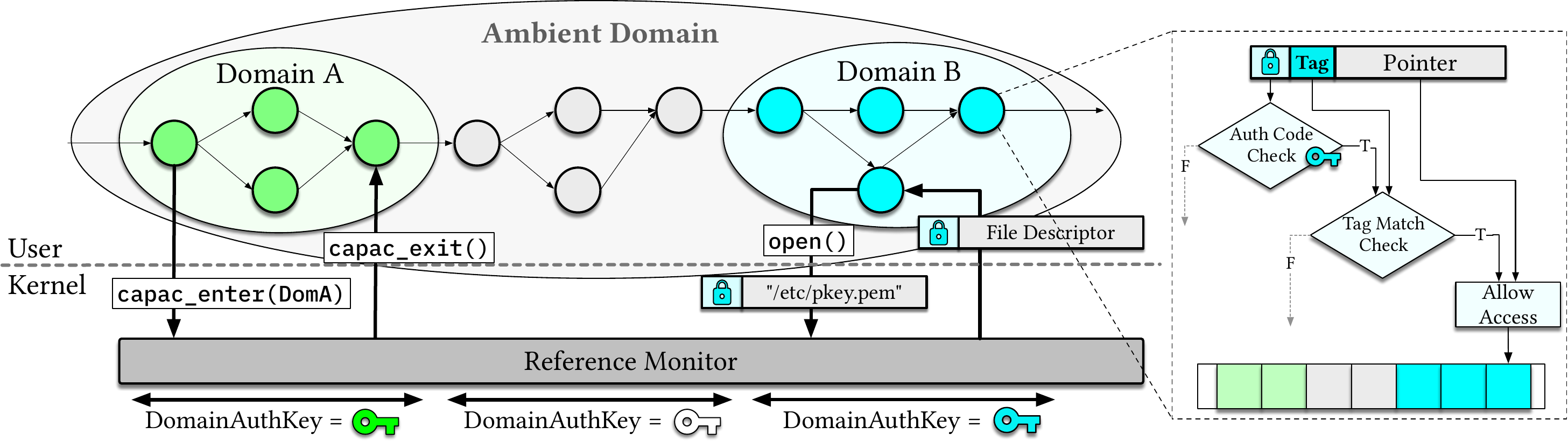}
	\caption{Overview of \thename's in-process domains and life-cycle objects protection. }
	\label{fig:overview}
\end{figure*}

\section{Background}

\subsection{Pointer Authentication (PA)}
\label{sec:background:pac}
\emph{\Glsentryfull{pa}}, introduced on ARMv8.3-A~\cite{arm-arch}, provides hardware acceleration and ISA extension for cryptographically authenticated pointers.
The design of \gls{pa} is inspired by previous works that protect pointers from corruption~\cite{ccfi,cpi}.
\gls{pa} allows attaching a \gls{pac} to a pointer in the unused bits [54:48] of a 64-bit address.
A \gls{pac} is generated using one of the five keys stored in the newly added registers that can only be accessed or modified with privileged instructions.
The five keys consist of two keys for signing code/instruction pointers (Instruction-\{A,B\}), another two for signing data (Data-\{A,B\}) pointers, and one general-purpose key (G).
We denote these keys as \key{IA},\key{IB}, \key{DA}, \key{DB} and \key{G}.
\Gls{pa} also introduces \texttt{pac} and \texttt{aut} families of instructions to \emph{sign} (i.e., calculate the \gls{pac} and attach it to the value in the operand register) and \emph{authenticate} a 64-bit value with an \emph{optional} modifier using the key indicated by the instruction name.
If authentication on a pointer succeeds, the authentication code in the pointer is cleared, making the pointer usable.
The \gls{pac} value is corrupted if the check fails, which triggers a segmentation fault when the pointer is dereferenced.
We denote the operations of \texttt{pac} and \texttt{aut} instructions that use the key \key{} to sign and authenticate a given data \emph{D} as \pac{}{\emph{D}}{\emph{mod}} and \aut{}{\emph{D}}{\emph{mod}}.

\subsection{Memory Tagging Extension (MTE)}
\label{sec:background:mte}
\emph{\Glsentryfull{mte}}~\cite{mte} is another hardware-backed security feature to ARMv8.5-A~\cite{arm-arch}.
\Gls{mte} adapts the principles of tagged memory in the existing research proposals and implementations in other architectures~\cite{timber-v,hw-tagged-memory,sparc-adi}. 
It facilitates memory safety by enabling the 4-bit tagging of pointers and 16-byte aligned address ranges.
When \gls{mte} is enabled, the processor raises an interrupt if the tags differ between the memory and the pointer, allowing vulnerabilities such as buffer overflow to be captured.
We denote \tagz{}{\emph{P}} as the tagging operation that first assigns the 4-bit tag \textbf{T} to the bits [59:56] of a pointer \emph{P}, then tag 16byte-aligned memory region between \emph{P} and \emph{P}$+$size with \textbf{T}.
In addition, \gls{pa} and \gls{mte} can be \emph{simultaneously} enabled.
Hence, a pointer can be tagged \emph{then} signed, and the resulting pointer would carry a tag in bits [60:56] and \gls{pac} in [54:48].

\section{Overview}
\label{sec:capability}
\label{sec:overview}

\autoref{fig:overview} illustrates an overview of \thename.
\thename's design demonstrates a comprehensive capability model for file and memory objects access control that can be applied to commodity ARM systems that support \gls{pa} and \gls{mte}. 
\thename's primary objective is to guarantee each domain's exclusive access rights to the sensitive domain-private objects. 
It associates each domain, called \emph{a \thename domain}, with a unique \gls{pa} cryptographic key and \emph{switches} the currently activated key using an in-kernel reference monitor before entering a domain. 

Consider the common pattern of sensitive object use shown in \autoref{fig:overview} (Domain B).
First, a file-system \emph{path reference} (\texttt{"/etc/key.pem"}) is used as an argument to the \texttt{open()} \gls{syscall}. 
A \emph{file descriptor} to the file object is then obtained. 
Finally, the object is loaded into a memory region (\eg, with the \texttt{read()} \gls{syscall}) so the program can interact with it through \emph{pointers} (\texttt{0x00004bfff}...).
In \thename, the creation and usage of three types of resource references are \emph{completely mediated} with a coherent \pamte-based authentication, provided by (1) a lightweight in-kernel reference monitor that efficiently validates path and file descriptor (\fd) references in \gls{syscall} arguments, and (2) \pamte-assisted tagged memory allocation and pointer authentication instrumentation that isolate domain-private memory and their references.

This section provides a high-level overview of \thename's capability model and its programming model. We will delve into the design and implementation of each of \thename's components that enforce its security in \cref{sec:design:impers}, \cref{sec:design:ref-mon}, and \cref{sec:design:instrumentation}.

\subsection{Threat model}
\label{sec:design:threat-model}

We assume that the domains are mutually distrusting. 
The program's vulnerabilities can potentially grant adversaries with arbitrary memory manipulation and control-flow subversion primitives, compromising an \thename domain or the \emph{ambient} domain (the rest of the program code not inside \thename domains).
Even so, a compromised domain must not access other domains' private objects.
We assume that the processor and the kernel are trusted and that modern OS security measures, such as the W{$\oplus$}X policy and ASLR, are in place.
Additionally, we deem the program initialization, including the initialization of \thename, free of attacker influence.
We exclude side-channel attacks and microarchitectural attacks from the scope of this paper.

\thename also incorporates the existing backward-edge and forward-edge \gls{cfi} in its implementation and is compatible with recent advances in \gls{pa}-assisted \gls{cfi} techniques.
\gls{pa}-based backward-edge \gls{cfi}~\cite{pac-it-up,qualcomm-pac,pacstack,llvm-ret-addr} authenticates the return addresses on the stack with \key{IB}, which render the traditional attacks (\eg, ROP attacks) that overwrite the return address on the stack infeasible.
\gls{pa}-based forward-edge \gls{cfi}~\cite{kernel-pac,pac-it-up} authenticates code pointers with \key{IA} with its LLVM \emph{ElementType} ID as the modifier, which restricts indirect calls to (1) valid entry points of functions (2) functions of the matching type.

\subsection{Security requirements}
\label{sec:overview:secreq}
The capability principles hold only when the common security requirements for capability are met.
In particular, we observe that \thename must satisfy the following security requirements:

\begin{enumerate}[label=\secreq{{R\arabic*}}]
	\item \emph{Non-impersonatable domains:} \thename must be able to identify and authenticate intra-process domains and prevent impersonation.
	\item \emph{Complete mediation:} all access to file and memory objects must be mediated by \thename.
	\item \emph{Non-reusable references:} \thename domain-private references are only valid within the domain that owns the object.
	\item \emph{Non-forgeable references:} \thename references must not be forged by the adversary.
\end{enumerate}

Throughout the rest of this paper, we use the above requirements to analyze and manifest the security guarantees of \thename.
Upholding the security requirements, therefore, is a challenge that must be addressed by \thename's design and implementation.

\subsection{Subjects and Objects}
\label{sec:overview:domains}

In \thename's capability model, the subjects are subgraphs of program execution that interact with sensitive objects called \emph{\thename domains}. 
Program code that does not belong to a domain is called the \emph{ambient} domain.
\thename identifies and authenticates each domain with a unique \gls{pa} key, called the \emph{domain key}. 
Each domain is also associated with an \gls{mte} tag for the tagging of the domain's private memory.
We use \key{DB} among the kernel-managed \gls{pa} keys as the domain key.
\key{DA} is the \emph{ambient key} used to sign and authenticate resources accessible by the entire program.
\key{\{IA,IB\}} are reserved for proposed \gls{pa}-based forward-edge and bardward-edge \gls{cfi} defenses~\cite{pac-it-up,pacstack,apple-pac,qualcomm-pac}.

\hdr{Domain switching}
A \thename domain is encapsulated between the APIs \texttt{capac\_enter} and \texttt{capac\_exit}.
Domain switching is performed with the help of a lightweight in-kernel reference monitor, as we will describe in \cref{sec:design:switch-auth}.
Upon entering a domain, the userspace program updates its current \gls{mte} tag and requests the reference monitor to switch the currently activating domain authentication key to that of the target domain.
A \emph{per-instance} \gls{pa} modifier is also maintained in both kernel and userspace to identify \emph{instances} of a domain invocation.
\texttt{capac\_exit()} restores the domain key, \gls{mte} tag and modifier to those of the ambient domain.
We do not support the nesting of domains in the current prototype. 

\begin{figure}[b] \centering \includegraphics[width=0.95\columnwidth]{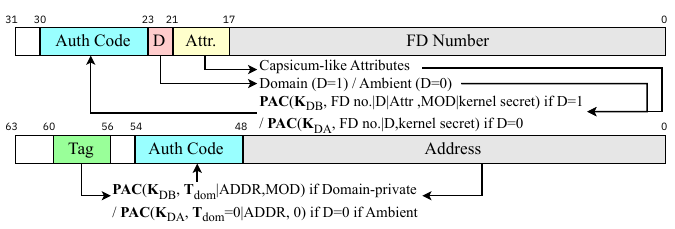}
	\caption{\thename's reference encapsulation of pointers and file descriptors with cryptographic authentication code (AC).}
	\label{fig:ref-encapsulation}
\end{figure}

\hdr{Object encapsulation}
As summarized by \autoref{fig:ref-encapsulation}, \thename facilities employ \gls{pa} to \emph{sign} private resource references with the domain key and the per-instance modifier (e.g., a unique session ID), engraving the reference ownership into its embedded \gls{ac}.
Before a reference is used to access some resource, \thename \emph{authenticates} its \gls{ac} with the currently activating domain key and modifier before granting the access.
Consequently, object references are turned into non-forgeable (\srforge) \emph{capability tokens} that are valid only when the context is within their owner domain.
By simply switching the authentication key, \thename prevents cross-domain reusing of both kernel and userspace object references (\srreuse).

\keypoint{
    We introduce the \emph{ambient objects}, non-private and accessible from all domains, \eg, global variables and non-sensitive files, to avoid the programming model becoming too restrictive.
}
All references to ambient objects are to be signed and authenticated with \key{DA}.
The ambient memory objects are given a \tagnum{Dom} of $0$, which makes all untagged memory pointers ambient memory references, and only the deliberately tagged pointers become references to domain-private memory.
To facilitate multi-domain interactions, \thename also supports the \emph{delegation} of references.
When a reference delegation request is received, \thename first authenticates the reference to prevent delegation of non-owned resources (\srimper), then \emph{re-signed} it with the target's domain key and instance modifier. 
Especially for memory reference delegation, \thename compares the pointee memory's tag with the pointer's tag before recoloring both to the target domain's memory.

\begin{table*}[t]
	\small
	\centering
	\begin{tabularx}{\textwidth}{llX}
		\hline
		 & \thead{API \& Compiler Annotation}                                      & \thead{Description}                                                                                        \\
		\hline
		\multirow{6}{*}[-0.5em]{\thead{\begin{sideways}API\end{sideways}}}
		 & \texttt{capac\_init(configurations)}                            & Initializes \thename facilities, assigns \cpath to domains, enables syscall authentication       \\
		 & \texttt{capac\_enter(target\_id, mod)}/\texttt{capac\_exit()}  & Enters the target domain/Exits to the ambient domain                                                       \\
		 & \texttt{capac\_limit\_fd(fd, cap\_mask)}                       & Limit the capabilities of an \fd                                                                           \\
		 & \texttt{capac\_delegate\_fd(fd, target\_id, mod, cap\_mask)}   & Limits the capabilities and delegate an \fd to the target domain                                           \\
		 & \texttt{capac\_malloc(size)}/\texttt{capac\_free(ptr)}         & Allocates/frees memory tagged with the currently active domain                                             \\
		 & \texttt{capac\_delegate\_ptr(ptr\_loc, size, target\_dom, mod)} & Delegates an in-memory pointer to \texttt{target\_dom} and also re-colors the memory\\
		\hline
		\multirow{1}{*}[0em]{\thead{\begin{sideways}Ann.\end{sideways}}}
		 & \texttt{DOM\_PRIV\_FUNC}                                       & Tags function’s stack frame with the currently executing domain’s tag (\tagnum{Dom})                       \\
		 & \texttt{DOM\_PRIV}                                             & Marks a \emph{source} domain-private pointer for taint analysis.\\
        \hline
	\end{tabularx}
	\caption{\thename APIs and program annotations}
	\label{tab:apis}
\end{table*}

\subsection{Programming model overview}
\label{sec:enforcement}
\label{sec:implementation}

\label{sec:design:prog-model}

\keypoint{
    A programmer interacts with \thename through a set of well-defined APIs provided by the runtime library and program \emph{annotations} that direct the program instrumentation, shown in \autoref{tab:apis}.
}
We demonstrate the programmer's perspective when retrofits intra-process capabilities into programs through a quintessential example of secret import and uses in \autoref{fig:code-example}.

\begin{figure}[t]
	\input{listings/capacity-code-example}
	\caption{Example of \thename's programming model, annotated with \thename references and their enforcement.}
	\label{fig:code-example}
\end{figure}

The programmer first conceptually defines in-program domains (\eg, \texttt{DOM\_CRYPTO}) within the program with respect to the domain-private resources.
The subprogram interval is enclosed by \texttt{capac\_enter()} and \texttt{capac\_exit()} (lines $4$ and  $7$), and each interval is paired with a domain key.
In the example, \texttt{ctx->id} contains the unique ID for the encryption context, which is set as the currently activated \gls{pa} modifier for \texttt{DOM\_CRYPTO}. 
A domain can span a few function calls, as shown in the example.
The domain entrance cannot be nested; however, \thename is designed to support the sharing of functions between domains.

The programmer then identifies domain-private objects and their references.
In the case of file(-like) objects, a list of domain-to-file mappings is provided as an argument to \texttt{capac\_init()} (line $1$) to notify the reference monitor of the domain ownership of file resources.
Afterward, the reference monitor automatically authenticates any path and \fd references in the \gls{syscall} arguments (lines $11$, $12$, $14$) and issues \fds signed with their owner domain when \fds are created. 
For memory objects, the programmer must mark the domain-private region containing the object and its domain-private pointer references.
To allocate domain-private memory, the programmer either changes their allocation site to use \texttt{capac\_malloc} (line $13$), which allocates memory tagged with the domain's tag and returns a tagged pointer, or uses \texttt{DOM\_PRIV\_FUNC} (line $17$) to direct the instrumentation to isolate the function's stack frame.
On \texttt{DOM\_PRIV\_FUNC}-annotated functions, the function is instrumented such that the stack frame is \emph{tagged} with the currently executing domain's tag number and reverted upon function return.
Then, the programmer applies \texttt{DOM\_PRIV} (line $18$) to the \emph{source} domain-private pointer. 
From then on, the pointer is \emph{taint-tracked} to mark all derived references to the object within the function.
\thename then instruments \gls{pa}-based signing/authentication with the domain key of tracked pointers, including the member variables (\eg, \texttt{ctx->secret\_key}) (lines 13, 14, 21).
\section{Domain Switching and Authentication}
\label{sec:design:impers}
\keypoint{
\thename's API design allows the introduction of flexible compartmentalization  boundaries to existing software through in-place annotations.
}
However, such a design must be robust against  \emph{domain impersonation} (\srimpers).
In \thename's threat model, domain impersonation happens when a compromised domain diverts the control flow into \thename APIs, \eg, calling the domain entry gates with arbitrary arguments, or corrupts \thename's critical values, \eg, \gls{mte} tag ID stored in memory to gain illegal access to domain-private resources.
\thename prevents such domain impersonation using two techniques.
First, \thename \emph{authenticates} the domain entry sites by requiring them to construct \emph{entry tokens}, which the reference monitor can verify before granting the domain switch.
Second, it introduces an in-userspace \emph{authentication} protocol that securely fetches the currently executing domain's \gls{mte} tag.
In both cases, after successful authentication, \thename securely places its critical values (the domain's memory tag and the per-instance modifier) in compiler-reserved registers and fetches them when needed to prevent illegal modifications.

\hdr{Secure domain switching}
\label{sec:design:libcapac}
\label{sec:design:switch-auth} 
The program enters a domain through the API \texttt{capac\_enter} (\autoref{tab:apis}) that internally invokes an \texttt{ioctl} syscall to the reference monitor.
The reference monitor then switches the currently activating \key{DB} to that of the target domain by writing into the dedicated \gls{pa} key registers with its kernel privilege. Hence, an attacker who can invoke \texttt{ioctl} calls with arbitrary values could achieve \emph{impersonation} and access domain-private references.
\keypoint{
	\thename protects the domain switches from impersonation by authenticating domain entry points with \key{G}.
}
Before a domain switching request, the call gate signs the domain ID argument in \texttt{capac\_enter(domID, mod)} with a \texttt{pacga} instruction, then passes the signed argument to \thename reference monitor.
Since there is no dedicated authentication instruction that uses \key{G}, the reference monitor uses \texttt{pacga} to generate the valid argument with the domain ID and compare it with the argument from the call gate. Domain entry is granted if the two values match.  
After the entry is granted, the call gate loads the per-instance modifier into a reserved floating-point register (\texttt{ModReg}), which is to be used by our instrumentation for pointer authentication demonstrated in \cref{sec:instrumentation}.
Finally, the call gate clears the token-containing register to prevent the leaking of the entry token.
\texttt{capac\_enter} is implemented as the following call gate, which is to be inlined to the domain switching sites:

\begin{lstlisting}[style=sslab-capacity-c]
    PACGA(token, target_id, 0)
    if(!ioctl(capac_fd, CAPAC_ENTER, token, modifier)) 
        exit();
    // Load per-instance modifier into ModReg register
    asm volatile("fmov ModReg, %[mod]" ::[mod] "r"(modifier) :);  
    // Clear entry token from register
\end{lstlisting}

\keypoint{
	\thename endows authenticity to the entry points by preventing attackers from diverting the control flow into arbitrary instructions with \gls{cfi} and ensuring that the instruction \texttt{pacga} is absent within the process but the domain entry sites.
}
\key{G} is currently unused in both deployed and proposed \gls{pa}-based defenses.
Therefore, it is unlikely that it would naturally appear under normal circumstances.
Also, since ARM is a RISC architecture, we can generally rule out the issue of unaligned, unintended occurrences of the instruction, an issue that makes preventing illegal instruction occurrences challenging on x86 architectures \cite{erim}.




\hdr{Authenticated domain ID retrieval}
\thename enables a trustworthy runtime domain ID fetching procedure.
This avoids hardcoding \gls{mte} tags into functions that require memory tagging since these functions might be used in confused deputy attacks to tag arbitrary memory regions. 
Moreover, fetching the domain ID at runtime allows a \thename-protected function to be called from multiple domains.
The procedure is also imperative in runtime domain memory and pointer tagging in \thename's instrumentation, as we will explain in \cref{sec:design:instrumentation}.
\thename's private heap memory allocator, \texttt{capac\_malloc()}, also uses the procedure to authenticate the callee's domain ID and tag the allocated memory.

To prepare for runtime domain ID retrieval, \thename first constructs a \gls{dst} that is filled with \emph{domain signatures} of each declared domain ({\tt DST[i] = }\pac{DBi}{0}{0}).
After initialization, the table remains read-only throughout program execution to prevent tampering.
Additionally, a global variable that stores the domain ID of the currently active domain, \texttt{int curr\_dom}, is updated on each \texttt{capac\_enter(domID)} and \texttt{capac\_exit()} invocation.

The following snippet is inlined to verify the current domain ID, such that it can be used as a \gls{mte} tag for the pending operation:
\begin{lstlisting}[style=sslab-capacity-c]
    domain_signature_t dom_sig = DST[curr_dom];
    AUTDB(dom_sig);
    assert(dom_sig == curr_dom);
    // Form a tag mask shifting left by 56 bits
    asm volatile("lsl %[tag], %[tag], 56" : [tag] "=r"(target_id)::); 
    // Load tag mask into TagReg register
    asm volatile("fmov TagReg, %[tag]" ::[tag] "r"(target_id) :);  
    // /*!\textcolor{clr-comment}{\tagnum{Dom}}!*/ is known from this point onward
\end{lstlisting}
The above procedure first fetches the current domain's signature from the \gls{dst}, then authenticates it with the domain key.
If authentication succeeds, a \emph{tag mask} is constructed, \eg \texttt{0x00ff00..00}, which can be applied to an untagged pointer with a bitwise \texttt{OR}.
The tag mask is stored in a reserved floating-point register (\texttt{TagReg}), so the memory tagging operations can efficiently retrieve it.

\chcomment{Begining paragraph of each section and last paragraph should highlight contribution, draw comparison with other works}
\chcomment{Discussions should be: (1) Coherency of the scheme, (2) System call filter vs. resource filter}

\section{File System Object Isolation}\label{sec:ref-mon}\label{sec:design:ref-mon}\label{sec:design:file-path}





\keypoint{
	\thename includes a reference monitor that verifies the \emph{arguments} of syscalls to enforce the complete mediation (\srcompmed) of system resource accesses.
	This capability-inspired authentication mechanism drastically differs from previous works implementing \gls{acl}-based reference monitors for in-process syscall filtering and file access control~\cite{erim,hodor,jenny,cerberus,pku-pitfall}.
}
In those systems, enforcing access control on domain-private sensitive file objects requires a separate logic for the monitor detached from the program semantics.
Moreover, \gls{acl}-based access control requires keeping track of the isolated objects and the domains that can access them at runtime~\cite{jenny}.
In contrast, the \fds in \thename are cryptographically secured; the proof of ownership and object attributes are attached to the reference itself, regardless of the number of subjects sharing a resource.

We implement the reference monitor in about $1000$ LoC. It provides domain key switching and \gls{syscall} interception through \gls{syscall} table hooking using a similar approach to a previous efficient reference monitor~\cite{jenny}.
To avoid the system monitor affecting all processes within the system, we set an unused flag in \texttt{struct thread\_info->flags} to mark \thename-enabled processes during initialization and check for the flag before performing syscall authentication.

\subsection{Enforcing domain-private file-system paths}
\label{sec:design:path}
\keypoint{
	\thename supports file path isolation to be compatible with existing programs without significant rewriting effort.
}
\thename's file path authentication utilizes the pre-existing methods for attaching policies to file system objects used by kernel security subsystems (e.g., SELinux~\cite{selinux}).

File paths protection is kickstarted by the \texttt{capac\_init} API call that sends a list of domain-private files and their owners to the reference monitor.
The reference monitor first resolves each path to obtain the corresponding \texttt{inode} structure.
It creates a per-process table in the file's \texttt{inode->f\_security}.
For each owner domain of a file path, it stores the \emph{domain signature}, \pac{DB}{\texttt{NULL}}{0}, in the table.
If the file is not domain-private, the table is left as \texttt{NULL}.
After, the reference monitor transitions the process into the \emph{protected mode}, where the reference monitor authenticates every path and \fd arguments used in syscalls.

Before authenticating a file object import, the requested file path is first resolved to obtain the underlying \texttt{inode} structure.
This is to avoid confusion when encountering relative paths and symbolic links.
If the table for the current process in \texttt{inode->f\_security} is \texttt{NULL}, the reference monitor recognizes that it is an \emph{ambient} object and can be accessed without further authentication.
If domain signatures are present, the reference monitor authenticates them using the corresponding domain key until a valid signature is found.
The reference monitor verifies that the current user context is in the object's owner domain if any valid signature is found while rejecting the file access otherwise.
Finally, the returned \fd for the requested file is signed with \key{DA} for ambient objects or the domain key for domain-private objects.




\subsection{Enforcing domain-private file descriptors}
\label{sec:design:fd}
\keypoint{
	\thename's reference monitor intercepts syscalls that generate \fds and create signed \fds before they are sent to the userspace.
	It also intercepts \fd-accepting syscalls (\eg, \texttt{read}) for \fd authentication.
}
Hence, all \fds issued to the process carry \glspl{ac} that are authenticated on use.
Additionally, we use a \emph{secret modifier} only known to the kernel as the \gls{pa} signing modifier, to prevent forging by signing a plain integer with \texttt{pac} instructions in the userspace (\srforge).
\thename's security model also requires that the integer for \fds are not reused (\srreuse), even after they are closed. 
We achieve this by changing the hook for the \texttt{close} to reserve the closed \fd in the thread's \fd table, preventing the OS from reusing it.

\hdr{In-process FD capabilities}
\keypoint{
	\thename's \fd access control scheme resembles that of Capsicum's capability-enabled \fds~\cite{capsicum} but is much more lightweight and is enforced at the in-process granularity.
}
\thename's \fd capabilities scheme requires no additional kernel metadata for the permissions of \fds since the information necessary to authorize an \fd use is engraved in the \fd itself.
Moreover, thanks to integrity protection provided by cryptographic authentication, we can securely embed fine-grained access control attributes into the file descriptors themself (\texttt{FD->Attr.} in \autoref{fig:ref-encapsulation}) without extensive bookkeeping.
A combination of \gls{pa}-assisted authentication on the \fd and a per-syscall check based on static policies is sufficient to grant or reject file descriptor access.

Our prototype supports four capabilities, represented as a bitmask of enabled capabilities.
The \texttt{CAP\_WRITE} and \texttt{CAP\_READ} capabilities allow read-related syscalls (\eg, \texttt{read}, \texttt{recvfrom}) and write-related syscalls (\eg, \texttt{write}) to use the \fd.
\texttt{CAP\_SOCKET} enables network-related syscalls, such as \texttt{accept} and \texttt{listen}.
\texttt{CAP\_DELEGATE} allows the \fd to be delegated to other domains.
The reference monitor uses a capability bitmask for each syscall to validate the \fd's attributes.
On the other hand, \texttt{cap\_limit\_fd()} instructs the reference monitor to remove the selected capabilities from an \fd and re-sign it.

\hdr{Signing and authentication of 32-bit \fds}
\autoref{fig:ref-encapsulation} demonstrates the signing strategy of file descriptors.
Since most file descriptors are 32-bit signed integers, we devised a 32-bit signing and authentication scheme.
To sign an \fd, the reference monitor first signs its 64-bit zero-extended value like a normal pointer.
Then, it attaches the PAC to bits[30:23] of the \fd.
The monitor uses bit[22] (\texttt{FD->D}) bit to distinguish between domain-private and ambient \fd, such that domain-private {\fd}s are authenticated using the domain key.
Bit[17:21] stores the bitmask of the previously described \fd capabilities.
bit[31] is unused to prevent the descriptor from being interpreted as an error code (e.g., \texttt{if (fd < 0)}).
\fd authentication on 32-bit \fds happens likewise.

\hdr{Handling special \fd values} During our implementation, we found that special \fd values, such as \texttt{STDIN}, \texttt{STDOUT}, or \texttt{AT\_FDCW}, need to be handled differently since they are often hard-coded into the application as integer values (e.g., 1 and 2). We currently make our reference monitor omit their authentication for compatibility, given that they are always ambient objects.

\chcomment{It is more efficient to change key vs modifier.}

\chcomment{PARTS does not create domains, and their implementation is not mature. Our implementation proves the possibility of complete mediation. Maybe bring up some example where parts would break}

\section{DOMAIN MEMORY ISOLATION}
\label{sec:design:instrumentation}
\label{sec:instrumentation}
\begin{figure*}[ht]
	\centering
	\noindent\begin{minipage}[t]{.27\textwidth}
		\begin{lstlisting}[style=sslab-capacity-c,language={[ARM]Assembler},keepspaces,numbers=none]{Name}
.prologue:
 // Sign return addr
 pacib  x30, sp 
 ... 
 // x9: DST[curr_dom]
 // Auth current domain get /*!\textcolor{clr-comment}{\tagnum{Dom}}!*/
 /*!\textbf{autdzb}!*/	x9    
 cmp    x9, x8 
 b.ne  .auth_failed
 // Build tag mask and save to ModReg
 lsl    x8, x8, #56 
 fmov   TagReg, x8
\end{lstlisting}
	\end{minipage}\hfill
	\noindent\begin{minipage}[t]{.20\textwidth}
		\begin{lstlisting}[style=sslab-capacity-c,language={[ARM]Assembler},keepspaces,numbers=none]{Name}
...
// Tag pointer
// x8: /*!\textcolor{clr-comment}{\tagnum{Dom}}!*/ tag mask 
fmov x8, TagReg
mov  x24, sp
orr	 x25, x24, x8 
// Tag stack frame
stg	 x25, [x25]  
...

\end{lstlisting}
	\end{minipage}\hfill
	\noindent\begin{minipage}[t]{.21\textwidth}
		\begin{lstlisting}[style=sslab-capacity-c,language={[ARM]Assembler},keepspaces,numbers=none]{Name}
// x0: domain-priv. ptr
fmov    x8, ModReg  
/*!\textbf{pacd}\underline{\textbf{b}}!*/   x0, x8   /*!\hfill\action{PTR-Sign}!*/
str	    x0, [mem] 
...
// x13: domain-priv. ptr
ldr     x13, [mem]
fmov    x8, ModReg
/*!\textbf{autd}\underline{\textbf{b}}!*/   x13, x8 /*!\hfill\action{PTR-Auth}!*/
... 
// x12: ambient ptr
/*!\textbf{autdz}\underline{\textbf{a}}!*/  x12   /*!\hfill\actionambient{PTR-Auth}!*/
\end{lstlisting}
	\end{minipage}\hfill
	\noindent\begin{minipage}[t]{.23\textwidth}
		\begin{lstlisting}[style=sslab-capacity-c,language={[ARM]Assembler},keepspaces,numbers=none]{Name}
.epilogue:
 mov	w0, #1
 // Untag/zero out stack 
 stzg	x24, [x24]
 // Clear tag mask register
 fmov TagReg, xzr
 ...
 // Auth return addr
 retab
\end{lstlisting}
	\end{minipage}
	\begin{minipage}[t]{.23\textwidth}
		\subcaption{Prologue: Domain ID auth.}
		\label{fig:asm-prologue}
	\end{minipage}\hfill
	\begin{minipage}[t]{.23\textwidth}
		\subcaption{Private stack tagging}
		\label{fig:asm-stack-tagging}
	\end{minipage}\hfill
	\begin{minipage}[t]{.23\textwidth}
		\subcaption{Domain-aware pointer auth.}
		\label{fig:asm-pa}
	\end{minipage}\hfill
	\begin{minipage}[t]{.23\textwidth}
		\subcaption{Epilogue: stack clean up}
		\label{fig:asm-epilogue}
	\end{minipage}
	\caption{\thename's instrumentation of an annotated function.}
	\label{fig:asm}
\end{figure*}



This section describes how \thename establishes its capability-inspired memory isolation model.
\thename enforces domain memory isolation through \gls{mte}-assisted segregation of domain-private memory regions and a \emph{complete mediation} (\srcompmed) of memory access through pointers that are made \emph{non-reusable} (\srreuse) and \emph{non-forgeable} (\srforge) capability tokens.

\thename first introduces facilities to allocate domain-private memory, including a tagged heap memory allocator and compiler-instrumented stack tagging.
\thename also present an instrumentation that mediates pointer uses with \gls{pa}-based \emph{on-load} pointer authentication.
It instruments the program to sign pointers before they are stored in memory and authenticate pointers as they are loaded from memory into registers.
\keypoint{
Different from PARTS~\cite{pac-it-up}, a previous work that also introduced an on-load pointer authentication scheme, \thename uses \key{DA} to sign/authenticate ambient pointers, and \emph{selectively} uses \key{DB} for domain-private pointers identified using a static taint analysis, making its instrumentation \emph{domain-aware}.
\thename also uses a user-defined, instance-specific modifier for domain-private pointers to prevent their reuses across domain instances.
}

\keypoint{
    \thename's instrumentation framework seeks to be compatible with complex user programs, as we will demonstrate through our evaluation (\cref{sec:evaluation}).
}
It must enforce \emph{complete mediation} of program pointer uses for whole-program pointers.
We initially used the existing implementation of PARTS~\cite{pac-it-up} but quickly found that it is inapplicable to large programs since even one incorrect handling would result in a crash.
Toward this end, our instrumentation scheme is developed from scratch to support large userspace programs reliably.
Our instrumentation handles \emph{stack spill} with a new \emph{pointer liveness tracking}  algorithm.
We also introduce solutions to compatibility issues we observed with whole-program \gls{pa} instrumentation.
The instrumentation framework is implemented on LLVM 14.0.0 \cite{llvm}, which includes built-in intrinsics for \gls{pa} and \gls{mte} features.
As of the time of writing, no available hardware supports MTE.
For this reason, we used a QEMU virtual machine (\texttt{-march=armv8.5-a+pauth+memtag}) that supports emulating MTE instructions as a testing target during implementation.

\subsection{Domain-private memory tagging}
%
\thename establishes domain-private memory regions by tagging each domain's private memory with the domain's integer domain ID (\tagnum{Dom}) obtained with domain authentication (\cref{sec:design:switch-auth}).

\hdr{Private heap memory allocation}
\libcapac manages a domain-private heap memory with its \texttt{dlmalloc} allocator that maintains a statically allocated, \gls{mte}-enabled memory pool.
Upon invocation, \texttt{capac\_malloc(size)} first performs \gls{dst}-based domain authentication so that it can determine the tag number of the currently executing domain as we explained in \cref{sec:design:switch-auth}.
Afterward, \texttt{capac\_malloc()} allocates memory from its memory pool, tags it with the domain ID, and returns a pointer tagged with the domain ID.
The returned pointer is not signed as it is delivered in the return value register, adhering to \thename instrumentation's on-store sign and on-load authentication policy.

\hdr{Private stack tagging}
\thename provides private \emph{stack} memory to domains through in-place tagging and untagging of the stack objects with the current domain ID (i.e., no explicit stack switching).
The \texttt{DOM\_PRIV\_STACK} annotation directive on a function notifies the instrumentation framework to instrument the function prologue and epilogue to prepare a domain-private stack before the function execution and to untag and zero out the tag before return.

\autoref{fig:asm} (a), (b), and (d) show the stack tagging instrumentation on an annotated function.
The instrumented prologue authenticates and obtains the currently executing domain's ID by consulting \gls{dst}, then loads the tagging mask into the reserved register.
Then, it retrieves the tagging mask from the reserved register and tags the function's stack frame when a stack variable is allocated.
The epilogue returns the stack frame to the ambient domain by zeroing out the stack contents and setting the tag on the memory region back to 0.
The instrumentation uses \texttt{stzg} instruction introduced in \gls{mte} that performs such operation efficiently in the hardware.



\subsection{Domain-aware pointer authentication}
\label{sec:design:ptrauth}




The instrumentation now inserts \texttt{pac} and \texttt{aut} instructions for pointer store and load sites and uses \key{DB} \emph{selectively} for domain-private pointers. 
Compared to \emph{on-use} pointer authentication schemes, such as the one used in \gls{pa}-based \gls{cfi}~\cite{pac-it-up,kernel-pac}, on-load authentication provides better efficiency and compatibility when adapted to whole-program data pointers~\cite{pac-it-up}.

The previous \gls{pa} on-load pointer instrumentation scheme~\cite{pac-it-up} instruments the program in LLVM \gls{mir} to handle cases where pointers are \emph{spilled} into memory by the compiler.
However, we found that it does not reliably detect and instrument such cases due to the use of ad-hoc heuristics.
To handle this issue, we also implement our instrumentation in \gls{mir} after most optimizations are already performed but introduce a vastly improved instrumentation algorithm.
Our instrumentation relies on embedded metadata from our IR-level taint analysis and built-in type metadata to inform the instrumentation decisions without modifying the instruction selection pipeline.
Moreover, instead of using heuristics, we introduce a liveness analysis that keeps track of sensitive pointers on the stack frame and inside the registers.
Those improvements allow us to reliably enforce the on-load authentication scheme while also avoiding intrusive changes to instruction selection and register allocation stages.



\hdr{Identifying domain-private pointer load/store}
\thename employs an intra-procedural data-flow analysis that aims to find all IR pointer-containing variables, where an annotated domain-private variable, or its members, may flow to.
The analysis takes two sources: variables that are explicitly annotated with \texttt{DOM\_PRIV} and pointers returned from \texttt{capac\_malloc()}.
We use a worklist algorithm starting from the taint source to visit all variables and instructions recursively.
It follows the def-use chains of LLVM IR instructions and adds all users of the visited instruction to the worklist.
When an instruction that addresses a pointer variable (\eg, \texttt{alloca}, \texttt{getelementptr}) is encountered along the def-use chain, the analysis attaches metadata to the LLVM \texttt{value} that the later instrumentation can retrieve.
Additionally, the analysis performs backward tracking whenever it encounters a store instruction since def-use analysis cannot track such data-flow.
It adds the instruction's \emph{destination} operand to the worklist in such cases.




%
\hdr{MIR instrumentation} 
Our \gls{mir} instrumentation is based on \emph{liveness analysis} in compiler designs~\cite{reg-alloc}.
The instrumentation keeps track of registers and stack frame locations that potentially contain pointers/sensitive pointers, called the \emph{live pointer set}.
At compile time, the pass scans and visits \gls{mir} instructions.
On load or store instructions, it tries to retrieve LLVM type and taint information to extract whether the instruction accesses a pointer/sensitive pointer.
If such information is unavailable, the instrumentation consults the live pointer set.
For a store instruction, the instrumentation checks whether the \emph{source} operand (e.g., ``\texttt{x1}'' in \verb|str x1, [sp, #8]|) is within the live pointer set, and whether the operand is sensitive.
For a load instruction, the same check is performed on the instruction's \emph{destination} operand (e.g., ``\verb|[sp, #8]|'' in \verb|ldr x8, [sp, #8]|).

Instrumentation is performed on the load and store instructions as shown in \autoref{fig:asm-pa}. 
On non-sensitive pointer operands, it instruments with \key{DA} signing/authentication. 
Before signing an ambient pointer, the instrumentation inserts a masking instruction that zeros out the tag to prevent pointer forging through ambient code signing (\srforge).
When instrumenting sensitive pointer operands, \key{DB} is used. 
An instruction that loads the modifier from \texttt{ModReg} into a spare register (\texttt{fmov}) is also inserted. 
The modifier is then used for signing/authentication of the pointer.
Finally, the instrumentation updates the live pointer set after visiting each instruction.
For instance, if an instruction \emph{kills} a register or overwrites a stack frame location with non-pointer data, the register/stack frame index is removed from the live pointer set.

%

\subsection{Handling compatibility issues}
We developed supporting transformations that can be optionally enabled to solve \gls{pa}-based instrumentation compatibility issues.

\hdr{Type-unsafe object initialization}
The following example demonstrates the type-unsafe object initialization pattern found in C applications, including NGINX and LibreSSL:
\begin{lstlisting}[style=sslab-capacity-c,linewidth=0.97\columnwidth]
obj_t* obj =  calloc(1, sizeof(obj_t));
//...
if (obj->ptr == NULL){ /* Perform initialization */ }
\end{lstlisting}
In this example, when the program loads an uninitialized pointer from memory at line $3$, the on-load authentication would fail since the pointer is not signed, making the branch condition evaluate incorrectly.
To solve this, we introduce a pass that transforms the program's NULL checks to cover the AUT failed NULL value (e.g., \texttt{ptr != NULL \&\& ptr != 0x20000000}).
We also explored inferring the type of the zero-initialized objects, then recursively signing all of its containing pointers, using type inference methods proposed in \cite{type-after-type}, but found that the approach is infeasible on complex programs without heavy source code processing.
\begin{figure}[t]
\small
\begin{equation*}
\text{visit($id,t$)} = 
  \begin{cases}
  \text{\{sign$(id)$\} $\cup$ visit$(*id, t_1)$} & \text{if $t = t_1*$}\\
  $$\bigcup\limits_{i=1}^{n} \text{visit$(id_i, t_i)$}$$
  &\text{if t = struct $\{id_1:t_1, ..., id_n:t_n\}$}\\
  $$\varnothing$$ &\text{if $t$ = int}
  \end{cases}
\end{equation*}
\caption{Type-based global constructor generation. sign($id$) signs an in-memory pointer indexed by LLVM variable $id$.}
\label{fig:glob}
\end{figure}

\hdr{Global pointers in complex structures}
The prototype of PARTS~\cite{pac-it-up} only scans and signs pointers in the global structures at the top-most level, which misses many cases where pointers are stored in the nested structs.
We tackle this problem more comprehensively with a type-assisted approach that recursively visits global variables, as shown in \autoref{fig:glob}.
The algorithm uses the same type syntax and notations as PtrSplit~\cite{ptrsplit}, where int represents an integer type, $t_1*$ represents a pointer type, and struct type contains a list of types for each member. 
\text{visit$(id,t)$} is invoked on every global variable $id$ with type $t$ to generate initialization functions. They are then inserted before the program logic to sign the global pointers.

\hdr{Issues with MIR instrumentation}
We also resolved numerous compatibility issues with the \gls{mir}-based instrumentation.
First, we found that the \gls{mir}-based instrumentation must be aware of register liveness.
When a pointer inside a register is signed and then stored in memory, the signed in-register pointer may still be \emph{live}.
This side effect of pointer signing can cause following legitimate memory access with the in-register pointer to cause a fault.
As a solution, our instrumentation inserts a \texttt{xpac} instruction to remove the \gls{pac} of the in-register value, immediately following the \texttt{pac} instruction if (1) the \texttt{pac}-ed register is used again in the function, or (2) the \texttt{pac}-ed register is passed to another function as an argument.
Another issue that we found was the cases of \gls{pa} instrumentation failing on load and store instructions that have the same register in source and destination operands (\eg, \texttt{ldr, x8, [x8]}).
To resolve this, we transform these instructions to use distinct operands (\eg, \texttt{ldr, x8, [x9]}) instead.


\section{EVALUATION}

\label{sec:evaluation}

\issue{this need rewriting}
In this section, we first conduct a set of microbenchmarks on the reference monitor and \libcapac to illustrate the overhead induced by individual operations of \thename (\cref{sec:evaluation:mon-microbench}).
We describe the adaptation of \thename in protecting sensitive resources of three real-world applications in \cref{sec:evaluation:porting-app}.
Finally, we evaluate the performance of \thename-adapted applications in \cref{sec:evaluation:macrobench} to show \thename's overall impact on performance.

\hdr{Evaluation method} We ensured the functional correctness of our implementation by using a QEMU~\cite{qemu} ARM virtual machine that supports both \gls{pa} and \gls{mte} and conducted performance evaluations using the Apple Mac Mini with an M1 processor running Asahi Linux~\cite{asahi-linux}.
The M1 processor includes the \gls{pa} extension; however, to our knowledge, there is no publicly available hardware with ARM MTE.
The QEMU's MTE emulation is mature to a point where it is used for developing a \gls{mte}-based security feature for the Linux kernel to be used in the near future~\cite{kasan}.
We follow the method from a previous work~\cite{hakc} that emulated the \emph{worst-case} performance impact of \gls{mte} by using \emph{\gls{mte} analogs}.
We insert the \gls{mte} analogs to replace the tagging instructions in our tagged heap memory allocator and also modify \thename's instrumentation to automatically insert the \gls{mte} analogs in places of LLVM's \gls{mte} intrinsics.

\begin{table}[t]
\setlength\tabcolsep{2pt}
\footnotesize
\begin{tabularx}{\columnwidth}{@{}c >{\tt}l >{$}Y<{$} >{$}Y<{$} >{$}Y<{$} >{\hspace{5pt}\tt}l >{$}Y<{$} >{$}Y<{$} >{$}Y<{$}}
  \toprule
  & \textnormal{\thead{Syscall\\ \& API}} & \text{\thead{Base.\\(ns)}} & \text{\thead{Cap.\\(ns)}} & \text{\thead{\underline{Ovh.}\\(\%)}} & \textnormal{\thead{Syscall\\ \& API}} & \text{\thead{Base.\\(ns)}} & \text{\thead{Cap.\\(ns)}} & \text{\thead{\underline{Ovh.}\\(\%)}} \\
  \midrule
  \multirow{8}{*}[-0.0em]{\thead{\begin{sideways}Syscall\end{sideways}}} 
  & socket & 718.5 &   722.5 & \underline{0.56}
  & setsockopt & 235.7 &   242.2 & \underline{2.75} \\
  & bind &  400.0 &   408.0 & \underline{1.98}
  & listen & 351.2 &  361.5 & \underline{2.91}  \\
  & accept4 & 1236 & 1248 & \underline{0.96}
  & recvfrom & 192.6 &  199.4 & \underline{3.56} \\
  &  openat & 471.5 &   486.9 & \underline{3.26}
  & read &247.1 &   254.2  & \underline{2.88} \\
  & pwrite64 & 320.6 &   328.2 & \underline{2.35}
  & pread64 &  231.8 &   239.0 & \underline{3.06} \\
  & fstat &  185.9 &   191.5 & \underline{3.03}
  & fcntl &  148.1 &   155.8 & \underline{5.20} \\
  & dup3 &  132.5 &   138.0  & \underline{4.22}
  & lseek & 149.7 & 154.8 & \underline{3.38} \\
  & mmap & 130.0 &   130.0 & \underline{0.01} & \textbf{getpid} & 118.6 & - & -\\
  \midrule
  \multirow{4}{*}[0em]{\begin{sideways}\thead{API}\end{sideways}} &
  \texttt{capac\_malloc} & 23.46 & 50.28 & \underline{114.6} & \texttt{capac\_free} & 14.7 & 37.4 & \underline{154.4} \\
  & \texttt{capac\_enter} & - & 167.7 & - & \texttt{capac\_exit} & - & 161.2 & - \\
  & \texttt{delegate\_ptr}\textsuperscript{*} & - &  327.7 & - & \texttt{delegate\_fd}\textsuperscript{*} & - & 182.9 & - \\
  & \texttt{limit\_fd}\textsuperscript{*} & - &  142.4 & -&&&& \\
  \bottomrule
\end{tabularx}
  \begin{tablenotes}
  \item \textsuperscript{*} {\small \texttt{capac\_} \ prefix is omitted}
  \end{tablenotes}
\caption{The latency of \thename system calls intervened by reference monitor and \libcapac APIs (\textbf{Cap.}), in comparision with the baseline (\textbf{Base.}). Latency of \texttt{getpid} is also measured for comparison.}
\label{tbl:micro}
\end{table}


\subsection{Microbenchmarks}
\label{sec:evaluation:mon-microbench}
We perform a set of isolated microbenchmarks on the syscalls intervened by the reference monitor, \libcapac API functions, and the instrumentation. The results are listed in \autoref{tbl:micro}.

\begin{table*}[t]
	\centering
	\footnotesize
	\newcommand*{\belowrulesepcolor}[1]{%
  \noalign{%
    \kern-\belowrulesep
    \begingroup
      \color{#1}%
      \hrule height\belowrulesep
    \endgroup
  }%
}
\newcommand*{\aboverulesepcolor}[1]{%
  \noalign{%
    \begingroup
      \color{#1}%
      \hrule height\aboverulesep
    \endgroup
    \kern-\aboverulesep
  }%
}
\begin{threeparttable}

  \begin{tabularx}{\textwidth}{@{}lp{1.4cm}p{3cm}ccYcccc}
    \toprule
    &\multirow{2}{*}{\thead{Domain}}                          &
    \multirow{2}{*}{\thead{Isolated Objects \{Ref. types\}}} &
    \multirow{2}{*}{\thead{Domain                                                                                                                                                                         \\Switches}} &
    \multirow{2}{*}{\thead{Auth.                                                                                                                                                                          \\Syscalls}} &
    \multirow{2}{*}{\thead{$\mathcal{A}$}} &
    \multicolumn{2}{c}{\thead{Executed PACs}}                &
    \multicolumn{2}{c}{\thead{Executed AUTs}}                                                                                                                                                             \\
      &   &   &   &   &   & \packey{DB} & \packey{DA} & \autkey{DB} & \autkey{DA} \\[1ex]
    \toprule
    \multirow{4}{*}[-2em]{\begin{sideways}\thead{NGINX + LibreSSL}\end{sideways}}
    &
    \domconn                                                 & \makecell[l]{Server socket \{\fd\}                                                                                                         \\Client connection socket \{\fd\}}   &
    $2$                                                      & $16$                                                         & $-$ & $-$ & $104$   & $-$         & $590$                                   \rule{0pt}{4.0ex}\\[2ex]
    &  
    \domhs                                                   & \makecell[l]{Server priv. key \{\cpath, \fd, \ptr\}                                                                                        \\TLS session key \{\ptr\}\\Client connection socket \{\fd\textsuperscript{\ddag}\}} &
    $3$                                                      & $18$                                                         & $4$ & $9$ & $10.1$K & $13$        & $134$K                                   \rule{0pt}{6.0ex}\\[2ex]

    & 
    
    \domses                                                  & \makecell[l]{TLS session key \{\ptr\textsuperscript{\ddag}\}                                                                               \\Client connection socket \{\fd\textsuperscript{\ddag}\} } &
    \makecell[c]{$7$/$8$/$10$/$13$                                                                                                                                                                        \\/$22$/$37$/$70$\textsuperscript{\dag}} &
    \makecell[c]{$14$/$16$/$20$/$28$                                                                                                                                                                      \\/$44$/$76$/$140$\textsuperscript{\dag}} &
    $3$                                                      &
    \makecell[c]{$43$/$49$/$61$/$85$                                                                                                                                                                      \\/$133$/$229$/$421$\textsuperscript{\dag}} &
    \makecell[c]{$818$/$852$/$917$/$1$K                                                                                                                                                                   \\/$1.3$K/$1.8$K/$2.9$K\textsuperscript{\dag}} &
    \makecell[c]{$3.4$K/$6.5$K/$12$K/$25$K                                                                                                                                                                \\/$50$K/$99$K/$199$K\textsuperscript{\dag}} &
    \makecell[c]{$7.7$K/$10$K/$14$K/$24$K                                                                                                                                                                 \\/$42$K/$79$K/$154$K\textsuperscript{\dag}}   \rule{0pt}{5.0ex}\\[2ex]
    & 
    \domamb                                                  & $-$                                                          &
    $-$                                                      &
    \makecell[c]{$94$/$99$/$100$/$109$                                                                                                                                                                    \\/$121$/$145$/$193$\textsuperscript{\dag}}  & $-$ & $-$ & $33.6$K & $-$   & $164$K \rule{0pt}{4.0ex} \\[2ex]
    \midrule
    \multirow{1}{*}[-1.1em]{\begin{sideways}\thead{SSH}\end{sideways}}
    &\textsc{PrivKey}                                                 & \makecell[l]{Client private key \{\cpath,\fd,\ptr\}}   &
    $1$                                                      & $72$                                                         & $55$ & $18$ & $6$K   & $24$         & $33.5$K                                 \rule{0pt}{2.0ex} \\[-2ex]
    &
    \domamb                                                  & $-$                                                          &
    $-$                & $37$                                      &
    $-$ & $-$ & $14.7$K & $-$   & $114$K \rule{0pt}{3.0ex}\\[1ex]
    \midrule
    \multirow{2}{*}[-0.9em]{\begin{sideways}\thead{wget}\end{sideways}}
    &\textsc{FileDownload}                                                 & \makecell[l]{Downloaded file \{\cpath,\fd,\ptr\}}   &
    $3$                                                      & $325$                                                         & $1$ & $129$ & $3$K   & $385$         & $8.6$K                                   \rule{0pt}{2.0ex}\\[-2ex]
    &
    \domamb                                                                                                      &
    $-$                                               & $-$ &     $131$ & $-$ & $-$ & $703$ & $-$   & $1.6$K \rule{0pt}{3.0ex}\\[1ex]
    \bottomrule
  \end{tabularx}

  \begin{tablenotes}
    \item  \textsuperscript{\dag}  {\small Measurements from \{16K,32K,64K,128K,256K,512K,1024K\} HTTPS file size configurations } \hspace{1.0em} \textsuperscript{\ddag}  {\small  Delegated references} \vspace{0.5em}
    \item {\small$ \mathcal{A}$ = Number of domain-private memory allocations}
  \end{tablenotes}

  \caption{\thename domains and protected assets in evaluated applications. Columns 3-9 show runtime measurements of \thename operations executed during a single iteration of \{HTTPS file transfer (NGINX) / SSH handshake / wget file download\}.}

  \label{tbl:porting-app}
  \label{tbl:app}
\end{threeparttable}

\end{table*}

\hdr{System call latency with reference monitor}
We measured the average latency of \thename-protected syscall used in our evaluated applications that authenticate file paths and \fds, and compared them against un-protected syscalls.
The overhead induced by \thename includes the latency from syscall hooking and \thename's reference authentication.
Our results reported an average of $2.65\%$ across the measured syscalls.
This shows the efficiency of \thename reference monitor design that fully utilizes hardware acceleration provided by \gls{pa}, not to mention the simplicity of capability where the reference alone is enough to make access control decisions without complex bookkeeping on domains and resources.
Also, this overhead is imposed only on the \thename-enabled processes.

\hdr{\libcapac APIs latencies}
The overhead of \texttt{capac\_malloc()} and \texttt{capac\_free()} originates from domain authentication and \gls{mte}-based memory tagging. 
Measuring the latency of domain private memory allocation and free on blocks of $1$KB, \texttt{capac\_malloc()} and \texttt{capac\_free()} reported approximately $114.6\%$ and $154.4\%$ .
\texttt{capac\_enter} and \texttt{capac\_exit}, \texttt{capac\_delegate\_ptr}, \texttt{capac\_delegate\_fd} and \texttt{capac\_limit\_fd} are simple \texttt{ioctl} calls to the in-kernel reference monitor, where only \texttt{capac\_limit\_fd} does not requires the key switching.
Their latencies are on par with general syscalls.
Besides a roundtrip to the kernel mode, we suspect their main source of overhead is from writing to the \gls{pa} key registers.

\begin{figure}[b]
	\centering
	\includegraphics[width=.95\columnwidth]{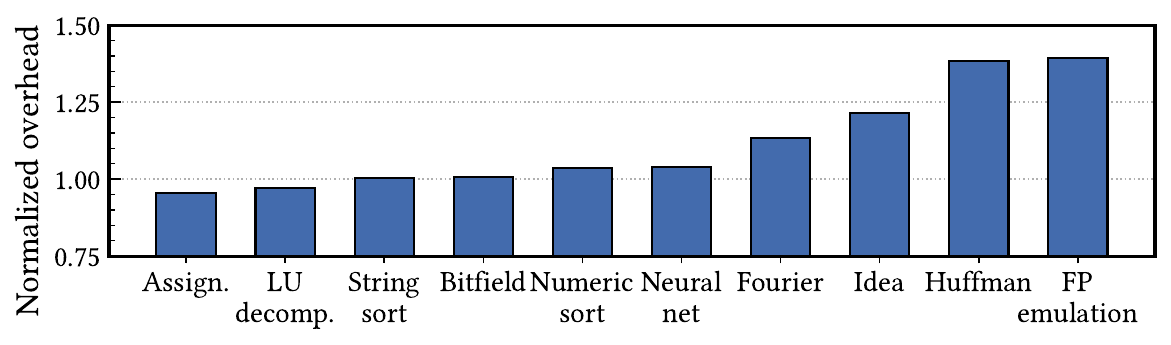}
	\caption{Overhead of instrumented nbench-byte, normalized to uninstrumented baseline and sorted by overhead.}
	\label{fig:nbench}
\end{figure}

\hdr{nbench-byte benchmark}
\label{sec:evaluation:nbench}
We measure the isolated overhead of \thename's instrumentation on the \emph{nbench-byte}~\cite{nbench} benchmark, which is also used by several previous works on \gls{pa}-based defenses~\cite{pactight,pac-it-up}.
nbench-byte consists of $11$ CPU and memory subsystem benchmarks that allow us to measure the overhead of \thename in general computations.
\autoref{fig:nbench} shows the results of nbench-byte compiled with \thename's instrumentation that enforces \gls{pa} on on-save signing and on-load authentication of all in-program pointers.
On average, \thename's instrumentation incurs $11.37\%$ performance degradation to the uninstrumented version.



\subsection{Adapting \thename to real-world programs}
We exemplify \thename's capability domains by adapting it to several open-source applications studied by previous isolation frameworks~\cite{erim,shreds,ptrsplit}. 
These applications include OpenSSH's \texttt{ssh} utility~\cite{openssh}, \texttt{wget} file download utility~\cite{wget}, and NGINX webserver~\cite{nginx} compiled along the TLS library LibreSSL~\cite{libressl}. 
Representative code examples of our modifications can be found in \autoref{appendix:example}.
The applications are also compiled with the \thename-instrumented \emph{musl} libc~\cite{musl} adapted to handle signed pointers in system call arguments.
\autoref{tbl:porting-app} summarizes the \thename domains in protected applications and their protected assets.


\hdr{OpenSSH's ssh (v9.3p1)}
We demonstrate \thename's program secrets protection throughout their life-cycles with \texttt{ssh}.    
We compile OpenSSH to use its built-in crypto library and use \thename to isolate the private key of \texttt{ssh}.
We introduce a new domain, \textsc{Privkey}, with exclusive access to the private key file and its in-memory buffer.
This domain spans from when the private key is loaded from the file system until after the login operation succeeds.
We then assign the static private key to the domain, allocate the private key buffer (\texttt{struct sshkey}) in private memory, annotate its pointers with \texttt{DOM\_PRIV}, and finally wrap functions that access the private key within domain entry gates.
These changes require about $50$ LoC changes over 13 functions across 5 files.

\begin{figure*}[t]
	\centering
	\includegraphics[width=\textwidth]{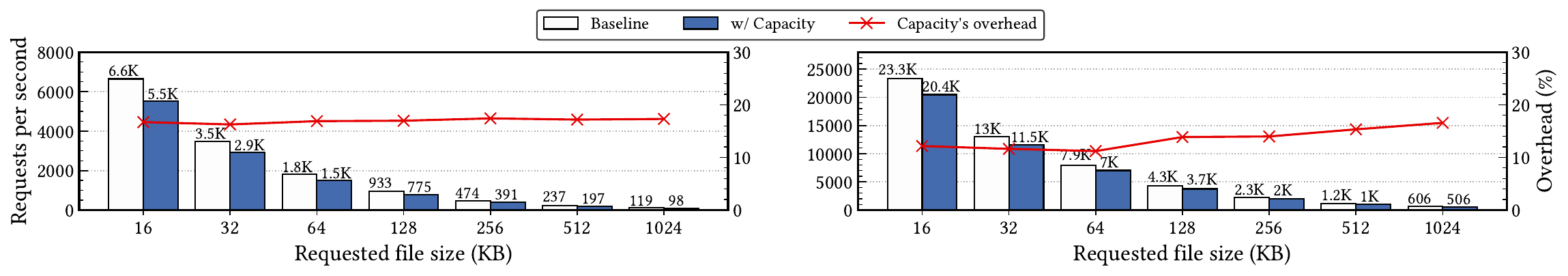}
	\begin{minipage}[t]{.49\linewidth}
		\centering
		\subcaption{Single-threaded}
		 \vspace{-1em}
	\end{minipage}%
	\begin{minipage}[t]{.49\linewidth}
		\centering
		\subcaption{Multi-threaded (8 Threads)}
		 \vspace{-1em}
	\end{minipage}
	\caption{\textsc{Baseline} webserver vs. \thename-enabled webserver throughput benchmark performed with local \texttt{ab} client.
	}
	\label{tab:exec-macro}
\end{figure*}

\hdr{\texttt{wget}  (v1.21.2)}
We use \thename to isolate the file received from the internet in \texttt{wget} to demonstrate least-privilege compartmentalization.
We create a domain called \textsc{FileDownload} with exclusive access to the downloaded buffer and the output file path/\fd. 
We then modify the initialization of \texttt{wget} to assign the output file access to \textsc{FileDownload}, and use \texttt{cap\_limit\_fd} to revoke all other attributes from the file's \fd except \texttt{CAP\_WRITE}. 
This ensures that the domain has write-only access to the output \fd \footnote{While \texttt{wget} already make the file handle write-only, \thename enables more fine-grained in-process access control that supports read and write domains.}.
The domain encapsulates the function \texttt{retrieve\_url}, which fetches the requested file from a URL into a \texttt{capac\_malloc}-allocated buffer before writing into the output file's \fd. 
We also annotate pointers variables that point to the downloaded file in memory with \texttt{CAPAC\_VAR}. 
The total modifications to \texttt{wget} is about $30$ LoC.

\begin{figure}[b]
	\centering	\includegraphics[width=0.98\columnwidth]{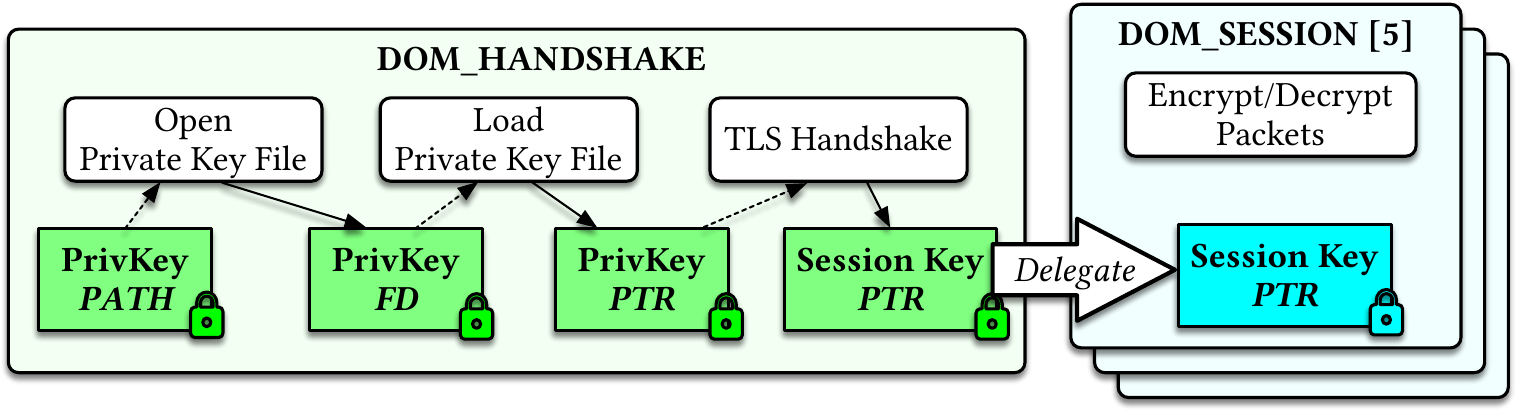}
	\caption{\thename enables life-cycle private key protection and per-instance isolation of session keys in the NGINX webserver.}
	\label{fig:webserver-domains}
\end{figure}


\label{sec:using-capacity}\label{sec:design:webserver}\label{sec:evaluation:porting-app}\label{sec:usage}

\hdr{NGINX (v1.23.0) + LibreSSL (v3.5.3)}
We use the NGINX webserver and its crypto library, LibreSSL, to demonstrate the multi-domain interaction of \thename. 
We create three domains in the webserver, namely \domhs, \domses, and \domconn to isolate the server private key, session keys, and connection sockets with the least privilege principle in mind.
\autoref{tbl:porting-app} displays a summary of the webserver domains and their isolated resources.
We modified $223$ LoC in LibreSSL and $114$ LoC in NGINX.


\autoref{fig:webserver-domains} illustrates the main operations and their interaction with domain-private objects in domain \domhs and \domses.
\domhs, which consists of $5$ NGINX and $2$ LibreSSL functions endowed with exclusive rights to the server private key.

\domses protects session keys by hosting $17$ functions,  mostly located inside LibreSSL's AES-GCM implementation.
We generate a unique instance ID and use it as the modifier for each session since a session's resources are mostly temporary and can be isolated. 
As \domhs execution finishes, it produces a TLS session key, which is to be \emph{delegated} to the corresponding \domses instance using the API \texttt{capac\_delegate\_ptr} (\autoref{tab:apis}).
The domain-private stack  is a noteworthy feature of \thename used in \domses.
When a function performs complex cryptographic operations, we use \texttt{DOM\_PRIV\_STACK} instead of tracking the data flow of sensitive data bouncing around in the function's local variables.

Finally, \domconn manages the webserver's sockets.
We wrap the functions that handle sockets inside its domain switches of \domconn, such that when the server requests its \emph{server socket} \fd using \texttt{sys\_socket}, the returned \fd is domain-private.
On client connection requests, \domconn invokes \texttt{sys\_accept4} with the server socket as the input and retrieves a \emph{client} socket \fd.
\domconn then delegate the returned \emph{client} socket \fd to \domhs, using \texttt{capac\_delegate\_fd}. 
After TLS handshaking, \domhs again delegates the socket \fd to \domses.
Notably, \texttt{CAP\_SOCKET} attribute is enabled by default when socket \fds are created, segregating socket and non-socket \fds in syscalls without the user's intervention.





\begin{table}[t]
    \centering
\small
\begin{tabular}{lccc}
\toprule
       \thead{Program} &\thead{Baseline (ms)} &\thead{w/ \thename (ms)}&\thead{Overhead (\%)}  \\
        \midrule
     \texttt{ssh}&$347.35$ & $348.62$  & $0.37$ \\
 \rowcolor[gray]{0.925}    \texttt{wget}&  $17.12$ & $18.14$&$5.95$ \\
     \bottomrule
\end{tabular}
    \caption{\thename overheads on  \texttt{ssh} and \texttt{wget}.}
    \label{tbl:sshwget}
\end{table}

\subsection{Application performance benchmark} \label{sec:evaluation:macrobench}
We perform performance benchmarks on the adapted applications.
These applications run with all design components of \thename explained thus far, including compile-time instrumentation, compatibility fixes, \libcapac, the kernel-level reference monitor, and backward and forward-edge CFI.
To accurately profile the source of \thename's performance overhead, we inserted probes into the reference monitor, \libcapac, and the instrumented program code to collect \emph{runtime} execution count of \thename operations during a single iteration of execution, as shown in~\autoref{tbl:porting-app}.


\hdr{\texttt{ssh} and \texttt{wget}} 
For \texttt{ssh}, we measure the average overhead of connecting to a local server $10,000$ times and compare it against the unmodified version.
We evaluate the protected \texttt{wget} by timing the average overhead in latency when requesting a 1MB file from a local http webserver $10,000$ times compared to the baseline.
\autoref{tbl:sshwget} displays the \thename-incured performance overheads in \texttt{ssh} and \texttt{wget}, which shows only minimal slowdown for both applications.


\hdr{NGINX + LibreSSL} We evaluate the throughput of the \thename-enabled NGINX webserver against its unmodified counterpart, which we shall call \textsc{Baseline}.
We use the Apache benchmarking tool, \texttt{ab}~\cite{apache-benchmark} locally (i.e., no network latency) to measure the number of requests processed per second, with varying file sizes, \{$16$K, $32$K ... $1$M\}, on also single-threaded and multi-threaded (8 threads) settings.
$100,000$ requests were performed, and the keep-alive option was used to avoid re-initializing the connections.
The cipher suite negotiated between the client and the server was \texttt{ECDHE-RSA-AES256-GCM-SHA384}.

The webserver benchmark results are displayed in \autoref{tab:exec-macro}, which shows the average performance degrades about $17\%$ for the single-threaded server and $13.54\%$ for the multi-threaded settings across all file size configurations.
The latency measurements for \thename operations from our microbenchmark (\autoref{tbl:micro}) allow us to reason that the performance overheads from the reference monitor and \libcapac would be very limited.
The varied file sizes allow us to capture the characteristics of \thename's performance overhead, which increases as the requested file sizes increase.
Along with the runtime statistic collected in \autoref{tbl:app}, it can be observed that \thename's overhead protecting pointers and memory is proportional to the amount of computation.
We can conclude that the performance overhead of \thename would be predominantly from the increased number of instructions due to instrumentation.



\section{SECURITY ANALYSIS \& DISCUSSION}
\label{sec:discussion}
In this section, we use the security requirements outlined in \cref{sec:overview:secreq} to analyze the security of \thename. 
We show that these security requirements are achieved throughout the life cycles of program resources.
We also discuss the possible mitigation for brute-forcing attacks at the end of the section.

\hdr{Non-impersonable domains (\srimper)}
The domain entry explained in \cref{sec:design:libcapac} renders invoking domain switches into arbitrary domains in unintended locations impossible. 
Also, our trusted domain ID fetching and in-register tag protection prevent the attacker from impersonating a domain by corrupting the currently activating domain ID spilled onto the stack or accessible globally.

\hdr{Complete mediation (\srcompmed)}
\thename's reference monitor ensures the use of \fd and \cpath in \glspl{syscall} is always authenticated, and all userspace \fds are signed.
Its whole-program instrumentation guarantees the signing and authentication of all ambient and domain-private pointers.
Bypassing \gls{pa} checks by jumping into non-entry locations of a function is infeasible with the presence of \gls{cfi}.
Hence, assuming no uninstrumented code, \thename achieves complete mediation of all resource usage.

\hdr{Non-reusable references (\srreuse)}
Under \thename's protection, a non-owner domain cannot access a protected file since the inode signature check will fail during file import.
A domain-private \fd leaked to another domain remains unusable thanks to our \fd authentication scheme using the domain key.
The reference monitor also prevents the OS from reusing \fd numbers after one is deallocated.
In addition, a domain compromised by the adversary cannot reuse in-memory pointers pointing to other domains' private memory.
Due to domain key mismatch, the authentication would fail at \texttt{pacdb} sites.
A pointer signed with \key{DB} would also fail to authenticate at pointer 
load sites where \key{DA} is used.

\hdr{Non-forgable references (\srforge)}
It is impossible to forge an illegal path reference to a protected file object since path string references are resolved to their in-kernel data structures.
Forging an \fd in the userspace is also prevented using a secret (to userspace) modifier, which allows the kernel to remain the sole issuer of unforgeable and protected file descriptors.

With two methods, the adversary may forge a tagged and signed pointer to access domain-private memory.
First, the adversary may overwrite an existing in-memory pointer through memory corruption.
Such attacks are thwarted by the \gls{pa}-based pointer authentication, whose primary purpose is to detect corruptions of in-memory pointers.
Second, through code reuse attacks, the adversary may sign arbitrarily manipulated data (e.g., 64-bit data with the victim’s domain tag and target sensitive memory address) using the available \texttt{pacda} and \texttt{pacdb} gadgets.
For \texttt{pacdb} gadgets, \thename PA contexts created by key switching force the attacker to use them within the target domain. 
We consider the isolated domains small enough so that it can be verified that the domain code is free of signing gadgets.
However, this is a substantial and realistic threat when completely safe TCB is not guaranteed \cite{civ, kernel-pac}, and a scanner to remove such gadgets could be utilized~\cite{kernel-pac}.
Given that no such convenient gadgets exist in \thename domains, the adversary may still launch an extremely sophisticated attack by finding gadgets that (1) leak the sensitive memory address, (2) create a pointer and add the appropriate tag to it, (3) use \texttt{pacda} to create an ambient pointer that points to domain-private area, and (4) pass the signed pointer to an \texttt{autda} site.
\thename's instrumentation currently zeros out the tag of ambient pointers before signing them, rendering such code reuse attacks infeasible.



\hdr{Brute-forcing atttacks}
An attacker can exploit certain ideal situations, such as infinite thread spawning~\cite{hacking-blind}, to try different \gls{ac} values until authentication succeeds.
This is an inherent weakness with \gls{pa}-based security.
Such attacks can be overcome by placing thresholds on the number of crashed threads due to \gls{pa} traps and systems call authentication failures~\cite{kernel-pac}.
\label{sec:discussion:sec}

\section{RELATED WORK}
\label{sec:related-work}
We explain the previous works that are closely related to our work in terms of security principle (capability), use of hardware primitives (ARM \gls{pa} and \gls{mte}), and objective (compartmentalization).

\hdr{Capability-based resource access control}
\label{sec:related-work:capability}
Capability-based access control has long been sought after due to its advantages in achieving the principle of least privilege~\cite{obj-cap,cap-addressing,eros,capsicum,cheri,cheri-risc,cheri-jni,jkernel,caja,framer,sel4}.
Previous works explored capabilities in OS access control and memory safety.
In the OS sphere, Capsicum~\cite{capsicum} introduced \emph{process-level} \fd capabilities to UNIX and has been adapted by FreeBSD~\cite{freebsd-capsicum}.
The SeL4 microkernel provides secure and fine-grained access to system resources through capability tokens~\cite{sel4}. 
Toward applying capabilities principles for memory safety, CHERI~\cite{cheri-risc} introduced a CPU architecture with built-in support for capability-based memory access control.
Subsequent works introduced extensions to the CHERI architecture with additional security features such as domain memory isolation~\cite{cheri,cheri-jni,cheriabi}.
Recently, Capstone~\cite{capstone} revised CHERI's base design to support revocable delegation and an extensible privilege hierarchy.

Unlike previous capability systems, \thename's capabilities can be consistently applied to file and memory object references.
Compared to Capsicum, \thename does not require isolating program components into processes; its capability model seeks to achieve least-privilege domains \emph{within} a process. 
Moreover, \thename's enforcement is much more lightweight thanks to hardware-assisted cryptographic authentication, 
Compared to capability architectures like CHERI, \thename's design must address the security requirements of capability on existing hardware features, \eg, complete mediation of pointer uses since the processor does not enforce them automatically.
Also, without hardware support, \thename lacks fine-grained permissions that CHERI provides over pointers, e.g., per-object read/write/execute capabilities.

\hdr{PA and MTE-based software defenses}
\label{sec:related-work:pa-mte-defenses}
Many \gls{pa}-based runtime defense mechanisms have been proposed~\cite{pac-it-up,pacsan,pacstack,pactight,kernel-pac,apple-pac,qualcomm-pac,ptauth} and deployed in commodity systems~\cite{apple-pac,qualcomm-pac}.
\gls{mte}-supported bug detecting tools have been introduced~\cite{memory-tagging,mte-kernel-support,mte-stack} with improved performance compared to software-only approaches.
Notably, PARTS~\cite{pac-it-up} introduces a \gls{dpi} policy that protects program-wide data and code pointers, but its instrumentation framework is not evaluated on real-world applications.
PacTight~\cite{pactight} enforces \emph{non-forgability}, \emph{non-copyability}, \emph{non-dangling} properties on only \emph{sensitive pointers} with an intricate signing and authentication scheme.
PTauth~\cite{ptauth} introduces a \gls{pa}-based use-after-free by authenticating pointers with their pointed-to object.

\thename introduces a robust framework for whole-program instrumentation of pointer authentication.
Moreover, previous \gls{pa}-based systems rely on modifiers to establish different authentication contexts.
On the other hand, \thename establishes authentication contexts through its key switching mechanism and uses the modifier to isolate references between domain instances.
Hence, different types of references can share a consistent authentication method in kernel and userspace without a complicated modifier assignment scheme.
However, \thename's use of \key{DB} for the domain key allows it to incorporate previous works on \gls{pa}-based \gls{cfi} into its design~\cite{pac-it-up,pacstack,kernel-pac,apple-pac}.


\hdr{Intra-process compartmentalization}
\label{sec:related-work:ipi}
Several works have leveraged x86 architecture's \gls{pku} to protect program domains and showed that \gls{pku}-based isolation significantly lower overhead compared to process-based and \gls{sfi}-based isolation~\cite{erim,hodor,cerberus,jenny,pkrusafe}.
Notably, ERIM~\cite{erim} and Hodor~\cite{hodor} laid the groundwork for utilizing \gls{pku} for intra-process isolation through a meticulously designed call gate between domains.
\thename is motivated by the underexplored design space for ARM-based in-process compartmentalization and access control.
Shred~\cite{shreds} proposed isolating the memory of in-process execution units with AArch32's Domain memory protection, but the feature has been removed in AArch64.

HAKC~\cite{hakc} is a framework for Linux device driver compartmentalization, also leveraging \gls{pa} and \gls{mte}.
Different from HAKC, \thename's use of \pamte is specialized for memory and system resources isolation in the userspace. In particular, with pre-defined \glspl{acl} as \gls{pa} modifiers, HAKC's pointer-use sites are tied to an access control policy,  regardless of the calling context. On the other hand, \thename's pointer authentication scheme checks if the reference is issued for the currently active domain, depending on the context (the effective \key{DB}).
This key difference brings implications that are pivotal in \thename's design. First, it allows \gls{pa} contexts to span across the user and kernel, enabling \thename's authentication scheme without a complex user-kernel modifier sharing scheme.
Second, functions can be called from multiple domains; the \texttt{AUT} authentication checks in the exact code location and uses different keys and therefore authenticates differently for each domain context.



Recently, there have been proposals for reference monitors better suited for in-process isolation~\cite{donky,jenny,cerberus,muswitch}.
Jenny~\cite{jenny} proposed a secure and efficient syscall reference monitor that delegates access control to userspace.
$\mu$Switch~\cite{muswitch} leverages implicit context switches to delay the kernel resource context switching until a syscall is performed.
While previous works on reference monitors require developers to write filtering rules to be applied to the monitor, \thename's cryptographically-secured \fd references are coherently built into the program's semantic, and their \gls{pa}-assisted authentication is both transparent and efficient.


\hdr{Isolation boundaries} 
\thename currently only provides intra-procedural sensitivity annotation propagation.
Automated whole-program and inter-procedural tracking of secret propagation is a field of its own, and previous works have proposed methods towards the objective~\cite{cryptompk,ptrsplit,dynpta,aces}. 
We expect that \thename can incorporate such methods in the future, although it is currently out of the scope of this paper, which focuses on introducing a new isolation mechanism.
On another note, a recent work~\cite{civ} discussed the perils of artificially drawn in-process boundaries and their interfaces. 
Such attack surface must be eliminated or minimized through validation and interface narrowing when porting existing programs to use \thename. 
If a program is redesigned or written from scratch with \thename, then a conscious effort can be made to have a secure interface by design.
The evaluated prototypes focus on showing the feasibility of \thename as an isolation mechanism, and we regard the validation process to be an orthogonal issue.

\section{Conclusion}
This paper proposed a novel design called \thename that enables the compartmentalization of in-process domains by employing hardware-accelerated and cryptographically-authenticated capabilities.
We presented complete mediation and authentication schemes that satisfy the security requirements of capability systems throughout the life cycle of sensitive domain objects and addressed compatibility issues when adapting \pamte-based instrumentation for large programs.
We evaluated \thename through microbenchmarks and real-world applications, including an NGINX webserver prototype in which the important resources are protected in secure domains.
The results show the efficacy of \thename the in terms of performance with an average webserver throughput overhead of $17\%$ for single-threaded and $13.54\%$ for multi-threaded experiments.

\balance

\begin{acks}
We deeply appreciate the anonymous reviewers for their constructive comments and feedback.
This work was supported by grants funded by the Korean government: the National Research Foundation of Korea (NRF) grant (NRF-2022R1C1C1010494),
Institute of Information \& Communications Technology Planning \& Evaluation (IITP) grants (No. 2022-0-00688, No. 2022-0-01199),
and Korea Internet \& Security Agency (KISA) grant (1781000009).

\end{acks}

\newpage

\bibliographystyle{ACM-Reference-Format} 
\bibliography{capacity}


\begin{thebibliography}{74}


\ifx \showCODEN    \undefined \def \showCODEN     #1{\unskip}     \fi
\ifx \showDOI      \undefined \def \showDOI       #1{#1}\fi
\ifx \showISBNx    \undefined \def \showISBNx     #1{\unskip}     \fi
\ifx \showISBNxiii \undefined \def \showISBNxiii  #1{\unskip}     \fi
\ifx \showISSN     \undefined \def \showISSN      #1{\unskip}     \fi
\ifx \showLCCN     \undefined \def \showLCCN      #1{\unskip}     \fi
\ifx \shownote     \undefined \def \shownote      #1{#1}          \fi
\ifx \showarticletitle \undefined \def \showarticletitle #1{#1}   \fi
\ifx \showURL      \undefined \def \showURL       {\relax}        \fi
\providecommand\bibfield[2]{#2}
\providecommand\bibinfo[2]{#2}
\providecommand\natexlab[1]{#1}
\providecommand\showeprint[2][]{arXiv:#2}

\bibitem[{Apple}(2021)]%
        {apple-pac}
\bibfield{author}{\bibinfo{person}{{Apple}}.} \bibinfo{year}{2021}\natexlab{}.
\newblock \bibinfo{title}{Apple Platform Security}.
\newblock \bibinfo{howpublished}{\url{ https://manuals.info.apple.com/MANUALS/1000/MA1902/en_US/apple-platform-security-guide.pdf }}.
\newblock
\newblock
\shownote{Last accessed May 05 , 2021,}.


\bibitem[{ARM Ltd}(2021)]%
        {arm-arch}
\bibfield{author}{\bibinfo{person}{{ARM Ltd}}.} \bibinfo{year}{2021}\natexlab{}.
\newblock \bibinfo{title}{{Arm Architecture Reference Manual Armv8, for Armv8-A architecture profile}}.
\newblock \bibinfo{howpublished}{\url{https://developer.arm.com/documentation/ddi0487/ga}}.
\newblock
\newblock
\shownote{{Last accessed Nov 18 , 2021,}}.


\bibitem[{ARM Ltd}(2022)]%
        {armv8-m}
\bibfield{author}{\bibinfo{person}{{ARM Ltd}}.} \bibinfo{year}{2022}\natexlab{}.
\newblock \bibinfo{title}{{Armv8-M Architecture Reference Manual}}.
\newblock \bibinfo{howpublished}{\url{https://developer.arm.com/documentation/ddi0553/bs}}.
\newblock
\newblock
\shownote{{Last accessed May 15 , 2022,}}.


\bibitem[{ARM Ltd.}(2023)]%
        {mte}
\bibfield{author}{\bibinfo{person}{{ARM Ltd.}}} \bibinfo{year}{2023}\natexlab{}.
\newblock \bibinfo{title}{{ARMv8.5-A Memory Tagging Extension}}.
\newblock \bibinfo{howpublished}{\url{https://developer.arm.com/-/media/Arm Developer Community/PDF/Arm_Memory_Tagging_Extension_Whitepaper.pdf}}.
\newblock
\newblock
\shownote{Last accessed March 10 , 2022,}.


\bibitem[Bittau et~al\mbox{.}(2014)]%
        {hacking-blind}
\bibfield{author}{\bibinfo{person}{Andrea Bittau}, \bibinfo{person}{Adam Belay}, \bibinfo{person}{Ali Mashtizadeh}, \bibinfo{person}{David Mazières}, {and} \bibinfo{person}{Dan Boneh}.} \bibinfo{year}{2014}\natexlab{}.
\newblock \showarticletitle{Hacking Blind}. In \bibinfo{booktitle}{\emph{2014 IEEE Symposium on Security and Privacy}}. \bibinfo{pages}{227--242}.
\newblock
\urldef\tempurl%
\url{https://doi.org/10.1109/SP.2014.22}
\showDOI{\tempurl}


\bibitem[Bittau et~al\mbox{.}(2008)]%
        {wedge}
\bibfield{author}{\bibinfo{person}{Andrea Bittau}, \bibinfo{person}{Petr Marchenko}, \bibinfo{person}{Mark Handley}, {and} \bibinfo{person}{Brad Karp}.} \bibinfo{year}{2008}\natexlab{}.
\newblock \showarticletitle{Wedge: Splitting Applications into Reduced-privilege Compartments}. In \bibinfo{booktitle}{\emph{Proceedings of the 5th USENIX Symposium on Networked Systems Design and Implementation}} (San Francisco, California) \emph{(\bibinfo{series}{NSDI'08})}. \bibinfo{publisher}{USENIX Association}, \bibinfo{address}{Berkeley, CA, USA}, \bibinfo{pages}{309--322}.
\newblock
\showISBNx{111-999-5555-22-1}


\bibitem[Brumley and Song(2004)]%
        {privtrans}
\bibfield{author}{\bibinfo{person}{David Brumley} {and} \bibinfo{person}{Dawn Song}.} \bibinfo{year}{2004}\natexlab{}.
\newblock \showarticletitle{Privtrans: Automatically Partitioning Programs for Privilege Separation}. In \bibinfo{booktitle}{\emph{Proceedings of the 13th Conference on USENIX Security Symposium - Volume 13}} (San Diego, CA) \emph{(\bibinfo{series}{SSYM'04})}. \bibinfo{publisher}{USENIX Association}, \bibinfo{address}{Berkeley, CA, USA}, \bibinfo{pages}{5--5}.
\newblock


\bibitem[Chen et~al\mbox{.}(2016)]%
        {shreds}
\bibfield{author}{\bibinfo{person}{Yaohui Chen}, \bibinfo{person}{Sebassujeen Reymondjohnson}, \bibinfo{person}{Zhichuang Sun}, {and} \bibinfo{person}{Long Lu}.} \bibinfo{year}{2016}\natexlab{}.
\newblock \showarticletitle{Shreds: Fine-Grained Execution Units with Private Memory}. In \bibinfo{booktitle}{\emph{2016 IEEE Symposium on Security and Privacy (SP)}}. \bibinfo{pages}{56--71}.
\newblock


\bibitem[Chisnall et~al\mbox{.}(2017)]%
        {cheri-jni}
\bibfield{author}{\bibinfo{person}{David Chisnall}, \bibinfo{person}{Brooks Davis}, \bibinfo{person}{Khilan Gudka}, \bibinfo{person}{David Brazdil}, \bibinfo{person}{Alexandre Joannou}, \bibinfo{person}{Jonathan Woodruff}, \bibinfo{person}{A.~Theodore Markettos}, \bibinfo{person}{J.~Edward Maste}, \bibinfo{person}{Robert Norton}, \bibinfo{person}{Stacey Son}, \bibinfo{person}{Michael Roe}, \bibinfo{person}{Simon~W. Moore}, \bibinfo{person}{Peter~G. Neumann}, \bibinfo{person}{Ben Laurie}, {and} \bibinfo{person}{Robert~N.M. Watson}.} \bibinfo{year}{2017}\natexlab{}.
\newblock \showarticletitle{CHERI JNI: Sinking the Java Security Model into the C}.
\newblock \bibinfo{journal}{\emph{SIGARCH Comput. Archit. News}} \bibinfo{volume}{45}, \bibinfo{number}{1} (\bibinfo{date}{apr} \bibinfo{year}{2017}), \bibinfo{pages}{569–583}.
\newblock
\showISSN{0163-5964}
\urldef\tempurl%
\url{https://doi.org/10.1145/3093337.3037725}
\showDOI{\tempurl}


\bibitem[Clements et~al\mbox{.}(2018)]%
        {aces}
\bibfield{author}{\bibinfo{person}{Abraham~A Clements}, \bibinfo{person}{Naif~Saleh Almakhdhub}, \bibinfo{person}{Saurabh Bagchi}, {and} \bibinfo{person}{Mathias Payer}.} \bibinfo{year}{2018}\natexlab{}.
\newblock \showarticletitle{{ACES}: Automatic Compartments for Embedded Systems}. In \bibinfo{booktitle}{\emph{27th USENIX Security Symposium (USENIX Security 18)}}. \bibinfo{publisher}{USENIX Association}, \bibinfo{address}{Baltimore, MD}, \bibinfo{pages}{65--82}.
\newblock
\showISBNx{978-1-939133-04-5}


\bibitem[Connor et~al\mbox{.}(2020)]%
        {pku-pitfall}
\bibfield{author}{\bibinfo{person}{R.~Joseph Connor}, \bibinfo{person}{Tyler McDaniel}, \bibinfo{person}{Jared~M. Smith}, {and} \bibinfo{person}{Max Schuchard}.} \bibinfo{year}{2020}\natexlab{}.
\newblock \showarticletitle{{PKU} Pitfalls: Attacks on PKU-based Memory Isolation Systems}. In \bibinfo{booktitle}{\emph{29th {USENIX} Security Symposium, {USENIX} Security 2020, August 12-14, 2020}}, \bibfield{editor}{\bibinfo{person}{Srdjan Capkun} {and} \bibinfo{person}{Franziska Roesner}} (Eds.). \bibinfo{publisher}{{USENIX} Association}, \bibinfo{pages}{1409--1426}.
\newblock


\bibitem[Davis et~al\mbox{.}(2019)]%
        {cheriabi}
\bibfield{author}{\bibinfo{person}{Brooks Davis}, \bibinfo{person}{Robert N.~M. Watson}, \bibinfo{person}{Alexander Richardson}, \bibinfo{person}{Peter~G. Neumann}, \bibinfo{person}{Simon~W. Moore}, \bibinfo{person}{John Baldwin}, \bibinfo{person}{David Chisnall}, \bibinfo{person}{Jessica Clarke}, \bibinfo{person}{Nathaniel~Wesley Filardo}, \bibinfo{person}{Khilan Gudka}, \bibinfo{person}{Alexandre Joannou}, \bibinfo{person}{Ben Laurie}, \bibinfo{person}{A.~Theodore Markettos}, \bibinfo{person}{J.~Edward Maste}, \bibinfo{person}{Alfredo Mazzinghi}, \bibinfo{person}{Edward~Tomasz Napierala}, \bibinfo{person}{Robert~M. Norton}, \bibinfo{person}{Michael Roe}, \bibinfo{person}{Peter Sewell}, \bibinfo{person}{Stacey Son}, {and} \bibinfo{person}{Jonathan Woodruff}.} \bibinfo{year}{2019}\natexlab{}.
\newblock \showarticletitle{CheriABI: Enforcing Valid Pointer Provenance and Minimizing Pointer Privilege in the POSIX C Run-Time Environment}. In \bibinfo{booktitle}{\emph{Proceedings of the Twenty-Fourth International Conference on Architectural Support for Programming Languages and Operating Systems}} (Providence, RI, USA) \emph{(\bibinfo{series}{ASPLOS '19})}. \bibinfo{publisher}{Association for Computing Machinery}, \bibinfo{address}{New York, NY, USA}, \bibinfo{pages}{379–393}.
\newblock
\showISBNx{9781450362405}
\urldef\tempurl%
\url{https://doi.org/10.1145/3297858.3304042}
\showDOI{\tempurl}


\bibitem[Dennis and Van~Horn(1966)]%
        {obj-cap}
\bibfield{author}{\bibinfo{person}{Jack~B. Dennis} {and} \bibinfo{person}{Earl~C. Van~Horn}.} \bibinfo{year}{1966}\natexlab{}.
\newblock \showarticletitle{Programming Semantics for Multiprogrammed Computations}.
\newblock \bibinfo{journal}{\emph{Commun. ACM}} \bibinfo{volume}{9}, \bibinfo{number}{3} (\bibinfo{date}{March} \bibinfo{year}{1966}), \bibinfo{pages}{143–155}.
\newblock
\showISSN{0001-0782}


\bibitem[Duck and Yap(2016)]%
        {new-lowfat-pointer}
\bibfield{author}{\bibinfo{person}{Gregory~J. Duck} {and} \bibinfo{person}{Roland H.~C. Yap}.} \bibinfo{year}{2016}\natexlab{}.
\newblock \showarticletitle{Heap Bounds Protection with Low Fat Pointers} \emph{(\bibinfo{series}{CC 2016})}. \bibinfo{publisher}{Association for Computing Machinery}, \bibinfo{address}{New York, NY, USA}, \bibinfo{pages}{132–142}.
\newblock
\showISBNx{9781450342414}
\urldef\tempurl%
\url{https://doi.org/10.1145/2892208.2892212}
\showDOI{\tempurl}


\bibitem[Eklektix(2022)]%
        {kasan}
\bibfield{author}{\bibinfo{person}{Eklektix}.} \bibinfo{year}{2022}\natexlab{}.
\newblock \bibinfo{title}{kasan: add hardware tag-based mode for arm64}.
\newblock \bibinfo{howpublished}{\url{https://lwn.net/Articles/831624/}}.
\newblock
\newblock
\shownote{Last accessed Jan 14 , 2022,}.


\bibitem[F5~Networks(2023)]%
        {nginx}
\bibfield{author}{\bibinfo{person}{Inc. F5~Networks}.} \bibinfo{year}{2023}\natexlab{}.
\newblock \bibinfo{title}{{Advanced Load Balancer, Web Server, \& Reverse Proxy}}.
\newblock \bibinfo{howpublished}{\url{https://www.nginx.com}}.
\newblock
\newblock
\shownote{Last accessed Jan 14 , 2022,}.


\bibitem[Fabry(1974)]%
        {cap-addressing}
\bibfield{author}{\bibinfo{person}{R.~S. Fabry}.} \bibinfo{year}{1974}\natexlab{}.
\newblock \showarticletitle{Capability-Based Addressing}.
\newblock \bibinfo{journal}{\emph{Commun. ACM}} \bibinfo{volume}{17}, \bibinfo{number}{7} (\bibinfo{date}{jul} \bibinfo{year}{1974}), \bibinfo{pages}{403–412}.
\newblock
\showISSN{0001-0782}
\urldef\tempurl%
\url{https://doi.org/10.1145/361011.361070}
\showDOI{\tempurl}


\bibitem[farkhani et~al\mbox{.}(2021)]%
        {ptauth}
\bibfield{author}{\bibinfo{person}{Reza~Mirzazade farkhani}, \bibinfo{person}{Mansour Ahmadi}, {and} \bibinfo{person}{Long Lu}.} \bibinfo{year}{2021}\natexlab{}.
\newblock \showarticletitle{PTAuth: Temporal Memory Safety via Robust Points-to Authentication}. In \bibinfo{booktitle}{\emph{30th {USENIX} Security Symposium ({USENIX} Security 21)}}. \bibinfo{publisher}{{USENIX} Association}.
\newblock


\bibitem[Felker(2022)]%
        {musl}
\bibfield{author}{\bibinfo{person}{Rich Felker}.} \bibinfo{year}{2022}\natexlab{}.
\newblock \bibinfo{title}{musl libc}.
\newblock \bibinfo{howpublished}{\url{https://musl.libc.org}}.
\newblock


\bibitem[Foundation(2023a)]%
        {wget}
\bibfield{author}{\bibinfo{person}{Free~Software Foundation}.} \bibinfo{year}{2023}\natexlab{a}.
\newblock \bibinfo{title}{{GNU Wget}}.
\newblock \bibinfo{howpublished}{\url{https://www.gnu.org/software/wget}}.
\newblock
\newblock
\shownote{Last accessed Jan 14 , 2022,}.


\bibitem[Foundation(2023b)]%
        {openssh}
\bibfield{author}{\bibinfo{person}{OpenBSD Foundation}.} \bibinfo{year}{2023}\natexlab{b}.
\newblock \bibinfo{title}{{OpenSSH}}.
\newblock \bibinfo{howpublished}{\url{https://www.openssh.com}}.
\newblock
\newblock
\shownote{Last accessed Jan 14 , 2022,}.


\bibitem[Foundation(2022)]%
        {apache-benchmark}
\bibfield{author}{\bibinfo{person}{The Apache~Software Foundation}.} \bibinfo{year}{2022}\natexlab{}.
\newblock \bibinfo{title}{{ab - Apache HTTP server benchmarking tool}}.
\newblock \bibinfo{howpublished}{\url{https://httpd.apache.org/docs/2.4/programs/ab.html}}.
\newblock
\newblock
\shownote{Last accessed Jan 14 , 2022,}.


\bibitem[Frascino(2020)]%
        {mte-kernel-support}
\bibfield{author}{\bibinfo{person}{Vincenzo Frascino}.} \bibinfo{year}{2020}\natexlab{}.
\newblock \bibinfo{title}{Memory Tagging Extension (MTE) in AArch64 Linux}.
\newblock \bibinfo{howpublished}{\url{ https://www.kernel.org/doc/html/latest/arm64/memory-tagging-extension.html }}.
\newblock
\newblock
\shownote{Last accessed March 10 , 2022,}.


\bibitem[Hedayati et~al\mbox{.}(2019)]%
        {hodor}
\bibfield{author}{\bibinfo{person}{Mohammad Hedayati}, \bibinfo{person}{Spyridoula Gravani}, \bibinfo{person}{Ethan Johnson}, \bibinfo{person}{John Criswell}, \bibinfo{person}{Michael~L Scott}, \bibinfo{person}{Kai Shen}, {and} \bibinfo{person}{Mike Marty}.} \bibinfo{year}{2019}\natexlab{}.
\newblock \showarticletitle{Hodor: Intra-process isolation for high-throughput data plane libraries}. In \bibinfo{booktitle}{\emph{2019 {USENIX} Annual Technical Conference ({USENIX}{ATC} 19)}}. \bibinfo{pages}{489--504}.
\newblock


\bibitem[{Intel Corporation}(2021)]%
        {intel-manual}
\bibfield{author}{\bibinfo{person}{{Intel Corporation}}.} \bibinfo{year}{2021}\natexlab{}.
\newblock \bibinfo{booktitle}{\emph{{Intel\textsuperscript{\textregistered} 64 and IA-32 Architectures Software Developer's Manual}}}.
\newblock Number 325462-075US.
\newblock


\bibitem[Ismail et~al\mbox{.}(2022)]%
        {pactight}
\bibfield{author}{\bibinfo{person}{Mohannad Ismail}, \bibinfo{person}{Andrew Quach}, \bibinfo{person}{Christopher Jelesnianski}, \bibinfo{person}{Yeongjin Jang}, {and} \bibinfo{person}{Changwoo Min}.} \bibinfo{year}{2022}\natexlab{}.
\newblock \bibinfo{title}{Tightly Seal Your Sensitive Pointers with PACTight}.
\newblock
\newblock
\urldef\tempurl%
\url{https://doi.org/10.48550/ARXIV.2203.15121}
\showDOI{\tempurl}


\bibitem[Jin et~al\mbox{.}(2022)]%
        {cryptompk}
\bibfield{author}{\bibinfo{person}{X. Jin}, \bibinfo{person}{X. Xiao}, \bibinfo{person}{S. Jia}, \bibinfo{person}{W. Gao}, \bibinfo{person}{H. Zhang}, \bibinfo{person}{D. Gu}, \bibinfo{person}{S. Ma}, \bibinfo{person}{Z. Qian}, {and} \bibinfo{person}{J. Li}.} \bibinfo{year}{2022}\natexlab{}.
\newblock \showarticletitle{Annotating, Tracking, and Protecting Cryptographic Secrets with CryptoMPK}. In \bibinfo{booktitle}{\emph{2022 2022 IEEE Symposium on Security and Privacy (SP) (SP)}}. \bibinfo{publisher}{IEEE Computer Society}, \bibinfo{address}{Los Alamitos, CA, USA}, \bibinfo{pages}{473--488}.
\newblock
\showISSN{2375-1207}
\urldef\tempurl%
\url{https://doi.org/10.1109/SP46214.2022.00028}
\showDOI{\tempurl}


\bibitem[Kilpatrick(2003)]%
        {privman}
\bibfield{author}{\bibinfo{person}{Douglas Kilpatrick}.} \bibinfo{year}{2003}\natexlab{}.
\newblock \showarticletitle{Privman: A Library for Partitioning Applications.}. In \bibinfo{booktitle}{\emph{USENIX Annual Technical Conference, FREENIX Track}} (2003-09-03). \bibinfo{publisher}{USENIX}, \bibinfo{pages}{273--284}.
\newblock
\showISBNx{1-931971-11-0}


\bibitem[Kirth et~al\mbox{.}(2022)]%
        {pkrusafe}
\bibfield{author}{\bibinfo{person}{Paul Kirth}, \bibinfo{person}{Mitchel Dickerson}, \bibinfo{person}{Stephen Crane}, \bibinfo{person}{Per Larsen}, \bibinfo{person}{Adrian Dabrowski}, \bibinfo{person}{David Gens}, \bibinfo{person}{Yeoul Na}, \bibinfo{person}{Stijn Volckaert}, {and} \bibinfo{person}{Michael Franz}.} \bibinfo{year}{2022}\natexlab{}.
\newblock \showarticletitle{PKRU-Safe: Automatically Locking down the Heap between Safe and Unsafe Languages}. In \bibinfo{booktitle}{\emph{Proceedings of the Seventeenth European Conference on Computer Systems}} (Rennes, France) \emph{(\bibinfo{series}{EuroSys '22})}. \bibinfo{publisher}{Association for Computing Machinery}, \bibinfo{address}{New York, NY, USA}, \bibinfo{pages}{132–148}.
\newblock
\showISBNx{9781450391627}
\urldef\tempurl%
\url{https://doi.org/10.1145/3492321.3519582}
\showDOI{\tempurl}


\bibitem[Klein et~al\mbox{.}(2009)]%
        {sel4}
\bibfield{author}{\bibinfo{person}{Gerwin Klein}, \bibinfo{person}{Kevin Elphinstone}, \bibinfo{person}{Gernot Heiser}, \bibinfo{person}{June Andronick}, \bibinfo{person}{David Cock}, \bibinfo{person}{Philip Derrin}, \bibinfo{person}{Dhammika Elkaduwe}, \bibinfo{person}{Kai Engelhardt}, \bibinfo{person}{Rafal Kolanski}, \bibinfo{person}{Michael Norrish}, \bibinfo{person}{Thomas Sewell}, \bibinfo{person}{Harvey Tuch}, {and} \bibinfo{person}{Simon Winwood}.} \bibinfo{year}{2009}\natexlab{}.
\newblock \showarticletitle{SeL4: Formal Verification of an OS Kernel}. In \bibinfo{booktitle}{\emph{Proceedings of the ACM SIGOPS 22nd Symposium on Operating Systems Principles}} (Big Sky, Montana, USA) \emph{(\bibinfo{series}{SOSP '09})}. \bibinfo{publisher}{Association for Computing Machinery}, \bibinfo{address}{New York, NY, USA}, \bibinfo{pages}{207–220}.
\newblock
\showISBNx{9781605587523}
\urldef\tempurl%
\url{https://doi.org/10.1145/1629575.1629596}
\showDOI{\tempurl}


\bibitem[Kuznetsov et~al\mbox{.}(2014)]%
        {cpi}
\bibfield{author}{\bibinfo{person}{Volodymyr Kuznetsov}, \bibinfo{person}{Laszlo Szekeres}, \bibinfo{person}{Mathias Payer}, \bibinfo{person}{George Candea}, \bibinfo{person}{R. Sekar}, {and} \bibinfo{person}{Dawn Song}.} \bibinfo{year}{2014}\natexlab{}.
\newblock \showarticletitle{{Code-Pointer} Integrity}. In \bibinfo{booktitle}{\emph{11th USENIX Symposium on Operating Systems Design and Implementation (OSDI 14)}}. \bibinfo{publisher}{USENIX Association}, \bibinfo{address}{Broomfield, CO}, \bibinfo{pages}{147--163}.
\newblock
\showISBNx{978-1-931971-16-4}


\bibitem[Kwon et~al\mbox{.}(2013)]%
        {lowfat-pointer}
\bibfield{author}{\bibinfo{person}{Albert Kwon}, \bibinfo{person}{Udit Dhawan}, \bibinfo{person}{Jonathan~M. Smith}, \bibinfo{person}{Thomas~F. Knight}, {and} \bibinfo{person}{Andre DeHon}.} \bibinfo{year}{2013}\natexlab{}.
\newblock \showarticletitle{Low-Fat Pointers: Compact Encoding and Efficient Gate-Level Implementation of Fat Pointers for Spatial Safety and Capability-Based Security}. In \bibinfo{booktitle}{\emph{Proceedings of the 2013 ACM SIGSAC Conference on Computer \& Communications Security}} (Berlin, Germany) \emph{(\bibinfo{series}{CCS '13})}. \bibinfo{publisher}{Association for Computing Machinery}, \bibinfo{address}{New York, NY, USA}, \bibinfo{pages}{721–732}.
\newblock
\showISBNx{9781450324779}
\urldef\tempurl%
\url{https://doi.org/10.1145/2508859.2516713}
\showDOI{\tempurl}


\bibitem[Lee et~al\mbox{.}(2018)]%
        {lotrx86}
\bibfield{author}{\bibinfo{person}{Hojoon Lee}, \bibinfo{person}{Chihyun Song}, {and} \bibinfo{person}{Brent~Byunghoon Kang}.} \bibinfo{year}{2018}\natexlab{}.
\newblock \showarticletitle{Lord of the X86 Rings: A Portable User Mode Privilege Separation Architecture on X86}. In \bibinfo{booktitle}{\emph{Proceedings of the 2018 ACM SIGSAC Conference on Computer and Communications Security}} (Toronto, Canada) \emph{(\bibinfo{series}{CCS '18})}. \bibinfo{publisher}{Association for Computing Machinery}, \bibinfo{address}{New York, NY, USA}, \bibinfo{pages}{1441–1454}.
\newblock
\showISBNx{9781450356930}


\bibitem[Lefeuvre et~al\mbox{.}(2023)]%
        {civ}
\bibfield{author}{\bibinfo{person}{Hugo Lefeuvre}, \bibinfo{person}{Vlad-Andrei B{\u{a}}doiu}, \bibinfo{person}{Yi Chen}, \bibinfo{person}{Felipe Huici}, \bibinfo{person}{Nathan Dautenhahn}, {and} \bibinfo{person}{Pierre Olivier}.} \bibinfo{year}{2023}\natexlab{}.
\newblock \showarticletitle{Assessing the Impact of Interface Vulnerabilities in Compartmentalized Software}. In \bibinfo{booktitle}{\emph{Proceedings 2023 Network and Distributed System Security Symposium. NDSS}}.
\newblock


\bibitem[Li et~al\mbox{.}(2022)]%
        {pacsan}
\bibfield{author}{\bibinfo{person}{Yuan Li}, \bibinfo{person}{Wende Tan}, \bibinfo{person}{Zhizheng Lv}, \bibinfo{person}{Songtao Yang}, \bibinfo{person}{Mathias Payer}, \bibinfo{person}{Ying Liu}, {and} \bibinfo{person}{Chao Zhang}.} \bibinfo{year}{2022}\natexlab{}.
\newblock \bibinfo{title}{PACSan: Enforcing Memory Safety Based on ARM PA}.
\newblock
\newblock
\urldef\tempurl%
\url{https://doi.org/10.48550/ARXIV.2202.03950}
\showDOI{\tempurl}


\bibitem[Liljestrand et~al\mbox{.}(2021)]%
        {pacstack}
\bibfield{author}{\bibinfo{person}{Hans Liljestrand}, \bibinfo{person}{Thomas Nyman}, \bibinfo{person}{Lachlan~J. Gunn}, \bibinfo{person}{Jan-Erik Ekberg}, {and} \bibinfo{person}{N. Asokan}.} \bibinfo{year}{2021}\natexlab{}.
\newblock \showarticletitle{PACStack: an Authenticated Call Stack}. In \bibinfo{booktitle}{\emph{30th {USENIX} Security Symposium ({USENIX} Security 21)}}. \bibinfo{publisher}{{USENIX} Association}.
\newblock


\bibitem[Liljestrand et~al\mbox{.}(2019)]%
        {pac-it-up}
\bibfield{author}{\bibinfo{person}{Hans Liljestrand}, \bibinfo{person}{Thomas Nyman}, \bibinfo{person}{Kui Wang}, \bibinfo{person}{Carlos~Chinea Perez}, \bibinfo{person}{Jan-Erik Ekberg}, {and} \bibinfo{person}{N. Asokan}.} \bibinfo{year}{2019}\natexlab{}.
\newblock \showarticletitle{{PAC} it up: Towards Pointer Integrity using {ARM} Pointer Authentication}. In \bibinfo{booktitle}{\emph{28th {USENIX} Security Symposium ({USENIX} Security 19)}}. \bibinfo{publisher}{{USENIX} Association}, \bibinfo{address}{Santa Clara, CA}, \bibinfo{pages}{177--194}.
\newblock
\showISBNx{978-1-939133-06-9}


\bibitem[Liu et~al\mbox{.}(2017)]%
        {ptrsplit}
\bibfield{author}{\bibinfo{person}{Shen Liu}, \bibinfo{person}{Gang Tan}, {and} \bibinfo{person}{Trent Jaeger}.} \bibinfo{year}{2017}\natexlab{}.
\newblock \showarticletitle{PtrSplit: Supporting General Pointers in Automatic Program Partitioning}. In \bibinfo{booktitle}{\emph{Proceedings of the 2017 ACM SIGSAC Conference on Computer and Communications Security}} (Dallas, Texas, USA) \emph{(\bibinfo{series}{CCS '17})}. \bibinfo{publisher}{Association for Computing Machinery}, \bibinfo{address}{New York, NY, USA}, \bibinfo{pages}{2359–2371}.
\newblock
\showISBNx{9781450349468}
\urldef\tempurl%
\url{https://doi.org/10.1145/3133956.3134066}
\showDOI{\tempurl}


\bibitem[Ltd.(2022)]%
        {mte-stack}
\bibfield{author}{\bibinfo{person}{Arm Ltd.}} \bibinfo{year}{2022}\natexlab{}.
\newblock \bibinfo{title}{-mmemtag-stack, -mno-memtag-stack}.
\newblock \bibinfo{howpublished}{\url{ https://developer.arm.com/documentation/100067/0612/armclang-Command-line-Options/-mmemtag-stack---mno-memtag-stack }}.
\newblock
\newblock
\shownote{Last accessed Jan 14 , 2022,}.


\bibitem[Martin et~al\mbox{.}(2022)]%
        {asahi-linux}
\bibfield{author}{\bibinfo{person}{Hector Martin}, \bibinfo{person}{Alyssa Rosenzweig}, \bibinfo{person}{Asahi Lina}, \bibinfo{person}{Dougall Johnson}, \bibinfo{person}{Sven Peter}, \bibinfo{person}{Mark Kettenis}, \bibinfo{person}{Martin Povišer}, {and} \bibinfo{person}{Janne Grunau}.} \bibinfo{year}{2022}\natexlab{}.
\newblock \bibinfo{title}{{Asahi Linux}}.
\newblock \bibinfo{howpublished}{\url{https://asahilinux.org}}.
\newblock
\newblock
\shownote{Last accessed March 08 , 2022,}.


\bibitem[Mashtizadeh et~al\mbox{.}(2015)]%
        {ccfi}
\bibfield{author}{\bibinfo{person}{Ali~Jose Mashtizadeh}, \bibinfo{person}{Andrea Bittau}, \bibinfo{person}{Dan Boneh}, {and} \bibinfo{person}{David Mazi\`{e }res}.} \bibinfo{year}{2015}\natexlab{}.
\newblock \showarticletitle{CCFI: Cryptographically Enforced Control Flow Integrity}. In \bibinfo{booktitle}{\emph{Proceedings of the 22nd ACM SIGSAC Conference on Computer and Communications Security}} (Denver, Colorado, USA) \emph{(\bibinfo{series}{CCS '15})}. \bibinfo{publisher}{Association for Computing Machinery}, \bibinfo{address}{New York, NY, USA}, \bibinfo{pages}{941–951}.
\newblock
\showISBNx{9781450338325}


\bibitem[McKee et~al\mbox{.}(2022)]%
        {hakc}
\bibfield{author}{\bibinfo{person}{Derrick McKee}, \bibinfo{person}{Yianni Giannaris}, \bibinfo{person}{Carolina~Ortega Perez}, \bibinfo{person}{Howard Shrobe}, \bibinfo{person}{Mathias Payer}, \bibinfo{person}{Hamed Okhravi}, {and} \bibinfo{person}{Nathan Burow}.} \bibinfo{year}{2022}\natexlab{}.
\newblock \showarticletitle{Preventing Kernel Hacks with HAKC}. In \bibinfo{booktitle}{\emph{Proceedings 2022 Network and Distributed System Security Symposium. NDSS}}, Vol.~\bibinfo{volume}{22}. \bibinfo{pages}{1--17}.
\newblock


\bibitem[Miller et~al\mbox{.}(2008)]%
        {caja}
\bibfield{author}{\bibinfo{person}{Mark~S. Miller}, \bibinfo{person}{Mike Samuel}, \bibinfo{person}{Ben Laurie}, \bibinfo{person}{Ihab Awad}, {and} \bibinfo{person}{Mike Stay}.} \bibinfo{year}{2008}\natexlab{}.
\newblock \showarticletitle{Caja: Safe active content in sanitized JavaScript}.
\newblock  (\bibinfo{date}{June 7} \bibinfo{year}{2008}).
\newblock


\bibitem[Nam et~al\mbox{.}(2019)]%
        {framer}
\bibfield{author}{\bibinfo{person}{Myoung~Jin Nam}, \bibinfo{person}{Periklis Akritidis}, {and} \bibinfo{person}{David~J. Greaves}.} \bibinfo{year}{2019}\natexlab{}.
\newblock \showarticletitle{{FRAMER:} a tagged-pointer capability system with memory safety applications}. In \bibinfo{booktitle}{\emph{Proceedings of the 35th Annual Computer Security Applications Conference, {ACSAC} 2019, San Juan, PR, USA, December 09-13, 2019}}, \bibfield{editor}{\bibinfo{person}{David Balenson}} (Ed.). \bibinfo{publisher}{{ACM}}, \bibinfo{pages}{612--626}.
\newblock
\urldef\tempurl%
\url{https://doi.org/10.1145/3359789.3359799}
\showDOI{\tempurl}


\bibitem[Oracle(2022)]%
        {sparc-adi}
\bibfield{author}{\bibinfo{person}{Oracle}.} \bibinfo{year}{2022}\natexlab{}.
\newblock \bibinfo{title}{{Using Application Data Integrity (ADI)}}.
\newblock \bibinfo{howpublished}{\url{https://docs.oracle.com/cd/E37838_01/html/E61059/gqajs.html}}.
\newblock
\newblock
\shownote{Last accessed March 02 , 2022,}.


\bibitem[Palit et~al\mbox{.}(2021)]%
        {dynpta}
\bibfield{author}{\bibinfo{person}{Tapti Palit}, \bibinfo{person}{Jarin Firose~Moon}, \bibinfo{person}{Fabian Monrose}, {and} \bibinfo{person}{Michalis Polychronakis}.} \bibinfo{year}{2021}\natexlab{}.
\newblock \showarticletitle{DynPTA: Combining Static and Dynamic Analysis for Practical Selective Data Protection}. In \bibinfo{booktitle}{\emph{2021 IEEE Symposium on Security and Privacy (SP)}}. \bibinfo{pages}{1919--1937}.
\newblock
\urldef\tempurl%
\url{https://doi.org/10.1109/SP40001.2021.00082}
\showDOI{\tempurl}


\bibitem[Peng et~al\mbox{.}(2023)]%
        {muswitch}
\bibfield{author}{\bibinfo{person}{D. Peng}, \bibinfo{person}{C. Liu}, \bibinfo{person}{T. Palit}, \bibinfo{person}{P. Fonseca}, \bibinfo{person}{A. Vahldiek-Oberwagner}, {and} \bibinfo{person}{M. Vij}.} \bibinfo{year}{2023}\natexlab{}.
\newblock \showarticletitle{$\mu$Switch: Fast Kernel Context Isolation with Implicit Context Switches}. In \bibinfo{booktitle}{\emph{2023 IEEE Symposium on Security and Privacy (SP)}}. \bibinfo{publisher}{IEEE Computer Society}, \bibinfo{address}{Los Alamitos, CA, USA}, \bibinfo{pages}{2956--2973}.
\newblock
\urldef\tempurl%
\url{https://doi.org/10.1109/SP46215.2023.10179284}
\showDOI{\tempurl}


\bibitem[Poletto and Sarkar(1999)]%
        {reg-alloc}
\bibfield{author}{\bibinfo{person}{Massimiliano Poletto} {and} \bibinfo{person}{Vivek Sarkar}.} \bibinfo{year}{1999}\natexlab{}.
\newblock \showarticletitle{Linear Scan Register Allocation}.
\newblock \bibinfo{journal}{\emph{ACM Trans. Program. Lang. Syst.}} \bibinfo{volume}{21}, \bibinfo{number}{5} (\bibinfo{date}{sep} \bibinfo{year}{1999}), \bibinfo{pages}{895–913}.
\newblock
\showISSN{0164-0925}
\urldef\tempurl%
\url{https://doi.org/10.1145/330249.330250}
\showDOI{\tempurl}


\bibitem[Project(2023)]%
        {freebsd-capsicum}
\bibfield{author}{\bibinfo{person}{FreeBSD Project}.} \bibinfo{year}{2023}\natexlab{}.
\newblock \bibinfo{title}{{FreeBSD} {Manual} {Pages}}.
\newblock \bibinfo{howpublished}{\url{https://www.freebsd.org/cgi/man.cgi?capsicum(4)}}.
\newblock


\bibitem[Project(2022a)]%
        {llvm-ret-addr}
\bibfield{author}{\bibinfo{person}{LLVM Project}.} \bibinfo{year}{2022}\natexlab{a}.
\newblock \bibinfo{title}{[AArch64] - return address signing}.
\newblock \bibinfo{howpublished}{\url{https://reviews.llvm.org/D49793}}.
\newblock
\newblock
\shownote{Last accessed May 05 , 2022,}.


\bibitem[Project(2022b)]%
        {llvm}
\bibfield{author}{\bibinfo{person}{LLVM Project}.} \bibinfo{year}{2022}\natexlab{b}.
\newblock \bibinfo{title}{{The LLVM Compiler Infrastructure}}.
\newblock \bibinfo{howpublished}{\url{https://llvm.org}}.
\newblock
\newblock
\shownote{Last accessed Jan 14 , 2022,}.


\bibitem[Project(2022c)]%
        {libressl}
\bibfield{author}{\bibinfo{person}{OpenBSD Project}.} \bibinfo{year}{2022}\natexlab{c}.
\newblock \bibinfo{title}{{LibreSSL}}.
\newblock \bibinfo{howpublished}{\url{https://www.libressl.org}}.
\newblock
\newblock
\shownote{Last accessed Jan 14 , 2022,}.


\bibitem[QEMU(2022)]%
        {qemu}
\bibfield{author}{\bibinfo{person}{QEMU}.} \bibinfo{year}{2022}\natexlab{}.
\newblock \bibinfo{title}{{QEMU: A generic and open source machine emulator and virtualizer}}.
\newblock \bibinfo{howpublished}{\url{https://www.qemu.org}}.
\newblock
\newblock
\shownote{Last accessed March 08 , 2022,}.


\bibitem[{QUALCOMM TECHNOLOGIES, INC}(2017)]%
        {qualcomm-pac}
\bibfield{author}{\bibinfo{person}{{QUALCOMM TECHNOLOGIES, INC}}.} \bibinfo{year}{2017}\natexlab{}.
\newblock \bibinfo{title}{{Pointer authentication on ARMv8.3}}.
\newblock \bibinfo{howpublished}{\url{ https://www.qualcomm.com/media/documents/files/whitepaper-pointer-authentication-on-armv8-3.pdf }}.
\newblock
\newblock
\shownote{{Last accessed Nov 15 , 2021,}}.


\bibitem[{Red Hat}(2021)]%
        {selinux}
\bibfield{author}{\bibinfo{person}{{Red Hat}}.} \bibinfo{year}{2021}\natexlab{}.
\newblock \bibinfo{title}{{What is SELinux}}.
\newblock \bibinfo{howpublished}{\url{https://www.redhat.com/en/topics/linux/what-is-selinux}}.
\newblock
\newblock
\shownote{Last accessed Apr 28 , 2021,}.


\bibitem[Roessler et~al\mbox{.}(2021)]%
        {muscope}
\bibfield{author}{\bibinfo{person}{Nick Roessler}, \bibinfo{person}{Lucas Atayde}, \bibinfo{person}{Imani Palmer}, \bibinfo{person}{Derrick McKee}, \bibinfo{person}{Jai Pandey}, \bibinfo{person}{Vasileios~P Kemerlis}, \bibinfo{person}{Mathias Payer}, \bibinfo{person}{Adam Bates}, \bibinfo{person}{Jonathan~M Smith}, \bibinfo{person}{Andre DeHon}, {et~al\mbox{.}}} \bibinfo{year}{2021}\natexlab{}.
\newblock \showarticletitle{$\mu$SCOPE: A Methodology for Analyzing Least-Privilege Compartmentalization in Large Software Artifacts}. In \bibinfo{booktitle}{\emph{24th International Symposium on Research in Attacks, Intrusions and Defenses}}. \bibinfo{pages}{296--311}.
\newblock


\bibitem[Schrammel et~al\mbox{.}(2022)]%
        {jenny}
\bibfield{author}{\bibinfo{person}{David Schrammel}, \bibinfo{person}{Samuel Weiser}, \bibinfo{person}{Richard Sadek}, {and} \bibinfo{person}{Stefan Mangard}.} \bibinfo{year}{2022}\natexlab{}.
\newblock \showarticletitle{Jenny: Securing Syscalls for {PKU-based} Memory Isolation Systems}. In \bibinfo{booktitle}{\emph{31st USENIX Security Symposium (USENIX Security 22)}}. \bibinfo{publisher}{USENIX Association}, \bibinfo{address}{Boston, MA}, \bibinfo{pages}{936--952}.
\newblock
\showISBNx{978-1-939133-31-1}
\urldef\tempurl%
\url{https://www.usenix.org/conference/usenixsecurity22/presentation/schrammel}
\showURL{%
\tempurl}


\bibitem[Schrammel et~al\mbox{.}(2020)]%
        {donky}
\bibfield{author}{\bibinfo{person}{David Schrammel}, \bibinfo{person}{Samuel Weiser}, \bibinfo{person}{Stefan Steinegger}, \bibinfo{person}{Martin Schwarzl}, \bibinfo{person}{Michael Schwarz}, \bibinfo{person}{Stefan Mangard}, {and} \bibinfo{person}{Daniel Gruss}.} \bibinfo{year}{2020}\natexlab{}.
\newblock \showarticletitle{Donky: Domain Keys {\textendash} Efficient {In-Process} Isolation for {RISC-V} and x86}. In \bibinfo{booktitle}{\emph{29th USENIX Security Symposium (USENIX Security 20)}}. \bibinfo{publisher}{USENIX Association}, \bibinfo{pages}{1677--1694}.
\newblock
\showISBNx{978-1-939133-17-5}


\bibitem[Serebryany et~al\mbox{.}(2018)]%
        {memory-tagging}
\bibfield{author}{\bibinfo{person}{Kostya Serebryany}, \bibinfo{person}{Evgenii Stepanov}, \bibinfo{person}{Aleksey Shlyapnikov}, \bibinfo{person}{Vlad Tsyrklevich}, {and} \bibinfo{person}{Dmitry Vyukov}.} \bibinfo{year}{2018}\natexlab{}.
\newblock \bibinfo{title}{{Memory Tagging and how it improves C/C++ memory safety}}.
\newblock
\newblock
\urldef\tempurl%
\url{https://doi.org/10.48550/ARXIV.1802.09517}
\showDOI{\tempurl}


\bibitem[Shapiro et~al\mbox{.}(1999)]%
        {eros}
\bibfield{author}{\bibinfo{person}{Jonathan~S. Shapiro}, \bibinfo{person}{Jonathan~M. Smith}, {and} \bibinfo{person}{David~J. Farber}.} \bibinfo{year}{1999}\natexlab{}.
\newblock \showarticletitle{EROS: A Fast Capability System}. In \bibinfo{booktitle}{\emph{Proceedings of the Seventeenth ACM Symposium on Operating Systems Principles}} (Charleston, South Carolina, USA) \emph{(\bibinfo{series}{SOSP '99})}. \bibinfo{publisher}{Association for Computing Machinery}, \bibinfo{address}{New York, NY, USA}, \bibinfo{pages}{170–185}.
\newblock
\showISBNx{1581131402}


\bibitem[{Uwe F. Mayer}(2017)]%
        {nbench}
\bibfield{author}{\bibinfo{person}{{Uwe F. Mayer}}.} \bibinfo{year}{2017}\natexlab{}.
\newblock \bibinfo{title}{{Linux/Unix nbench}}.
\newblock \bibinfo{howpublished}{\url{https://www.math.utah.edu/~mayer/linux/bmark.html}}.
\newblock
\newblock
\shownote{Last accessed March 08 , 2022,}.


\bibitem[Vahldiek-Oberwagner et~al\mbox{.}(2019)]%
        {erim}
\bibfield{author}{\bibinfo{person}{Anjo Vahldiek-Oberwagner}, \bibinfo{person}{Eslam Elnikety}, \bibinfo{person}{Nuno~O Duarte}, \bibinfo{person}{Michael Sammler}, \bibinfo{person}{Peter Druschel}, {and} \bibinfo{person}{Deepak Garg}.} \bibinfo{year}{2019}\natexlab{}.
\newblock \showarticletitle{ERIM: Secure, efficient in-process isolation with protection keys ({ MPK})}. In \bibinfo{booktitle}{\emph{28th {USENIX} Security Symposium ({USENIX} Security 19)}}. \bibinfo{pages}{1221--1238}.
\newblock


\bibitem[van~der Kouwe et~al\mbox{.}(2018)]%
        {type-after-type}
\bibfield{author}{\bibinfo{person}{Erik van~der Kouwe}, \bibinfo{person}{Taddeus Kroes}, \bibinfo{person}{Chris Ouwehand}, \bibinfo{person}{Herbert Bos}, {and} \bibinfo{person}{Cristiano Giuffrida}.} \bibinfo{year}{2018}\natexlab{}.
\newblock \showarticletitle{Type-After-Type: Practical and Complete Type-Safe Memory Reuse}. In \bibinfo{booktitle}{\emph{Proceedings of the 34th Annual Computer Security Applications Conference}} (San Juan, PR, USA) \emph{(\bibinfo{series}{ACSAC '18})}. \bibinfo{publisher}{Association for Computing Machinery}, \bibinfo{address}{New York, NY, USA}, \bibinfo{pages}{17–27}.
\newblock
\showISBNx{9781450365697}
\urldef\tempurl%
\url{https://doi.org/10.1145/3274694.3274705}
\showDOI{\tempurl}


\bibitem[von Eicken et~al\mbox{.}(1999)]%
        {jkernel}
\bibfield{author}{\bibinfo{person}{Thorsten von Eicken}, \bibinfo{person}{Chi-Chao Chang}, \bibinfo{person}{Grzegorz Czajkowski}, \bibinfo{person}{Chris Hawblitzel}, \bibinfo{person}{Deyu Hu}, {and} \bibinfo{person}{Dan Spoonhower}.} \bibinfo{year}{1999}\natexlab{}.
\newblock \bibinfo{booktitle}{\emph{J-Kernel: A Capability-Based Operating System for Java}}.
\newblock \bibinfo{publisher}{Springer Berlin Heidelberg}, \bibinfo{address}{Berlin, Heidelberg}, \bibinfo{pages}{369--393}.
\newblock
\showISBNx{978-3-540-48749-4}
\urldef\tempurl%
\url{https://doi.org/10.1007/3-540-48749-2_17}
\showDOI{\tempurl}


\bibitem[Voulimeneas et~al\mbox{.}(2022)]%
        {cerberus}
\bibfield{author}{\bibinfo{person}{Alexios Voulimeneas}, \bibinfo{person}{Jonas Vinck}, \bibinfo{person}{Ruben Mechelinck}, {and} \bibinfo{person}{Stijn Volckaert}.} \bibinfo{year}{2022}\natexlab{}.
\newblock \showarticletitle{You Shall Not (by)Pass! Practical, Secure, and Fast PKU-Based Sandboxing}. In \bibinfo{booktitle}{\emph{Proceedings of the Seventeenth European Conference on Computer Systems}} (Rennes, France) \emph{(\bibinfo{series}{EuroSys '22})}. \bibinfo{publisher}{Association for Computing Machinery}, \bibinfo{address}{New York, NY, USA}, \bibinfo{pages}{266–282}.
\newblock
\showISBNx{9781450391627}
\urldef\tempurl%
\url{https://doi.org/10.1145/3492321.3519560}
\showDOI{\tempurl}


\bibitem[Wang et~al\mbox{.}(2020b)]%
        {monguard}
\bibfield{author}{\bibinfo{person}{Xiaoguang Wang}, \bibinfo{person}{SengMing Yeoh}, \bibinfo{person}{Pierre Olivier}, {and} \bibinfo{person}{Binoy Ravindran}.} \bibinfo{year}{2020}\natexlab{b}.
\newblock \showarticletitle{Secure and Efficient In-Process Monitor (and Library) Protection with Intel MPK}. In \bibinfo{booktitle}{\emph{Proceedings of the 13th European Workshop on Systems Security}} (Heraklion, Greece) \emph{(\bibinfo{series}{EuroSec '20})}. \bibinfo{publisher}{Association for Computing Machinery}, \bibinfo{address}{New York, NY, USA}, \bibinfo{pages}{7–12}.
\newblock
\showISBNx{9781450375238}
\urldef\tempurl%
\url{https://doi.org/10.1145/3380786.3391398}
\showDOI{\tempurl}


\bibitem[Wang et~al\mbox{.}(2020a)]%
        {seimi}
\bibfield{author}{\bibinfo{person}{Zhe Wang}, \bibinfo{person}{Chenggang Wu}, \bibinfo{person}{Mengyao Xie}, \bibinfo{person}{Yinqian Zhang}, \bibinfo{person}{Kangjie Lu}, \bibinfo{person}{Xiaofeng Zhang}, \bibinfo{person}{Yuanming Lai}, \bibinfo{person}{Yan Kang}, {and} \bibinfo{person}{Min Yang}.} \bibinfo{year}{2020}\natexlab{a}.
\newblock \showarticletitle{Seimi: Efficient and secure smap-enabled intra-process memory isolation}. In \bibinfo{booktitle}{\emph{2020 IEEE Symposium on Security and Privacy (SP)}}. IEEE, \bibinfo{pages}{592--607}.
\newblock


\bibitem[Watson et~al\mbox{.}(2010)]%
        {capsicum}
\bibfield{author}{\bibinfo{person}{Robert N.~M. Watson}, \bibinfo{person}{Jonathan Anderson}, \bibinfo{person}{Ben Laurie}, {and} \bibinfo{person}{Kris Kennaway}.} \bibinfo{year}{2010}\natexlab{}.
\newblock \showarticletitle{Capsicum: Practical Capabilities for {UNIX}}. In \bibinfo{booktitle}{\emph{19th {USENIX} Security Symposium, Washington, DC, USA, August 11-13, 2010, Proceedings}}. \bibinfo{publisher}{{USENIX} Association}, \bibinfo{pages}{29--46}.
\newblock


\bibitem[Watson et~al\mbox{.}(2015)]%
        {cheri}
\bibfield{author}{\bibinfo{person}{Robert N.~M. Watson}, \bibinfo{person}{Jonathan Woodruff}, \bibinfo{person}{Peter~G. Neumann}, \bibinfo{person}{Simon~W. Moore}, \bibinfo{person}{Jonathan Anderson}, \bibinfo{person}{David Chisnall}, \bibinfo{person}{Nirav~H. Dave}, \bibinfo{person}{Brooks Davis}, \bibinfo{person}{Khilan Gudka}, \bibinfo{person}{Ben Laurie}, \bibinfo{person}{Steven~J. Murdoch}, \bibinfo{person}{Robert~M. Norton}, \bibinfo{person}{Michael Roe}, \bibinfo{person}{Stacey~D. Son}, {and} \bibinfo{person}{Munraj Vadera}.} \bibinfo{year}{2015}\natexlab{}.
\newblock \showarticletitle{{CHERI:} {A} Hybrid Capability-System Architecture for Scalable Software Compartmentalization}. In \bibinfo{booktitle}{\emph{2015 {IEEE} Symposium on Security and Privacy, {SP} 2015, San Jose, CA, USA, May 17-21, 2015}}. \bibinfo{publisher}{{IEEE} Computer Society}, \bibinfo{pages}{20--37}.
\newblock


\bibitem[Weiser et~al\mbox{.}(2019)]%
        {timber-v}
\bibfield{author}{\bibinfo{person}{Samuel Weiser}, \bibinfo{person}{Mario Werner}, \bibinfo{person}{Ferdinand Brasser}, \bibinfo{person}{Maja Malenko}, \bibinfo{person}{Stefan Mangard}, {and} \bibinfo{person}{Ahmad-Reza Sadeghi}.} \bibinfo{year}{2019}\natexlab{}.
\newblock \showarticletitle{TIMBER-V: Tag-Isolated Memory Bringing Fine-grained Enclaves to RISC-V}. In \bibinfo{booktitle}{\emph{Proceedings 2019 - Network and Distributed System Security Symposium (NDSS)}}. \bibinfo{publisher}{Internet Society}.
\newblock
\urldef\tempurl%
\url{https://doi.org/10.14722/ndss.2019.23068}
\showDOI{\tempurl}


\bibitem[Woodruff et~al\mbox{.}(2014)]%
        {cheri-risc}
\bibfield{author}{\bibinfo{person}{Jonathan Woodruff}, \bibinfo{person}{Robert N.~M. Watson}, \bibinfo{person}{David Chisnall}, \bibinfo{person}{Simon~W. Moore}, \bibinfo{person}{Jonathan Anderson}, \bibinfo{person}{Brooks Davis}, \bibinfo{person}{Ben Laurie}, \bibinfo{person}{Peter~G. Neumann}, \bibinfo{person}{Robert~M. Norton}, {and} \bibinfo{person}{Michael Roe}.} \bibinfo{year}{2014}\natexlab{}.
\newblock \showarticletitle{The {CHERI} capability model: Revisiting {RISC} in an age of risk}. In \bibinfo{booktitle}{\emph{{ACM/IEEE} 41st International Symposium on Computer Architecture, {ISCA} 2014, Minneapolis, MN, USA, June 14-18, 2014}}. \bibinfo{publisher}{{IEEE} Computer Society}, \bibinfo{pages}{457--468}.
\newblock


\bibitem[Yoo et~al\mbox{.}(2021)]%
        {kernel-pac}
\bibfield{author}{\bibinfo{person}{Sungbae Yoo}, \bibinfo{person}{Jinbum Park}, \bibinfo{person}{Seolheui Kim}, \bibinfo{person}{Yeji Kim}, {and} \bibinfo{person}{Taesoo Kim}.} \bibinfo{year}{2021}\natexlab{}.
\newblock \bibinfo{title}{In-Kernel Control-Flow Integrity on Commodity OSes using ARM Pointer Authentication}.
\newblock
\newblock
\urldef\tempurl%
\url{https://doi.org/10.48550/ARXIV.2112.07213}
\showDOI{\tempurl}


\bibitem[Yu et~al\mbox{.}(2023)]%
        {capstone}
\bibfield{author}{\bibinfo{person}{Jason~Zhijingcheng Yu}, \bibinfo{person}{Conrad Watt}, \bibinfo{person}{Aditya Badole}, \bibinfo{person}{Trevor~E. Carlson}, {and} \bibinfo{person}{Prateek Saxena}.} \bibinfo{year}{2023}\natexlab{}.
\newblock \showarticletitle{Capstone: A Capability-based Foundation for Trustless Secure Memory Access}. In \bibinfo{booktitle}{\emph{32nd USENIX Security Symposium (USENIX Security 23)}}. \bibinfo{publisher}{USENIX Association}, \bibinfo{address}{Anaheim, CA}, \bibinfo{pages}{787--804}.
\newblock
\showISBNx{978-1-939133-37-3}
\urldef\tempurl%
\url{https://www.usenix.org/conference/usenixsecurity23/presentation/yu-jason}
\showURL{%
\tempurl}


\bibitem[Zeldovich et~al\mbox{.}(2008)]%
        {hw-tagged-memory}
\bibfield{author}{\bibinfo{person}{Nickolai Zeldovich}, \bibinfo{person}{Hari Kannan}, \bibinfo{person}{Michael Dalton}, {and} \bibinfo{person}{Christos Kozyrakis}.} \bibinfo{year}{2008}\natexlab{}.
\newblock \showarticletitle{Hardware Enforcement of Application Security Policies Using Tagged Memory}. In \bibinfo{booktitle}{\emph{8th {USENIX} Symposium on Operating Systems Design and Implementation, {OSDI} 2008, December 8-10, 2008, San Diego, California, USA, Proceedings}}, \bibfield{editor}{\bibinfo{person}{Richard Draves} {and} \bibinfo{person}{Robbert van Renesse}} (Eds.). \bibinfo{publisher}{{USENIX} Association}, \bibinfo{pages}{225--240}.
\newblock


\end{thebibliography}

\newpage
\appendix
\section{Examples of adapting \thename to real-world applications} \label{appendix:example}
This section shows examples of adapting \thename to applications described in \cref{sec:evaluation:porting-app} that were omitted for brevity.
\autoref{fig:wget}, \autoref{fig:ssh} and \autoref{fig:libre} demonstrate the representative modifications that were made to \texttt{wget}, \texttt{ssh} and the LibreSSL library used by NGINX. We also annotated the code examples with source comments to describe the functions' behavior and the enforcement of \thename.

\begin{figure}[b]
    \centering
\begin{lstlisting}[style=sslab-capacity-c,keepspaces]
// Modified to allocate private key buffer in private memory  
struct sshkey * sshkey_new(int type) {
	struct sshkey *k;
/*!\textcolor{magenta}{+}!*/   if ((k = capac_malloc(sizeof(*k))) == NULL)
	   return NULL;
    // Initialize sshkey
    // ...	
	return k;
}
// Load open the private key and load it into an in-memory buffer. 
// The path "filename" is assigned to DOM_PRIVKEY at initialization
int sshkey_load_private_type(int type, const char *filename, 
    const char *passphrase, 
/*!\textcolor{magenta}{+}!*/   DOM_PRIV struct sshkey **keyp, 
    char **commentp) {
    // ...
    if ((fd = open(filename, O_RDONLY)) == -1)  /*!\hfill\action{PATH-Auth}/\action{FD-Sign}!*/
	   return SSH_ERR_SYSTEM_ERROR;
    // ...
    // Load private key into fd
	r = sshkey_load_private_type_fd(fd, type, passphrase, keyp, commentp); /*!\hfill\action{FD-Auth}/\action{PTR-Sign}/\action{PTR-Auth}!*/
    // ...
} 
\end{lstlisting} 
\caption{Modifications for private key isolation in \texttt{ssh}}
    \label{fig:ssh}
\end{figure}

\vspace*{\fill}

\begin{figure}[b]
    \centering
\begin{lstlisting}[style=sslab-capacity-c,keepspaces]
int main (int argc, char **argv) {
    // ...
    // Issue the output document to DOM_FILEDOWNLOAD
    // opt.output_document: argument for -O option
    // e.g., "wget <url>/<file> -O output.text"
/*!\textcolor{magenta}{+}!*/   struct capac_fd_issue issue[] = {{
        .pathname = opt.output_document,
        .owner_id = DOM_FILEDOWNLOAD,
    }};
/*!\textcolor{magenta}{+}!*/   capac_init(issue, ...);
    // ...
/*!\textcolor{magenta}{+}!*/   capac_enter(DOM_FILEDOWNLOAD, 0);
    // ...
    retrieve_url (...);
    // ...
/*!\textcolor{magenta}{+}!*/   capac_exit();
}
// Called by retrieve_url to obtain an FD for the output file
static uerr_t open_output_stream (struct http_stat *hs, int count, FILE **fp) {
    // ...
    // Limit the fd attribute to make it write-only
    int fd = /*!\hfill\action{FD-Sign}!*/
        open (hs->local_file, ...); /*!\hfill\action{PATH-Auth}!*/
/*!\textcolor{magenta}{+}!*/   fd = capac_limit_fd(CAP_READ | CAP_DELEGATE);
    *fp = fdopen(fd, "wb");  /*!\hfill\action{FD-Auth}!*/
    // ...
}
// Called by retrieve_url to obtain the file from the socket, 
// and write it into the output file
int fd_read_body (...) {
    // ...
    // Allocate download buffer in private memory
    // and annotate its pointer as domain-private
    int dlbufsize = max(BUFSIZ, 8 * 1024);
/*!\textcolor{magenta}{+}!*/   DOM_PRIV char *dlbuf = capac_malloc(dlbufsize);
    // ...
}

\end{lstlisting}
    \caption{Modifications for downloaded file isolation in \texttt{wget}.}
    \label{fig:wget}
\end{figure}

\begin{figure}[b]
    \centering
\begin{lstlisting}[style=sslab-capacity-c,keepspaces]
// Modified to allocate private key buffer in private memory 
EVP_PKEY * evp_pkey_new(void) {
	evp_pkey *ret;
/*!\textcolor{magenta}{+}!*/   ret = capac_malloc(sizeof(evp_pkey));
    // Initialize evp_pkey
    // ...
	return (ret);
}
// Called by dom_handshake to set the tls session's private ikey
static int ssl_set_pkey(ssl_cert *c,
/*!\textcolor{magenta}{+}!*/   dom_priv evp_pkey *pkey) {
    // ...
	c->pkeys[i].privatekey = pkey;  /*!\hfill\action{PTR-Sign}!*/
	c->key = &(c->pkeys[i]);  /*!\hfill\action{PTR-Sign}!*/
    // ...
}
// Called by dom_handshake to sign key exchange messages with priv. key
int ssl3_send_server_key_exchange(ssl *s) {
/*!\textcolor{magenta}{+}!*/   dom_priv evp_pkey *pkey;
    // these pointers are marked as sensitive by taint analysis
    evp_pkey_ctx *pctx;
    evp_md_ctx *md_ctx = null;
    // ...
}
// Invoked by dom_handshake at the end of key exchange
// modified to allocate session key inside private memory
// and delegate the session key to dom_session  
static int aead_aes_gcm_init(
/*!\textcolor{magenta}{+}!*/   dom_priv evp_aead_ctx *ctx, 
    const unsigned char *key, size_t key_len,
    size_t tag_len, capac_modifier session_id)
{
    // Marked as sensitive by taint analysis
    evp_pkey_ctx *pctx;
 	struct aead_aes_gcm_ctx *gcm_ctx;
    // ...
/*!\textcolor{magenta}{+}!*/   if ((gcm_ctx = capac_malloc(sizeof(struct aead_aes_gcm_ctx))) == null)
		return 0;
    // ... 
	gcm_ctx->ctr =  /*!\hfill\action{PTR-Sign}!*/
        aes_gcm_set_key(&gcm_ctx->ks.ks, &gcm_ctx->gcm, key, key_len);
	ctx->aead_state = gcm_ctx; /*!\hfill\action{PTR-Sign}!*/
    // ...
/*!\textcolor{magenta}{+}!*/   capac_delegate_ptr(ctx->aead_state, &ctx->aead_state,
        sizeof(struct aead_aes_gcm_ctx), dom_session, session_id);
/*!\textcolor{magenta}{+}!*/   capac_delegate_ptr(gcm_ctx->gcm.key, &gcm_ctx->gcm.key,
        sizeof(gcm_ctx->gcm.key), dom_session, session_id);
    return 1;
}
\end{lstlisting}
    \caption{Modifications for private key and session key isolation in LibreSSL.}
    \label{fig:libre}
\end{figure}
\vspace*{\fill}
\end{document}